\documentclass[aps,prc,superscriptaddress,showpacs,twocolumn,floatfix]{revtex4}

\usepackage{graphicx}
\usepackage{amsmath}

\def\gtorder{\mathrel{\raise.3ex\hbox{$>$}\mkern-14mu
 \lower0.6ex\hbox{$\sim$}}}
\def\ltorder{\mathrel{\raise.3ex\hbox{$<$}\mkern-14mu
 \lower0.6ex\hbox{$\sim$}}}

\def\etal{\textit{et al.}}

\begin{document}

\title{Phenomenological extraction of two-photon exchange amplitudes from elastic electron-proton scattering cross section data} 

\author{I.~A.~Qattan}
\affiliation{Khalifa University of Science and Technology, Department
of Physics, P.O. Box 127788, Abu Dhabi, UAE}

\date{\today} 


\begin{abstract}
\begin{description}

\item[Background:] 
The inconsistency in the results obtained from the Rosenbluth separation method 
and the high-$Q^2$ recoil polarization results on the ratio $\mu_p G_E^p/G_M^p$ implies a systematic difference 
between the two techniques. Several studies suggested that missing higher order radiative corrections to
elastic electron-proton scattering cross section $\sigma_R(\varepsilon,Q^2)$ and in particular hard 
two-photon-exchange (TPE) effect contributions could account for the discrepancy.  

\item[Purpose:] 
In this work, I improve on and extend to low- and high-$Q^2$ values the extractions of the $\varepsilon$ 
dependence of the real parts of the TPE amplitudes, relative to the magnetic form factor, as well 
as the ratio $P_{l}/P_{l}^{\mbox{Born}}(\varepsilon,Q^2)$ using world data on $\sigma_R(\varepsilon,Q^2)$ 
with an emphasis on precise new data covering the low-momentum region which is sensitive to the large-scale 
structure of the nucleon.

\item[Method:]
I combine cross section and polarization measurements of elastic electron-proton scattering to extract 
the TPE amplitudes. Because the recoil polarization data were confirmed {\emph{``experimentally''}} to 
be essentially independent of $\varepsilon$, I constrain the ratio $P_{t}/P_{l}(\varepsilon,Q^2)$ to its 
$\varepsilon$-independent term (Born value) by setting the TPE contributions to zero.
That allows for the amplitude $Y_M(\varepsilon,Q^2)$ and $\sigma_R(\varepsilon,Q^2)$ to be expressed 
in terms of the remaining two amplitudes $Y_E(\varepsilon,Q^2)$ and $Y_3(\varepsilon,Q^2)$ 
which in turn were parametrized as second-order polynomials in $\varepsilon$ and $Q^2$ to reserve as possible
the linearity of $\sigma_R(\varepsilon,Q^2)$ as well as to account for possible nonlinearities in the TPE 
amplitudes. Further, I impose the Regge limit which ensures the vanishing of the TPE contributions to 
$\sigma_R(\varepsilon,Q^2)$ and the TPE amplitudes in the limit $\varepsilon \rightarrow 1$.

\item[Results:]
I provide simple parametrizations of the TPE amplitudes, along with an estimate of the fit 
uncertainties. The extracted TPE amplitudes are compared to previous phenomenological extractions and TPE 
calculations. The $P_{l}/P_{l}^{\mbox{Born}}$ ratio is extracted using the new parametrizations of the TPE amplitudes 
and compared to previous extractions, TPE calculations, and direct measurements at $Q^2 =$ 2.50 (GeV/c)$^2$.  

\item[Conclusions:]
The extracted TPE amplitudes are on the few-percentage-points level, and behave roughly linearly 
with increasing $Q^2$ where they become nonlinear at high $Q^2$. On the contrary to $Y_M$ which is 
influenced mainly by elastic contributions, I find $Y_E$ to be influenced by inelastic contributions at large $Q^2$ 
values. While $Y_E$ and $Y_3$ differ in magnitude, they have opposite sign and tend to partially cancel each other. 
This suggests that the TPE correction to $\sigma_R(\varepsilon,Q^2)$ is driven mainly by $Y_M$ and to a lesser 
extent by $Y_3$ in agreement with previous phenomenological extractions and hadronic TPE calculations.

\end{description}
\end{abstract}


\maketitle

\section{Introduction}

The electromagnetic form factors of the nucleon $G_{E}^{(p,n)}$ and $G_{M}^{(p,n)}$, known as the Sachs form factors 
\cite{sachs62}, are fundamental quantities in nuclear and elementary particle physics since they 
are used to parameterize the internal structure of the nucleon. Such a parameterization is our way to describe the 
deviation of the nucleon from the point-like particle picture. Clearly, precise knowledge of these form factors is important since 
they are an essential ingredient in many calculations and analyses. The Sachs 
form factors are functions of the four-momentum transfer squared, $Q^2$, only. They can be expressed in terms of the 
Dirac, $F_{1}^{(p,n)}(Q^2)$, and Pauli, $F_{2}^{(p,n)}(Q^2)$, form factors as
\begin{eqnarray} \label{eq:SachsFFs}
G_{E}^{(p,n)}(Q^2) = F_{1}^{(p,n)}(Q^2) - \tau F_{2}^{(p,n)}(Q^2),\nonumber \\
G_{M}^{(p,n)}(Q^2) = F_{1}^{(p,n)}(Q^2) + F_{2}^{(p,n)}(Q^2),~
\end{eqnarray}
where $\tau = Q^2/4M_N^2$, and $M_N$ is the mass of the nucleon. 

In elastic electron-proton scattering, the reduced cross section $\sigma_{R}$, which is simply the measured differential cross section multiplied by 
a kinematic factor, can be expressed in terms of $G_E^p$ and $G_M^p$ as
\begin{equation} \label{eq:reduced1}
\sigma_{R} = \frac{d\sigma}{d\Omega} \frac{(1+\tau)\varepsilon}{\tau \sigma_{ns}} = \Bigg[(G_M^{p}(Q^2))^2 + \frac{\varepsilon}{\tau} (G_{E}^{p}(Q^2))^2 \Bigg],
\end{equation}
where $\sigma_{ns}$ is the non-structure cross section, $\varepsilon$ is the virtual photon longitudinal polarization parameter defined as $\varepsilon^{-1} = \Big[ 1 + 2 (1+\tau) \tan^{2}({\frac{\theta_{e}}{2}})\Big]$, and $\theta_{e}$ is the electron scattering angle.

Equation (\ref{eq:reduced1}) is known as the Rosenbluth formula~\cite{rosenbluth50} or the 
Longitudinal-Transverse (LT) separation method. It is derived in the Born approximation based on the 
assumption that the electron and proton interact through the exchange of one photon (OPE). 
The reduced cross section $\sigma_{R}$ is measured at several $\varepsilon$ points for a fixed $Q^{2}$, 
and a linear fit of $\sigma_{R}$ to $\varepsilon$ gives $(G_M^{p})^{2}$ as the intercept and 
$(G_{E}^{p})^{2}/\tau$ as the slope allowing for the ratio $R= G_E^p/G_M^p$ to be determined
at that fixed $Q^{2}$ point.

The ratio $G_E^p/G_M^p$ can also be determined using 
the recoil polarization or polarized target (PT) method~\cite{dombey69,akhiezer74,arnold81}, which requires measurement of
the spin-dependent cross section. A longitudinally polarized beam of electrons is scattered elastically, 
transferring their polarization to the unpolarized protons (target). The two polarization transfer observables 
of interest here are the transverse, $P_{t}$, and longitudinal, $P_{l}$, components of the transfered polarization. 
The normal component, $P_{n}$, does not exist in elastic scattering in OPE. 
Simultaneous measurements of $P_{t}$ and $P_{l}$ allows for the determination of the ratio $R$ as
\begin{equation} \label{eq:ratio}
R = \frac{G_E^p}{G_M^p} = - \frac{P_{t}}{P_{l}} \frac{(E+E')}{2M_{p}} \tan({\frac{\theta_{e}}{2}}),
\end{equation}
where $E$, $E'$, and $\theta_{e}$ are the incident energy, final energy, and scattered angle of
the electron.

The two methods yield strikingly different results~\cite{qattan05,arrington07a,perdrisat07}, 
with values of $\mu_{p}G_E^p/G_M^p$ differing almost by a factor of three at high $Q^2$. Here 
$\mu_p$ is the magnetic moment of the proton. In the LT separation
method, the ratio shows approximate form factor scaling, $\mu_p G_E^p/G_M^p
\approx 1$, albeit with large uncertainties at high $Q^2$ values. The recoil
polarization method yields a ratio that decreases roughly
linearly with increasing $Q^2$, with some hint of flattening out above 5 (GeV/c)$^2$.

\section{Two-Photon Exchange Contributions}  \label{visit_interference}

To reconcile these measurements, several studies suggested that missing higher order radiative corrections to
the electron-proton elastic scattering cross sections, in particular two-photon exchange (TPE) diagrams,
may explain the discrepancy~\cite{guichon03, arrington03a, arrington04a}. 
We account for TPE contributions to $\sigma_{R}$ by adding the real function $F(\varepsilon,Q^2)$ to the Born
reduced cross section
\begin{equation} \label{eq:reduced3}
\sigma_{R} = \big(G_M^p\big)^2\big[1 + \frac{\varepsilon}{\tau} R^2\big] + F(\varepsilon,Q^2),
\end{equation}
where $R=G_E^p/G_M^p$ is the recoil polarization ratio.
The role of TPE effects was
studied extensively both theoretically~\cite{blunden03,blunden05a,kondratyuk05,kondratyuk07,kivel09,kivel11,kivel13,tomalak15a,tomalak15b,lorenz15,chen04,afanasev05,bystritskiy07,gustafsson05,borisyuk06,borisyuk07,borisyuk08,borisyuk09,borisyuk12,borisyuk14,borisyuk15,zhou07,zhou09,zhou15,graczyk15,graczyk13,graczyk11,braun06} and phenomenologically~\cite{guichon03,qattanphd,tvaskis06,borisyuk07b,borisyuk11,chen07,arrington05,guttmann11,rekalo04,qattan11a,qattan11b,qattan12,qattan14,qattan15}. Most calculations suggested that the TPE corrections
are relatively small, but have a significant angular dependence which mimics the effect of a larger
value of $G_E^p$. Detailed reviews of the role of the TPE effect in electron-proton scattering can be
found in~\cite{arrington11b,carlson07}.

Experimentally, several measurements were performed to verify the discrepancy~\cite{christy04,qattan05}
and to try and measure or constrain TPE contributions. Precise examinations of the $\varepsilon$
dependence of $\sigma_{R}$~\cite{qattanphd,tvaskis06,chen07} found no deviation from the linear
behaviour predicted in the OPE approximation. Another measurement was performed to look for TPE
effects by extracting $\mu_{p}G_E^p/G_M^p$ at fixed $Q^2$ as a function of scattering angle~\cite{meziane11}. In
the Born approximation, the result should be independent of scattering angle, and no deviation from the
OPE prediction was observed.

Based on the observations above, it is possible to try and extract the TPE contributions based on the
observed discrepancy between the LT and PT results. Assuming that the TPE contributions are linear in
$\varepsilon$ and that the PT results do not depend on $\varepsilon$, and knowing that the TPE
contribution must vanish in the forward limit ($\varepsilon \to 1$)~\cite{chen07,rekalo04}, it is possible to
extract the TPE contribution to the unpolarized cross section in a combined analysis of LT and PT
data~\cite{guichon03,qattanphd,tvaskis06,borisyuk07b,borisyuk11,chen07,arrington05,guttmann11,rekalo04,qattan11a,qattan11b,qattan12,qattan14,qattan15}. Where polarization data exist as a
function of $\varepsilon$, it is possible to attempt to extract the TPE amplitudes with fewer
assumptions~\cite{guttmann11,borisyuk11}, though with relatively large uncertainties.

The most direct technique for measuring TPE is the comparison of electron-proton and positron-proton
scattering. The is done by measuring the ratio $R_{e^{+} e^{-}}(\varepsilon,Q^2) = \frac{1-\delta_{2\gamma}}{1+\delta_{2\gamma}} \approx 1-2\delta_{2\gamma}$ after correcting for the electron-proton Bremsstrahlung interference term 
and the conventional charge-independent radiative corrections. Here $\delta_{2\gamma}$ is the fractional TPE 
correction for electron-proton scattering. The leading TPE contribution comes from the interference of the OPE 
and TPE amplitudes, and so has the opposite sign for positron and electron scattering. The only other first-order 
radiative correction which depends on the lepton sign is the interference between diagrams with Bremsstrahlung 
from the electron and proton, and this contribution is generally small. Thus, after correcting the measured ratio 
for the Bremsstrahlung interference term, the comparison of positron and electron scattering allows for the most
direct measurement of  TPE contributions. Until recently, there was only limited evidence for any non-zero TPE 
contribution from such comparisons~\cite{arrington04b}, as data were limited to low $Q^2$ or large $\varepsilon$, 
where the TPE contributions appear to be small.  In addition, the details of the radiative corrections applied to 
these earlier measurements are not always available, and it is not clear if the charge-even corrections were 
applied in all cases.  New measurements~\cite{adikaram15,rachek15} have found more significant indications of 
TPE contributions at low $\varepsilon$ and moderate $Q^2$, which are consistent with hadronic TPE 
calculations~\cite{blunden05a}.

To account for the exchange of two or more photons, 
Guichon and Vanderhaeghen~\cite{guichon03} expressed the hadronic vertex
function in terms of three independent complex amplitudes (generalized form factors) which depend on both 
$Q^2$ and $\varepsilon$: $\tilde{G}_E^p(\varepsilon,Q^2)$, $\tilde{G}_M^p(\varepsilon,Q^2)$, and $\tilde{F_{3}}(\varepsilon,Q^2)$. These generalized form factors can be broken into the usual Born (OPE) and the TPE contributions as: 
$\tilde{G}_{E,M}^p(\varepsilon,Q^2)= G_{E,M}^p(Q^2) + \delta G_{E,M}^p(\varepsilon,Q^2)$ with $Y_{2\gamma}(\nu,Q^2)$ 
defined as $\Re \Big( \frac{\nu \tilde{F_{3}}} {M^2_{p}|G_M^p|} \Big)$, where $\Re$ stands for the real part, and 
$\nu =M^2_{p}\sqrt{(1+\varepsilon)/(1-\varepsilon)} \sqrt{\tau(1+\tau)}$. In the Born approximation,
$\tilde{G}_{E,M}^p(\varepsilon,Q^2)=G_{E,M}^p(Q^2)$, which are the real electric and magnetic Sachs form factors and
$\tilde{F_{3}}=0$. With the inclusion of the TPE effects, the reduced cross section can be expressed as
\begin{equation} \label{eq:Guichon1}
\sigma_{R} = |\tilde{G}_M^p|^2 \Bigg[ 1 + \frac{\varepsilon}{\tau} \frac{|\tilde{G}_E^p|^2}{|\tilde{G}_M^p|^2} + 2\varepsilon \Bigg(1 + \frac{|\tilde{G}_E^p|}{\tau |\tilde{G}_M^p|}\Bigg) Y_{2\gamma} \Bigg],
\end{equation}
and the ratio of the transverse, $P_{t}$, to the longitudinal, $P_{l}$, components of the recoil proton polarization as
\begin{equation} \label{eq:Guichon2}
\frac{P_{t}}{P_{l}} = - \sqrt{\frac{2\varepsilon}{\tau(1+\varepsilon)}} \Bigg[ \frac{|\tilde{G}_E^p|}{|\tilde{G}_M^p|} + \Bigg(1 - \frac{2\varepsilon}{1+\varepsilon} \frac{|\tilde{G}_E^p|}{|\tilde{G}_M^p|} \Bigg) Y_{2\gamma}\Bigg],     
\end{equation}
allowing for the proton form factor ratio to be written for the L-T separation (Rosenbluth) as
\begin{equation} \label{eq:Guichon3}
R_{LT}^2 = \Bigg(\frac{|\tilde{G}_E^p|}{|\tilde{G}_M^p|}\Bigg)^2 + 2\Bigg(\tau + \frac{|\tilde{G}_E^p|}{|\tilde{G}_M^p|}\Bigg) Y_{2\gamma},
\end{equation} 
and for the recoil-polarization as
\begin{equation} \label{eq:Guichon4}
R_{PT} = \frac{|\tilde{G}_E^p|}{|\tilde{G}_M^p|} + \Bigg(1 - \frac{2\varepsilon}{1+\varepsilon} \frac{|\tilde{G}_E^p|}{|\tilde{G}_M^p|}\Bigg) Y_{2\gamma}.
\end{equation}

Phenomenologically and based on the formalism of Guichon and Vanderhaeghen, Eqs.~(\ref{eq:Guichon1}) and (\ref{eq:Guichon2}), Guttmann and collaborators \cite{guttmann11} determined the $\varepsilon$ dependence of the TPE amplitudes around $Q^2=$ 2.50 (GeV/c)$^2$ using the high precision data on the ratios $-\mu_{p}\sqrt{\tau(1+\varepsilon)/(2\varepsilon)} P_{t}/P_{l}$ and $P_{l}/P_{l}^{\mbox{Born}}$ determined by the GEp-2$\gamma$ experiment 
\cite{meziane11}, and the reduced cross sections $\sigma_{R}$ from the Super-Rosenbluth experiment \cite{qattan05}. 
For convenience, the reduced cross section $\sigma_{R}/G^2_{Mp}$, the ratio $-\mu_{p}\sqrt{\tau(1+\varepsilon)/(2\varepsilon)} P_{t}/P_{l}$, 
and the ratio $P_{l}/P_{l}^{\mbox{Born}}$ were expressed in terms of the ratio $G_E^p/G_M^p$ and the real parts of the TPE amplitudes relative to the magnetic form factor or: $Y_{M}(\varepsilon,Q^2) = \Re (\delta \tilde{G}_M^p/G_M^p)$, 
$Y_{E}(\varepsilon,Q^2) = \Re (\delta \tilde{G}_E^p/G_M^p)$, and $Y_{3}(\varepsilon,Q^2) = (\nu/M^2_{p}) \Re (\tilde{F_{3}}/G_M^p)$ as

\begin{subequations}
\begin{eqnarray} \label{eq:Guttmann1}
\frac{\sigma_{R}}{(G_M^p)^2} = 1+ \frac{\varepsilon}{\tau} \Big(\frac{G_E^p}{G_M^p}\Big)^2 + 2Y_M 
+ \frac{2\varepsilon}{\tau} \frac{G_E^p}{G_M^p} Y_E \nonumber \\
+ 2\varepsilon \Big(1+\frac{G_E^p}{\tau G_M^p}\Big)Y_3 + O(e^4), \\
-\sqrt{\frac{\tau(1+\varepsilon)}{2\varepsilon}} \frac{P_{t}}{P_{l}} = \frac{G_E^p}{G_M^p} + Y_E - \frac{G_E^p}{G_M^p} Y_M~~~~~~~ \nonumber\\
+ \Big(1-\frac{2\varepsilon}{1+\varepsilon} \frac{G_E^p}{G_M^p}\Big)Y_3 + O(e^4),\\
\frac{P_{l}}{P_{l}^{\mbox{Born}}} = 1-2\varepsilon \Big(1+ \frac{\varepsilon}{\tau} \Big(\frac{G_E^p}{G_M^p}\Big)^{2}\Big)^{-1}~~~~~~~ \nonumber\\  
\times \Big\lbrace \Big[\frac{\varepsilon}{1+\varepsilon}\Big(1- \frac{1}{\tau}\Big(\frac{G_E^p}{G_M^p}\Big)^{2}\Big) +
\frac{G_E^p}{\tau G_M^p}\Big]Y_3 \nonumber \\
+ \frac{G_E^p}{\tau G_M^p}\Big[Y_E - \frac{G_E^p}{G_M^p}Y_M\Big]\Big\rbrace + O(e^4).
\end{eqnarray}
\end{subequations}

The ratio 
$-\mu_{p}\sqrt{\tau(1+\varepsilon)/(2\varepsilon)} P_{t}/P_{l}$ does not show any $\varepsilon$ dependence and a fit in the
form of: $\mu_{p} R + B\varepsilon^{c}(1-\varepsilon)^{d}$ was performed. Here $\mu_{p} R = \mu_{p}G_E^p/G_M^p$ is 
the Born value (OPE), with $B$, $c$, and $d$ being constants. For several values of $c$, and $d$, the constant $B$
was effectively zero, and the ratio $P_{t}/P_{l}$ was fit to its Born value of
$\mu_{p} R = \mu_{p}G_E^p/G_M^p = 0.693 \pm 0.006_{\mbox{stat}} \pm 0.010_{\mbox{sys}}$. The ratio $P_{l}/P_{l}^{\mbox{Born}}$ shows a decrease for
$\varepsilon \rightarrow$ 0. Although in qualitative agreement with perturbative QCD (pQCD) \cite{borisyuk09,kivel09}, the ratio 
$P_{l}/P_{l}^{\mbox{Born}}$ at $Q^2=$ 2.50 (GeV/c)$^2$ falls-off faster than the pQCD prediction. Therefore, the ratio 
$P_{l}/P_{l}^{\mbox{Born}}$ was fit to two different functional forms: $P_{l}/P_{l}^{\mbox{Born}} = 1 + a_{l}\varepsilon^{4}(1-\varepsilon)^{1/2}$ (Fit I), and $P_{l}/P_{l}^{\mbox{Born}} = 1 + a_{l}\varepsilon~\mbox{ln}(1-\varepsilon)(1-\varepsilon)^{1/2}$ (Fit II), giving
a value of $a_{l} = 0.11 \pm 0.03_{\mbox{stat}} \pm 0.06_{\mbox{sys}}$ for (Fit I), and $a_{l} = -0.032 \pm 0.008_{\mbox{stat}} \pm 0.020_{\mbox{sys}}$
for (Fit II). To find the $\varepsilon$ dependence of $\sigma_{R}$ at $Q^2=$ 2.64 (GeV/c)$^2$ and due to the experimentally 
observed linearity of the Rosenbluth plots, a fit of $\sigma_{R}/(\mu_{p}G_{D})^2$ to $\varepsilon$ using the form $(a + b \varepsilon)$ was done. Here $G_D(Q^2)$ is the dipole parametrization: $G_D(Q^2)=[1+Q^2/(0.71(\mbox{GeV/c})^2)]^{-2}$.
This fit yields a value of $a = 1.106 \pm 0.006$ and
$b = 0.160 \pm 0.009$. Using the assumption that for $\varepsilon \rightarrow$ 1, Regge limit, the TPE correction to 
$\sigma_{R}$ vanishes, $\sigma_{R}$ can now be expressed in the OPE approximation as: $\sigma_{R}/(\mu_{p}G_{D})^2 = [(G_M^p)^2 + (G_E^p)^2/\tau]/(\mu_{p}G_{D})^2 = (a + b)$. Using the obtained values of $a$, $b$, and 
$\mu_{p} R = \mu_{p}G_E^p/G_M^p$, the ratio $(G_M^p/\mu_{p}G_{D})^2$ was found to be $(G_M^p/\mu_{p}G_{D})^2 = 1.168 \pm 0.010$. 

Using the $\varepsilon$-dependence forms obtained for $\sigma_{R}/(G_M^p)^2$, $-\mu_{p}\sqrt{\tau(1+\varepsilon)/(2\varepsilon)} P_{t}/P_{l}$, and $P_{l}/P_{l}^{\mbox{Born}}$ along with the values extracted for
$(G_M^p/\mu_{p}G_{D})^2$ and $\mu_{p} R = \mu_{p}G_E^p/G_M^p$ in Eqs.~(9), the three TPE amplitudes $Y_{M}$, $Y_{E}$, and $Y_{3}$ were extracted as a function of 
$\varepsilon$ at $Q^2=$ 2.50 (GeV/c)$^2$. The amplitude $Y_{M}$ is on the 
few percent level and is mainly driven by the TPE effect in $\sigma_{R}$ which is to a good approximation given by 
$\sigma^{2\gamma}_{R} \sim (Y_{M} + \varepsilon Y_{3})$. The amplitude $Y_{M}$ rises approximately linearly in 
$\varepsilon$ and starts showing deviation from linearity as $\varepsilon \rightarrow$ 1. 
On the other hand, the amplitudes $Y_{E}$ and $Y_{3}$ are mainly driven by the polarization data. They are on the 2-3\% level and have opposite signs in the region constrained by the polarization data ($\varepsilon \geq 0.6$). 
Therefore, they tend to partially cancel each other in the polarization transfer ratio. The leading TPE
correction to the ratio $P_{t}/P_{l}$ is approximately given by $(P_{t}/P_{l})^{2\gamma} = (Y_{E} + Y_{3})$. Finally,
the amplitude $Y_{3}$ is driven by the $P_{l}$ data and the TPE correction to $P_{l}$ is given by 
$P_{l}^{2\gamma} \approx -2\varepsilon^2/(1+\varepsilon)Y_{3}$. 

Based on the framework of~\cite{guichon03}, Arrington~\cite{arrington05}
performed a global analysis where he extracted the TPE amplitudes $\Delta
G_{E,M}^p$ and $Y_{2\gamma}$.  He assumed that the amplitudes were
$\varepsilon$-independent and took $\Delta G_E^p=0$. Values for
$Y_{2\gamma}(Q^2)$ were extracted from the difference between polarization and
Rosenbluth measurements, taking into account the uncertainties in both data
sets. Based on the high-$\varepsilon$ constraints from the comparison of
positron and electron scattering~\cite{arrington04b} the amplitude $\Delta
G_{Mp}$ was determined by requiring that its contribution to $\sigma_{R}$ at
$\varepsilon=1$ cancelled the contribution of $Y_{2\gamma}$.  The extracted
TPE amplitudes and their estimated uncertainties are then parametrized as a
function of $Q^2$, and used to apply TPE corrections to the form factors
obtained from a global Rosenbluth analysis~\cite{arrington04a} and the new
recoil polarization data.  
The $Y_{2\gamma}$ amplitude was determined in the range of 0.6 $\le Q^2 \le$ 6 (GeV/c)$^2$ 
and then parametrized according to: $Y_{2\gamma}=0.035\big[1-\exp(-Q^2/1.45)\big]$. 
In contrast to~\cite{guichon03}, the true ratio $G_{E}^p/G_{M}^p$ obtained was actually 
corrected for all TPE amplitudes $\delta G_{E,M}^p$ and $Y_{2\gamma}$ and it was well 
parametrized by $\mu_{p}G_{E}^p/G_{M}^p = (1-0.158Q^2)$.

In both of these analysis, there is not enough information to directly
determine the amplitudes, so assumptions have to be made about the relative
importance and the $\varepsilon$ dependence of the three TPE amplitudes.
Common to these analyses are the assumption that the correction is close to
linear in $\varepsilon$, as no non-linearities have been
observed~\cite{qattanphd, tvaskis06, gustafsson05, chen07}.  If one also
neglects the TPE correction to the polarization data, which is significantly
smaller at high $Q^2$, then it is not necessary to work in terms of the
polarization amplitude; one can simply parametrize the TPE contributions
to the reduced cross section, taking a linear (or nearly linear) $\varepsilon$
dependence.

The analysis of Borisyuk and Kobushkin~\cite{borisyuk11} takes a similar
approach, although they use a different linear combination of amplitudes than
Ref.~\cite{guichon03}.  Again, the $\varepsilon$ dependence of $P_t/P_l$ is
taken to be zero~\cite{meziane11}, and the correction to the cross section
is taken to be linear~\cite{tvaskis06}. Data from available electron-proton
scattering cross sections in the range of 2.20 $\le Q^2 \le$ 2.80 (GeV/c)$^2$
were interpolated to $Q^2 =$ 2.50~(GeV/c)$^2$ to extract the amplitudes at a fixed
$Q^2$ value.  This yields an extraction in terms of a single amplitude.

Theoretically, different approaches were adapted to calculate TPE corrections to electron-proton scattering observables such as: 
the hadronic~\cite{guichon03,blunden03,blunden05a,kondratyuk05,kondratyuk07,tomalak15b,lorenz15,borisyuk06,borisyuk07,borisyuk12,borisyuk14,borisyuk15,zhou07,zhou09,zhou15}, partonic models (GPDs)~\cite{chen04,afanasev05}, dispersion relations (DRs)~\cite{borisyuk08,tomalak15a}, and pQCD~\cite{kivel09,kivel11,kivel13,braun06,borisyuk09} 
based calculations. 
However, calculations of the TPE amplitudes or generalized form factors were done mainly using hadronic- and  
pQCD-based calculations. In the hadronic approach, which is mainly valid at small to moderate $Q^2$ values, 
the TPE processes are mediated by the production of 
virtual hadrons and/or hadronic resonances in the intermediate state. Therefore, the TPE amplitudes 
can be broken down into contributions according to the hadronic intermediate state involved such as: 
the elastic contributions where only a pure nucleon is considered, and inelastic contributions coming from multi-particle 
states such as $p \pi$, $p \pi \pi$, and in particular, emphasis was placed on hadronic intermediate states containing the prominent $\Delta(1232)$ 
resonance, Ropper resonance, and $\pi N$ (pion + nucleon). The $\pi N$ contribution can be 
further split into contributions coming from different partial waves or channels such as the $P_{33}$ channel 
where the $\Delta(1232)$ resonance resides ($\pi N$ intermediate state with quantum numbers of the $\Delta(1232)$ resonance) as it has 
100\% $\pi N$ content, and contributions from higher total angular momentum of $J =$ 1/2 and 3/2 
(eight different channels). However, these contributions are to be viewed as the first terms of an infinite expansion of the 
total $\pi N$ contribution in the intermediate state as it is not clear when such a series will eventually converge.  
The second is the quark approach where the nucleon is treated as system of interacting quarks (partons) 
in the intermediate state with their interactions described by perturbative Quantum Chromodynamics (pQCD).
       
The most important and well-studied contribution in the hadronic-type approach is the elastic contribution 
which influences mainly the magnetic form factor. Kondratyuk~\emph{et al.} calculated the effect of adding 
the $\Delta(1232)$ resonance~\cite{kondratyuk05} and other several light resonances~\cite{kondratyuk07} 
on the cross section. The overall effect is smaller than the elastic contribution
with the $\Delta(1232)$ resonance yielding the largest contribution, and the contributions from the other resonances 
partially cancelling each other. 

Borisyuk and Kobushkin~\cite{borisyuk12} evaluated the $\Delta(1232)$ resonance
contribution to both the cross section and the TPE amplitudes. The $\Delta(1232)$ resonance affected mainly
the generalized electric form factor ($\delta G_E/G_M$), while the elastic contribution influenced mainly the 
magnetic form factor ($\delta G_M/G_M$). The effect of the $\Delta(1232)$ resonance was found to grow in magnitude 
with $Q^2$ where it became sizable and far exceeded that of the elastic contribution at large $Q^2$. 
This implies a relatively large correction to the recoil polarization ratio $\mu_p G_E^p/G_M^p$ at high $Q^2$. The total correction
to the recoil polarization ratio $\delta R$ at high $Q^2$, $Q^2 >$ 3.0 (GeV/c)$^2$), including both the elastic 
and $\Delta$ contributions or $\delta R = (\delta^{el} + \delta^{\Delta})$ was found to be much larger than the 
experimentally quoted systematic uncertainty. When the correction $\delta R$ is applied to polarization measurements, the ratio $R$ becomes negative at $Q^2 =$ 8.5 (GeV/c)$^2$. 

The two hadronic calculations of the 
$\Delta(1232)$ resonance contribution discussed above assumed zero-width for the resonances (widths assumed 
negligibly small). Such an assumption is rather strong as the width of the $\Delta(1232)$ resonance 
($\Gamma_\Delta \approx$ 110 MeV) is comparable to the distance from threshold ($M_\Delta \approx$ 160 MeV). 

Borisyuk and Kobushkin~\cite{borisyuk14} evaluated the effect of the $\pi N$ (pion + nucleon) hadronic intermediate state with emphasis on 
the $P_{33}$ channel where the $\Delta(1232)$ resonance resides. They included a realistic resonance width and shape 
and corresponding background. While the $\Delta(1232)$ resonance contribution is negligible compared to the elastic 
one at low $Q^2$, the correction to the $\delta G_E/G_M$ amplitude is large and grows rapidly in magnitude for 
$Q^2 \geq 2.50$ exceeding that of the elastic intermediate state in agreement with their previous results
which assumed zero-width~\cite{borisyuk12}. Consequently, the recoil polarization ratio $\mu_p G_E^p/G_M^p$ is 
affected at high $Q^2$. However, the magnitude of the correction is $\sim$ 30\% smaller than that obtained 
assuming zero-width. 

Calculations of the TPE amplitudes taking into account hadronic intermediate state containing the
$\pi N$ system with higher total angular momentum ($J =$ 1/2 and 3/2 for eight $\pi N$ different channels) 
assuming finite resonance width, realistic resonance shape and form factors, as well as nonresonant background 
were also performed~\cite{borisyuk15}. It was found that the largest contributions came from the channels with quantum 
numbers of the lightest resonances dominated mainly by the contribution of the $P_{33}$ channel. On the other hand, 
the correction to the recoil polarization ratio $\delta R$ at high $Q^2$ is smaller 
but still sizable and grows roughly linearly with $Q^2$ due to a cancellation between the different channels 
contributions.  

Zhou and Yang~\cite{zhou09,zhou15} calculated the $\Delta(1232)$ 
resonance contribution to $\sigma_R$ where they used a correct vertex function for 
$\gamma N \rightarrow \gamma N \Delta$, realistic $\gamma N \Delta$ form factors, and coupling constants. 
In their calculations, the TPE correction $\delta_\Delta$ showed a rising and decreasing behaviour as 
$\varepsilon \rightarrow$ 1 in disagreement with predictions of other hadronic calculations which 
suggested the vanishing of the correction in the limit $\varepsilon \rightarrow$ 1. Such disagreement was 
attributed to the asymptotic behaviour of the TPE amplitudes at $s \rightarrow \infty$ which was assumed to 
vanish in other calculations. While their calculations agreed reasonably well with 
measured $\sigma_R$ from Ref.~\cite{andivahis94}, substantial discrepancy remained when $\sigma_R$ data 
from Refs.~\cite{walker94,qattan05} were used. They concluded that their model should be restricted 
to low energy, $W \leq$ 3--4 GeV, and to low $Q^2$ experimental data in that energy range. 

Lorenz~\emph{et al.}~\cite{lorenz15} calculated TPE corrections to $\sigma_R$ including elastic and $\Delta$-resonance 
intermediate states using phenomenological information on the vertices. They included Coulomb contribution and 
updated photocoupling values for the $\gamma N \delta$-vertices, and used data on the $Q^2$ dependence of the 
nucleon-$\Delta$ transition from electroproduction of nucleon resonances in terms of helicity amplitudes. 
The results showed strong dependence on the choice of elastic nucleon and nucleon-$\Delta$ transition form 
factors used as input, in disagreement with previous calculations. At low $Q^2$, the $\Delta$ contribution is 
much smaller than the elastic contribution. On the other hand, the contribution at high $Q^2$ is 
comparable to that of the elastic.  

At high $Q^2$ values, the hadronic approach becomes inadequate and the TPE corrections are 
calculated mainly within the framework of GPDs~\cite{chen04,afanasev05} and pQCD~\cite{kivel09,kivel11,kivel13,braun06,borisyuk09}. Afanasev \emph{et al.}~\cite{afanasev05} calculated 
TPE corrections to $\sigma_R$ using formalism of GPDs. The authors doubt the applicability of pQCD for 
the currently existing data. Borisyuk and Kobushkin~\cite{borisyuk09} calculated TPE corrections to $\sigma_R$
(the $\delta G_M/G_M$ amplitude) within the framework of pQCD for the proton target using wave functions based on
QCD sum rules. The $\delta G_M/G_M$ amplitude has linear $\varepsilon$ dependence, and grows logarithmically
with $Q^2$ reaching 3.5\% of the Born amplitude at $Q^2 =$ 30 (GeV/c)$^2$. At lower $Q^2$, a smooth connection 
with hadronic calculations assuming an elastic intermediate state is possible. However, at high $Q^2$, the two calculations 
yield different results suggesting the inadequacy of the hadronic approach for $Q^2 \geq$ 3.0 (GeV/c)$^2$. Kivel and
Vanderhaeghen~\cite{kivel13} calculated TPE corrections to $\sigma_R$ at moderately large $Q^2$ values of 2.64, 3.20, 
and 4.10 (GeV/c)$^2$ using a QCD factorization approach within the framework of the soft-collinear effective theory 
(SCET) arising from both the soft- and hard-spectator scattering contributions. The TPE corrections to $\sigma_R$ are 
linear in $\varepsilon$ but small in magnitude. However, for $Q^2 >$ 2.5--3.0 (GeV/c)$^2$, the description of 
$\sigma_R$ as a linear function in $\varepsilon$ for the whole $\varepsilon$ range is no longer valid as deviation 
from linearity at small $\varepsilon$ is seen at large $Q^2$. While the effect of nonlinearity is relatively small 
at $Q^2 =$  2.64 (GeV/c)$^2$, it is sizable at $Q^2 =$ 4.10 (GeV/c)$^2$. Such observation is also in agreement with 
hadronic-type calculations which reported similar behaviour at moderate $Q^2$ values. The TPE amplitudes $Y_M$ and $Y_3$ 
were also calculated and both showed weak $Q^2$ dependence with opposite $\varepsilon$ dependence. 
The absolute values of the amplitudes are much smaller than those extracted phenomenologically in 
Ref.~\cite{guttmann11} and with non-vanishing value as $\varepsilon \rightarrow$ 1. 
The ratio $P_l/P_{l}^{\mbox{Born}}$ is dominated by the soft-spectator contribution and with non-vanishing value 
at $\varepsilon =$ 1.

\section{Extraction of the TPE Amplitudes} \label{extraction}

In this section I discuss the procedure used to extract the three TPE amplitudes $Y_M$, $Y_E$, and $Y_3$ 
(generalized form factors) as a function of $\varepsilon$ at fixed $Q^2$ value based on the formalism of 
Guichon and Vanderhaeghen, Eqs.~(9). The procedure followed 
by Guttmann~\etal~\cite{guttmann11} to extract the $\varepsilon$ dependence of the TPE amplitudes at $Q^2 =$ 2.50 
(GeV/c)$^2$ was outlined in Sec.(\ref{visit_interference}). They used the measurements of 
the $\varepsilon$ dependence of the ratios $\mu_{p} R$ and $P_l/P_{l}^{\mbox{Born}}$ at $Q^2 =$ 2.50 (GeV/c)$^2$ 
from Ref.~\cite{meziane11}, along with the $\varepsilon$ dependence of the cross section at $Q^2=$ 2.64 (GeV/c)$^2$ 
from Ref.~\cite{qattan05} to constrain the three TPE amplitudes. However, the parametrizations of the
$\varepsilon$ dependence of the ratio $P_l/P_{l}^{\mbox{Born}}$ used in their analysis, Fits I and II, are not
well motivated by the experimental data. 

In this article I do the following:

(i) I show in principle that the TPE amplitudes under certain 
assumptions and constraints can be extracted at fixed $Q^2$ value using combined cross section and polarization 
measurements of elastic electron-proton scattering. 

(ii) I extend the previous extractions~\cite{guttmann11}  
to cover both low- and high-$Q^2$ values, and provide simple parametrizations of the TPE amplitudes.
I compare my results to different phenomenological extractions and direct TPE calculations. 

(iii) I use my parametrizations of the TPE amplitudes to calculate the ratio $P_l/P_l^{\mbox{Born}}$ 
for a range of low- and high-$Q^2$ values, and then compare the results to recent fits, theoretical 
calculations, and direct measurements at $Q^2 =$ 2.50 (GeV/c)$^2$. 

The procedure, together with the constraints and assumptions used is outlined below:

(1) I assume that the TPE correction is responsible mainly for the discrepancy between the cross section and 
polarization data measurements.

(2) Because the recoil polarization data were confirmed {\emph{``experimentally''}} to be essentially independent of 
$\varepsilon$, I constrain the ratio $-\sqrt{\tau(1+\varepsilon)/(2\varepsilon)} P_{t}/P_{l}$ in Eq.~(9b)
to its $\varepsilon$-independent term (Born value) or $R = G_{E}^p/G_{M}^p$ by setting the TPE contributions to 
zero. Therefore, I will use the recent improved parametrization of the ratio $R$ along with its associated 
uncertainty~\cite{qattan15} from polarization measurements at both low- and high-$Q^2$ values
\begin{equation} \label{eq:JRA_RptNew}
\mu_p R = \frac{1}{1+0.1430Q^2-0.0086Q^4+0.0072Q^6},
\end{equation} 
with  an absolute uncertainty in the fit given by: $\delta^2_R(Q^2) = \mu_p^{-2}[(0.006)^2 +
(0.015\mbox{ln}(1+Q^2))^2]$, with $Q^2$ in (GeV/c)$^2$. Therefore, the amplitude $Y_M(\varepsilon,Q^2) $ can be 
expressed in terms of the remaining $Y_E(\varepsilon,Q^2)$ and $Y_3(\varepsilon,Q^2)$ amplitudes as
\begin{eqnarray} \label{eq:constrain1}
Y_M(\varepsilon,Q^2)  = \frac{Y_E(\varepsilon,Q^2)  + \Big(1-\frac{2\varepsilon}{1+\varepsilon} R\Big)Y_3(\varepsilon,Q^2) }{R}. 
\end{eqnarray}

(3) Using the constraint on $Y_M(\varepsilon,Q^2)$ from Eq.~(\ref{eq:constrain1}), I can now express $\sigma_{R}/(G_M^p)^2$ in Eq.~(\ref{eq:Guttmann1}) in terms of the $Y_E$ and $Y_3$ amplitudes as
\begin{eqnarray} \label{eq:constrain2}
\frac{\sigma_{R}}{(G_M^p)^2} = 1+ \frac{\varepsilon}{\tau} R^2 +\Big[\frac{2}{R} + \frac{2\varepsilon R}{\tau}\Big]Y_E(\varepsilon,Q^2) \nonumber\\
+\Big[\frac{2}{R}\Big(1-\frac{2\varepsilon R}{1+\varepsilon}\Big) + 2\varepsilon \Big(1+\frac{R}{\tau}\Big)\Big]Y_3(\varepsilon,Q^2).
\end{eqnarray}

(4) Because the TPE amplitudes are functions of $\varepsilon$ and $Q^2$, I expand each 
of the amplitudes $Y_E$ and $Y_3$ as a polynomial of degree $n$ as
\begin{eqnarray} \label{eq:constrain3}
Y_E(\varepsilon,Q^2) = \sum_{k=0}^{n} \alpha_{k}(Q^2)\varepsilon^{k},  \nonumber\\
Y_3(\varepsilon,Q^2) = \sum_{k=0}^{n} \beta_{k}(Q^2)\varepsilon^{k},~
\end{eqnarray}   
where the coefficients $\alpha_{k}$ and $\beta_{k}$ $(k=0,1,\cdots,n)$ are functions of $Q^2$ only.  

(5) Because of the experimentally observed linearity of the Rosenbluth plots~\cite{qattanphd,tvaskis06,gustafsson05,chen07} where 
$\sigma_R$ exhibits no (or weak) nonlinearity in $\varepsilon$, I truncate the series at $n =$ 2 to reserve as possible the linearity  
of $\sigma_R$ as well as to account for any possible nonlinearities in the TPE amplitudes 
\begin{eqnarray} \label{eq:constrain4}
Y_E(\varepsilon,Q^2) = \alpha_{0} + \alpha_{1}\varepsilon + \alpha_{2}\varepsilon^{2},~~~~~~~~~~~~~~~~~~ \nonumber\\
\mbox{and}~~~~~~~~~~~~~~~~~~~~~~~~~~~~~~~~~~~~~~~~~~~~~~~~~~~~~~~~~~~~~~ \nonumber\\
Y_3(\varepsilon,Q^2) = \beta_{0}  + \beta_{1}\varepsilon +  \beta_{2}\varepsilon^{2}.~~~~~~~~~~~~~~~~~~~~
\end{eqnarray} 

(6) Substituting Eqs.~(\ref{eq:constrain4}) in Eq.~(\ref{eq:Guttmann1}), and imposing the Regge limit where  
the TPE correction to $\sigma_{R}$ vanishes in the limit $\varepsilon \rightarrow 1$, I obtain 
$Y_E(1,Q^2)= - Y_3(1,Q^2)$ or simply: $(\alpha_{0} + \alpha_{1}+ \alpha_{2}) = -(\beta_{0} + \beta_{1} + \beta_{2})$. 
Further, to ensure the correct behaviour of the TPE amplitudes as $\varepsilon \rightarrow 1$ where each amplitude
must go to zero (Regge limit), I obtain the following constraints on the coefficients: 
$\alpha_{0} = - (\alpha_{1}+ \alpha_{2})$ and $\beta_{0} = - (\beta_{1} + \beta_{2})$.    

(7) Using the constraints on $\alpha_{0}$ and $\beta_{0}$ derived above, $\sigma_{R}$ can now be expressed as
\begin{eqnarray} \label{eq:constrain5}
\frac{\sigma_{R}}{(G_M^p)^2} = 1+ \frac{\varepsilon}{\tau} R^2 +\Big[\frac{2}{R} + \frac{2\varepsilon R}{\tau}\Big]\Big[\alpha_{1}(\varepsilon-1)+ \alpha_{2}(\varepsilon^2-1)\Big]\nonumber\\
+\Big[\frac{2}{R}\Big(1-\frac{2\varepsilon R}{1+\varepsilon}\Big) + 2\varepsilon \Big(1+\frac{R}{\tau}\Big)\Big]
\Big[\beta_{1}(\varepsilon-1)+\beta_{2}(\varepsilon^2-1)\Big]\nonumber\\.
\end{eqnarray}
with $(G_M^p)^2$, $\alpha_{1}$, $\alpha_{2}$, $\beta_{1}$, and $\beta_{2}$ are functions of $Q^2$ only, and
can in principle be determined by fitting $\sigma_{R}$ to $\varepsilon$ for a fixed $Q^2$ value.

(8) In order to utilize Eq.~(\ref{eq:constrain5}) above, and for a fixed $Q^2$ value, $\sigma_{R}$ 
must be measured at a minimum of six $\varepsilon$ points. However, the number of fitting parameters can be reduced 
by fixing the value of $(G_M^p)^2$ and making use of the assumption that for $\varepsilon \rightarrow 1$, the TPE correction to $\sigma_{R}$ vanishes or: $\sigma_{R}(\varepsilon=1, Q^2) = \big[(G_M^p)^2+ (G_E^p)^2/\tau\big]$.
In addition, because of the experimentally observed linearity of the Rosenbluth plots where $\sigma_{R}$ data show 
a linear behaviour in $\varepsilon$ suggesting the fit: $\sigma_{R} = [a(Q^2) + \varepsilon b(Q^2)]$. 
Therefore, for a fixed $Q^2$ value, I linear fit $\sigma_{R}$ to $\varepsilon$ and extract the constants 
$a(Q^2)$ and $b(Q^2)$. Equating the two expressions for $\sigma_{R}(\varepsilon=1, Q^2)$ yields
\begin{equation} \label{eq:GMp2Extract}
(G_M^p(Q^2))^2 = \frac{a(Q^2)+b(Q^2)}{(1+\frac{R^2}{\tau})}.
\end{equation} 

(9) I constrain $R$ and $(G_M^p)^2$ in Eq.~(\ref{eq:constrain5}) to their values as given by 
Eqs.~(\ref{eq:JRA_RptNew}) and~(\ref{eq:GMp2Extract}), respectively, and fit $\sigma_{R}$ to $\varepsilon$
with $\alpha_{i}$ and $\beta_{i}$ ($i=1,2$) being the parameters of the fit. The coefficients $\alpha_{0}$ and 
$\beta_{0}$ are then determined using the constraints: $\alpha_{0} = - ( \alpha_{1}+ \alpha_{2})$ and 
$\beta_{0} = - (\beta_{1} + \beta_{2})$. Finally, for a fixed $Q^2$ value, the three TPE amplitudes 
and the ratio $P_{l}/P_{l}^{\mbox{Born}}$ are determined as a function of $\varepsilon$ 
using Eqs.~(\ref{eq:constrain1}) and (\ref{eq:constrain4}), and Eq.~(9c), respectively.

\section{Results and Discussion} \label{results}

I extract the values of $(G_M^p(Q^2))^2$ and the TPE amplitudes coefficients $\alpha_{k}$ and $\beta_{k}$ $(k=1,2)$ 
following the procedure outlined in Sec.~(\ref{extraction}). I fit the world data on $\sigma_R$ used in 
the analysis of Ref.~\cite{qattan15} to extract $(G_M^p(Q^2))^2$ first following Eq.~(\ref{eq:GMp2Extract}). 
By fixing the values of $(G_M^p(Q^2))^2$ and the recoil 
polarization ratio $R$, I then fit $\sigma_R$ to Eq.~(\ref{eq:constrain5}), and extract the TPE amplitudes 
coefficients. In this analysis, 93 $Q^2$ points up to $Q^2 =$ 5 (GeV/c)$^2$ were used with $\sigma_R$ measured 
at a minimum of five $\varepsilon$ points. For the high $Q^2$ points, $Q^2 \gtorder$ 1 (GeV/c)$^2$, the majority of 
$\sigma_R$ measurements were made at a limited number of $\varepsilon$ points (below 5 points), and therefore, only 
a handful set of these $Q^2$ points could be used in the analysis. However, this was not the case for the low-$Q^2$ 
measurements.

\begin{figure}[!htbp]
\begin{center}
\begin{tabular}{c c}
\includegraphics*[width=4.25cm]{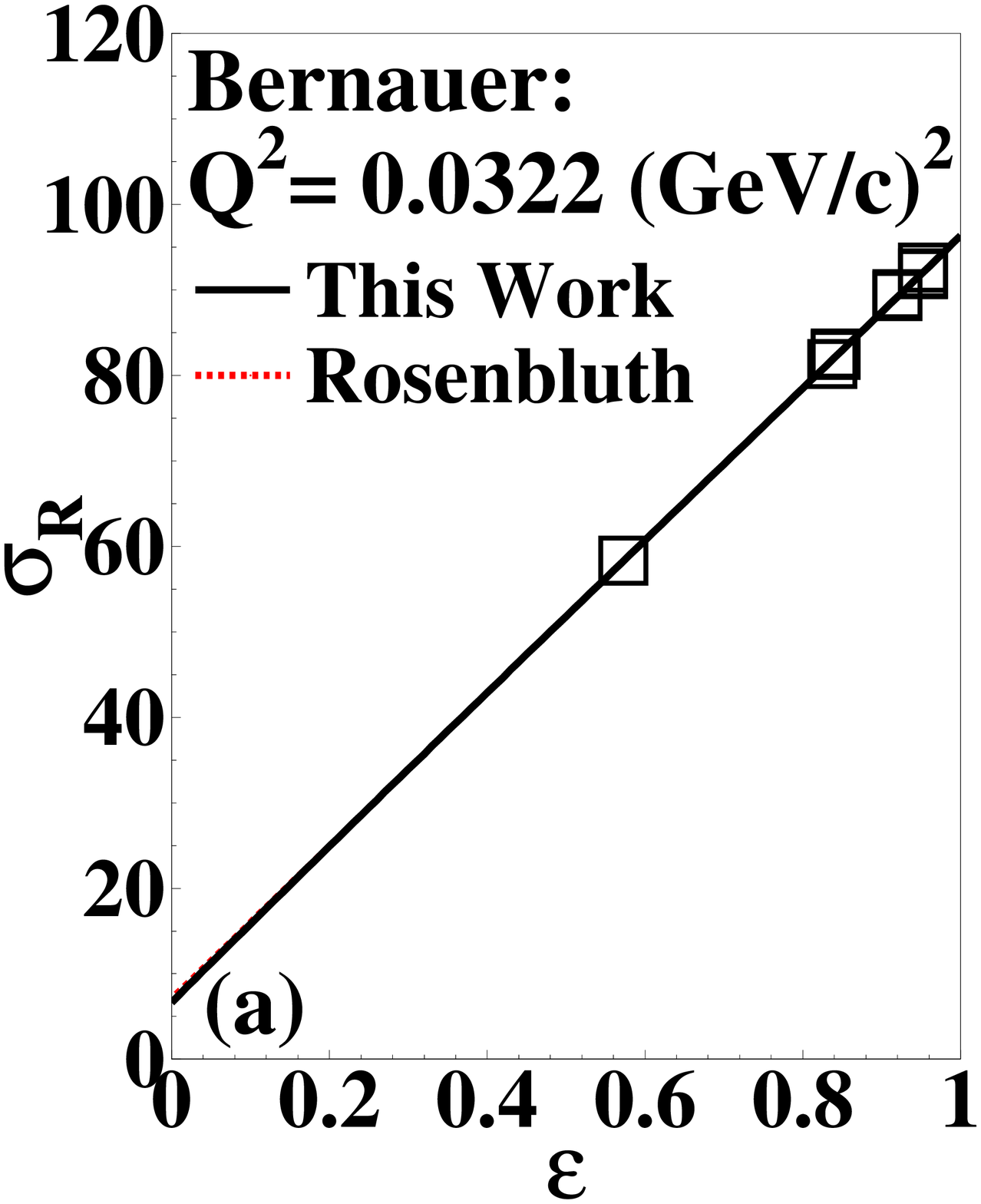} &
\includegraphics*[width=4.25cm]{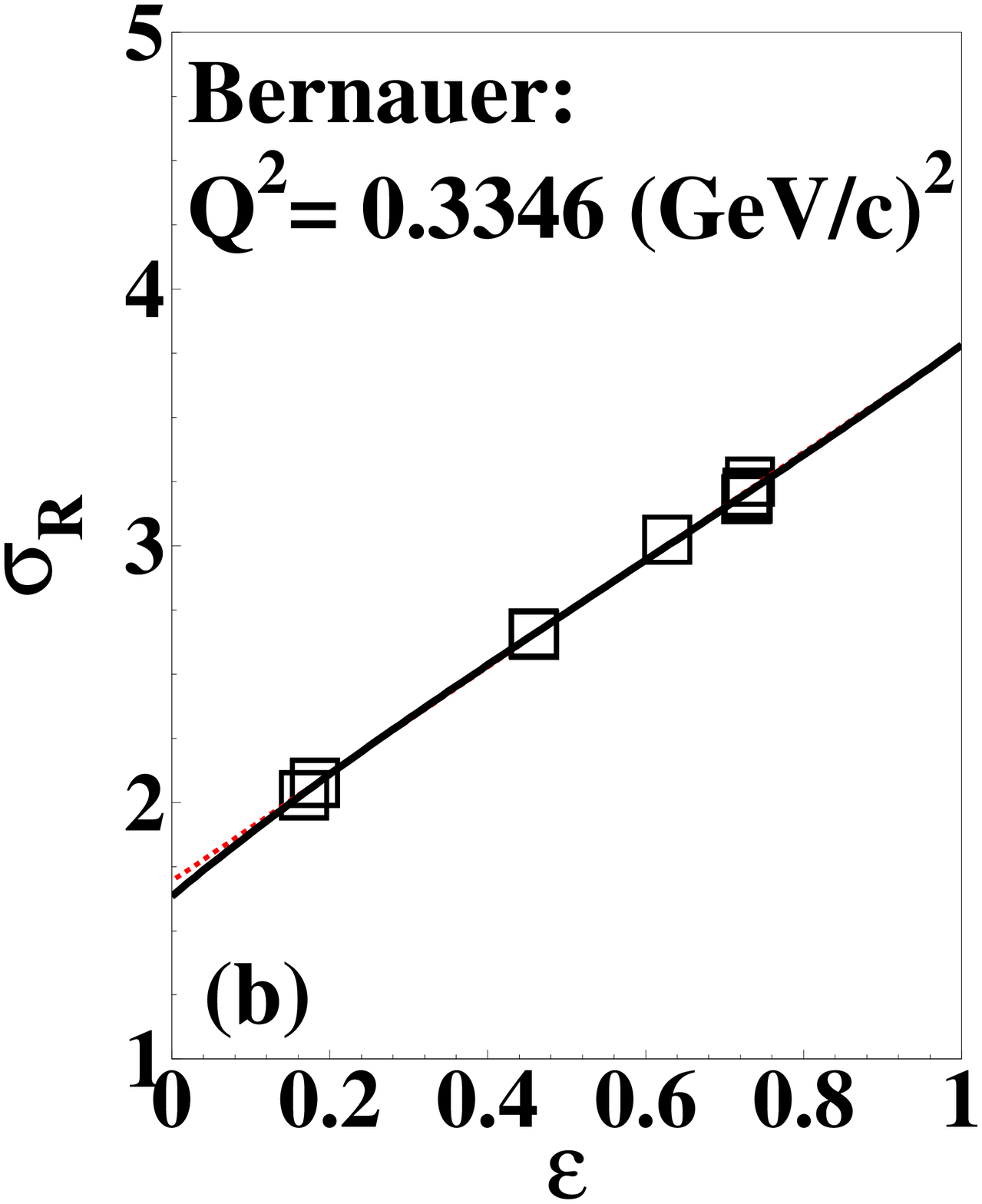} \\ 
\includegraphics*[width=4.25cm]{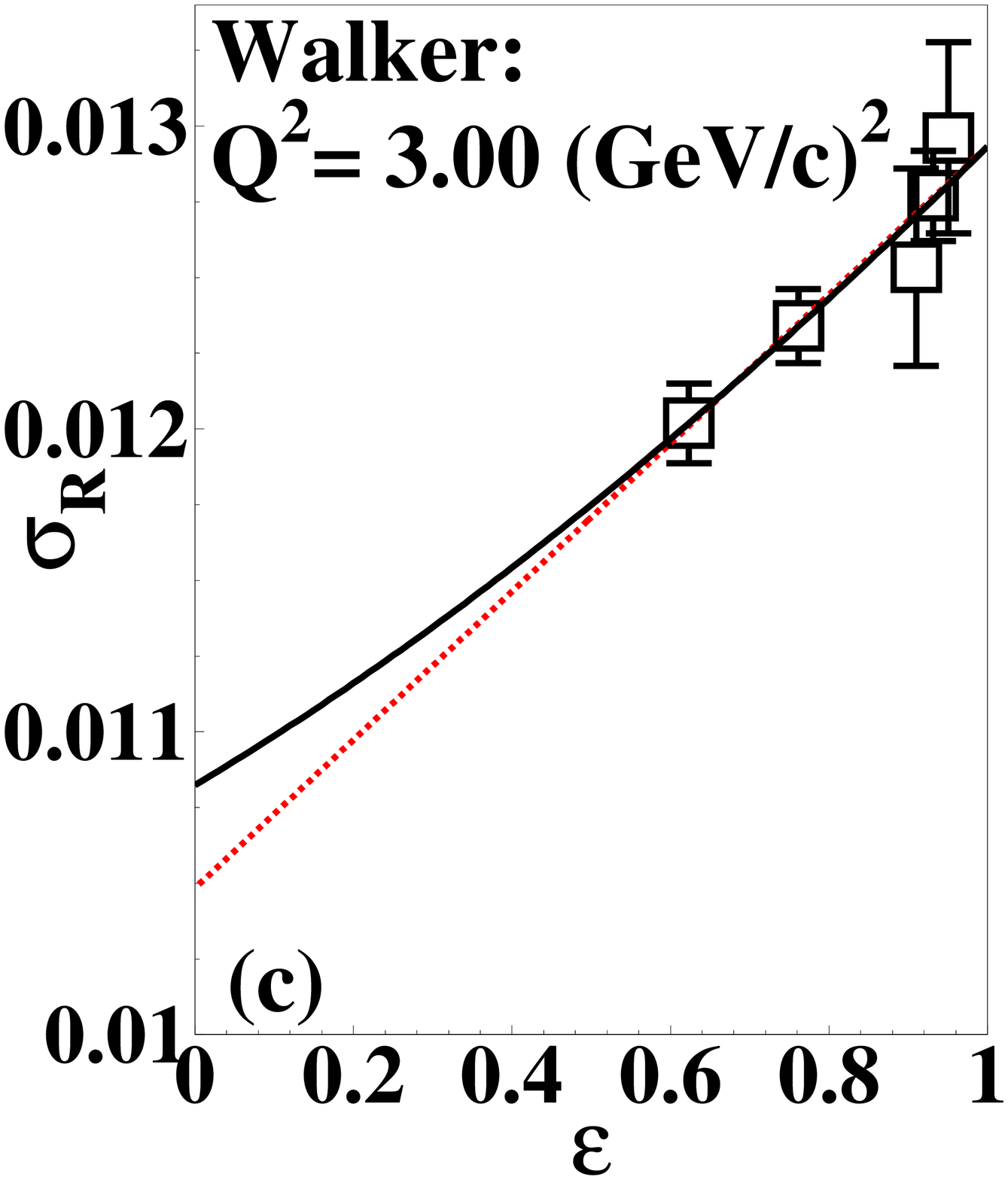} &
\includegraphics*[width=4.25cm]{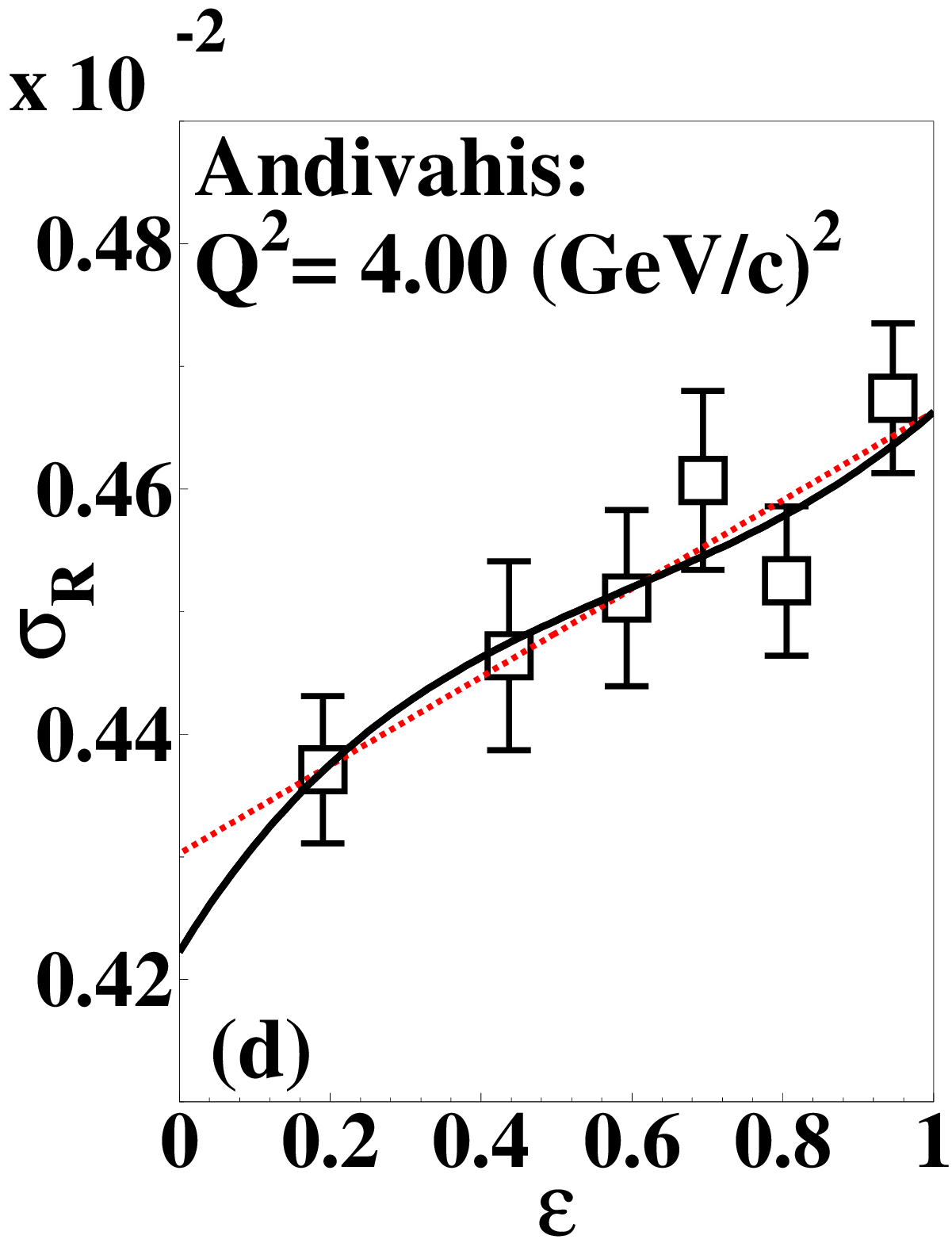} \\
\vspace{-0.5cm}
\end{tabular}
\end{center}
\caption{(Color online) The reduced cross section $\sigma_R$ as a function of $\varepsilon$ 
for a sample of low- and high-$Q^2$ data points from the data of Refs.~\cite{bernauer14,walker94,andivahis94}. 
Also shown my new fit based on Eq.~(\ref{eq:constrain5}) (solid black line), and the Rosenbluth fit 
(dashed red line) for comparison.}
\label{fig:SigR}
\end{figure}

Figure~\ref{fig:SigR} shows the result of my $\sigma_R$ fit for a sample of low- and high-$Q^2$ data points 
along with the Rosenbluth fit for comparison. My fit describes the data remarkably well with some 
deviation from linearity at low $\varepsilon$ for the high $Q^2$ points. Such nonlinearity was also observed in
several hadronic- and pQCD-based calculations in this range. See discussion in Sec.~\ref{visit_interference} for
details.
While most $\sigma_R$ fits yielded reasonable reduced $\chi^2$ 
values ranging from $0.03 < \chi_{\nu}^2 < 1.80$,  fits to 17 low-$Q^2$ data points in the range $0.0195 < Q^2 < 0.779$ 
(GeV/c)$^2$ from Refs.~\cite{bernauer14,janssens66} yielded $\chi_{\nu}^2 > 1.80$.

I also determined the values of $(G_E^p(Q^2))^2$ 
using the improved parametrization of the ratio $R$ and its associated uncertainty~\cite{qattan15}. The values of 
$(G_E^p/G_D(Q^2))^2$ and $(G_M^p/\mu_pG_D(Q^2))^2$ along with the fit results for the TPE amplitudes coefficients 
are included in the online Supplemental Material~\cite{suppl2016}.  

\begin{figure}[!htbp]
\begin{center}
\includegraphics*[width=8.2cm]{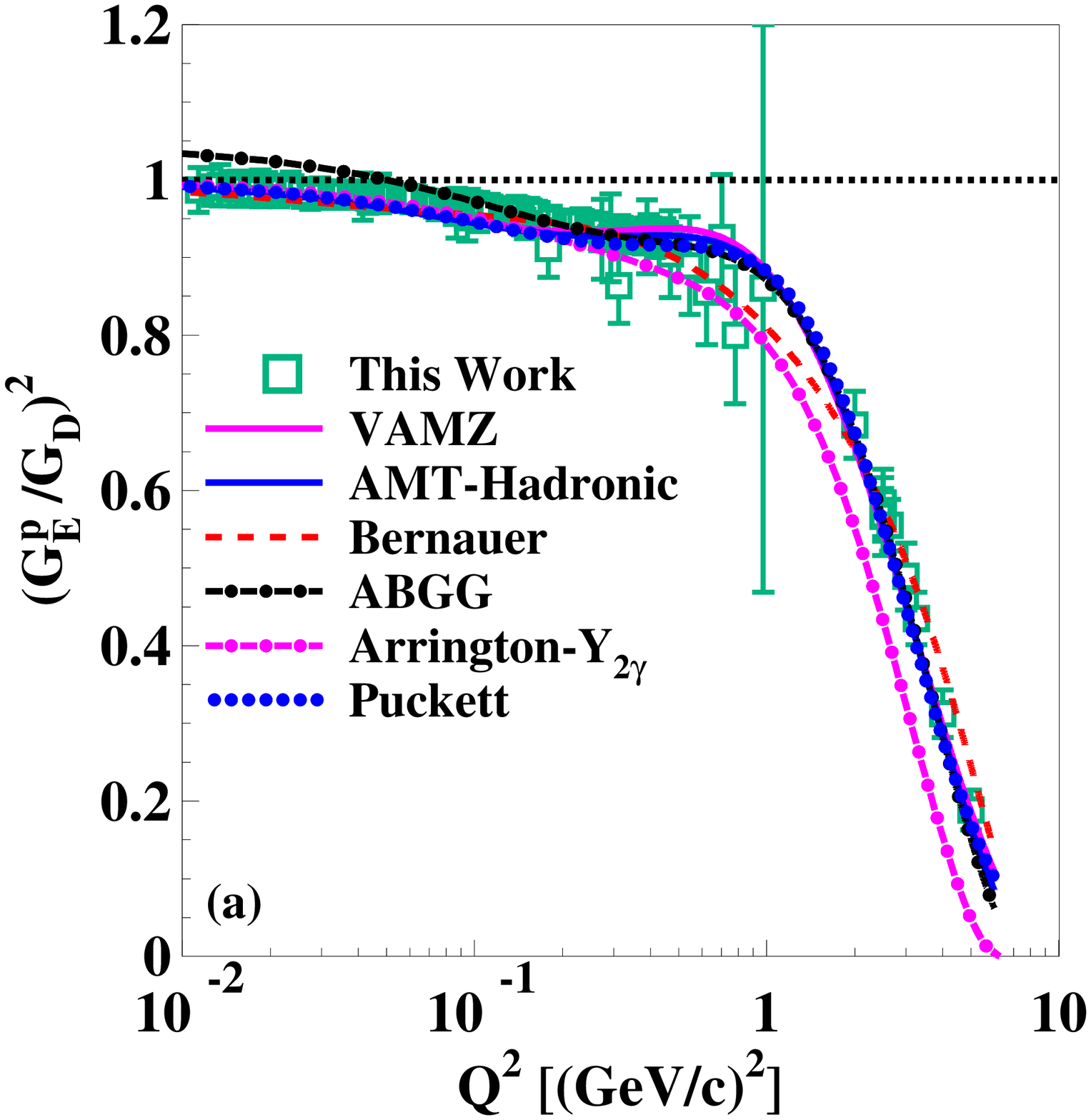} 
\includegraphics*[width=8.2cm]{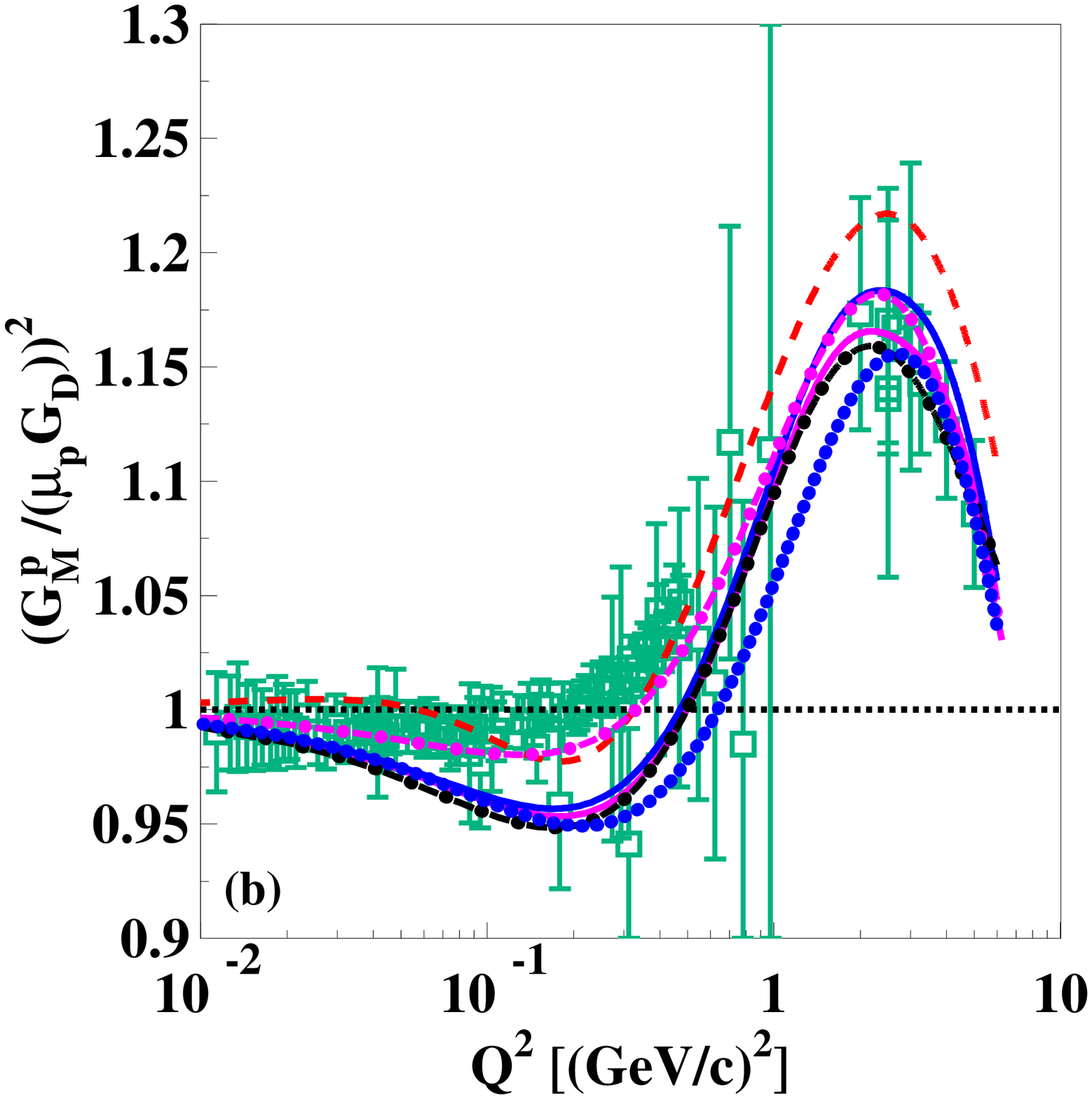} 
\vspace{-0.5cm}
\end{center}
\caption{(Color online) $(G_E^p/G_D(Q^2))^2$ (top) and $(G_M^p/\mu_pG_D(Q^2))^2$ (bottom) as obtained using 
Eq.~(\ref{eq:GMp2Extract}) and the parametrization of the ratio $R$ from Eq.~(\ref{eq:JRA_RptNew}) 
(Open dark-green squares). In addition,
I compare the results to the extractions from several previous TPE calculations and phenomenological fits:
AMT~\cite{arrington07} (solid blue line), VAMZ~\cite{venkat11} (solid magenta line), 
Bernauer~\cite{bernauer14} (long-dashed red line), ABGG~\cite{alberico09} (dashed-dotted black line),
Arrington $Y_{2\gamma}$~\cite{arrington05} (dashed-dotted magenta line), and 
Puckett (large-dotted blue line) (the fit labeled ``new'' in Ref.~\cite{puckett12}).}
\label{fig:FFs2}
\end{figure}

Figure~\ref{fig:FFs2} shows the values of $(G_E^p/G_D(Q^2))^2$ and $(G_M^p/\mu_pG_D(Q^2))^2$ extracted from
this work. I also compare my results to the values as extracted based on hadronic calculations, labeled 
``AMT-Hadronic''~\cite{arrington07} and ``VAMZ''~\cite{venkat11}. In addition, I show fits from 
previous phenomenological analyses labeled: ``ABGG''~\cite{alberico09}, ``Bernauer''~\cite{bernauer14}, 
``Arrington-$Y_{2\gamma}$''~\cite{arrington05}, and ``Puckett''~\cite{puckett12}.
    
For $(G_E^p/G_D(Q^2))^2$, my results in general are in reasonably good
agreement with extractions using calculated TPE corrections and phenomenological-based fits. However,
the ``Bernauer'' and the ``Arrington-$Y_{2\gamma}$'' fits are very different at large $Q^2$. The 
``Arrington-$Y_{2\gamma}$'' fit, on the other hand, is the only analysis that allows
for TPE contributions to the recoil polarization data, though the extraction of these terms is extremely
model dependent. 

For $(G_M^p/\mu_pG_D(Q^2))^2$ and at low $Q^2$, my results are significantly above most previous fits. 
This reflects the discrepancy between the recent Mainz data which yields values of $G_M^p$ which are 
systematically 2--5\% larger than previous world data~\cite{bernauer14}. At low $Q^2$, this corresponds 
to only a small difference in the cross section at large scattering angle, but for the larger $Q^2$ values 
of the Mainz experiment, this corresponds to a significant difference in the measured cross sections. 
Note that except for the Bernauer result, most of the previous phenomenological extractions of the form factors 
and TPE contributions were focused on large $Q^2$ values, and so did not always worry about how well the 
parameterizations of $R$ reproduced low $Q^2$ data.

%

\begin{figure}[!htbp]
\begin{center}
\begin{tabular}{c c}
\includegraphics*[width=4.25cm]{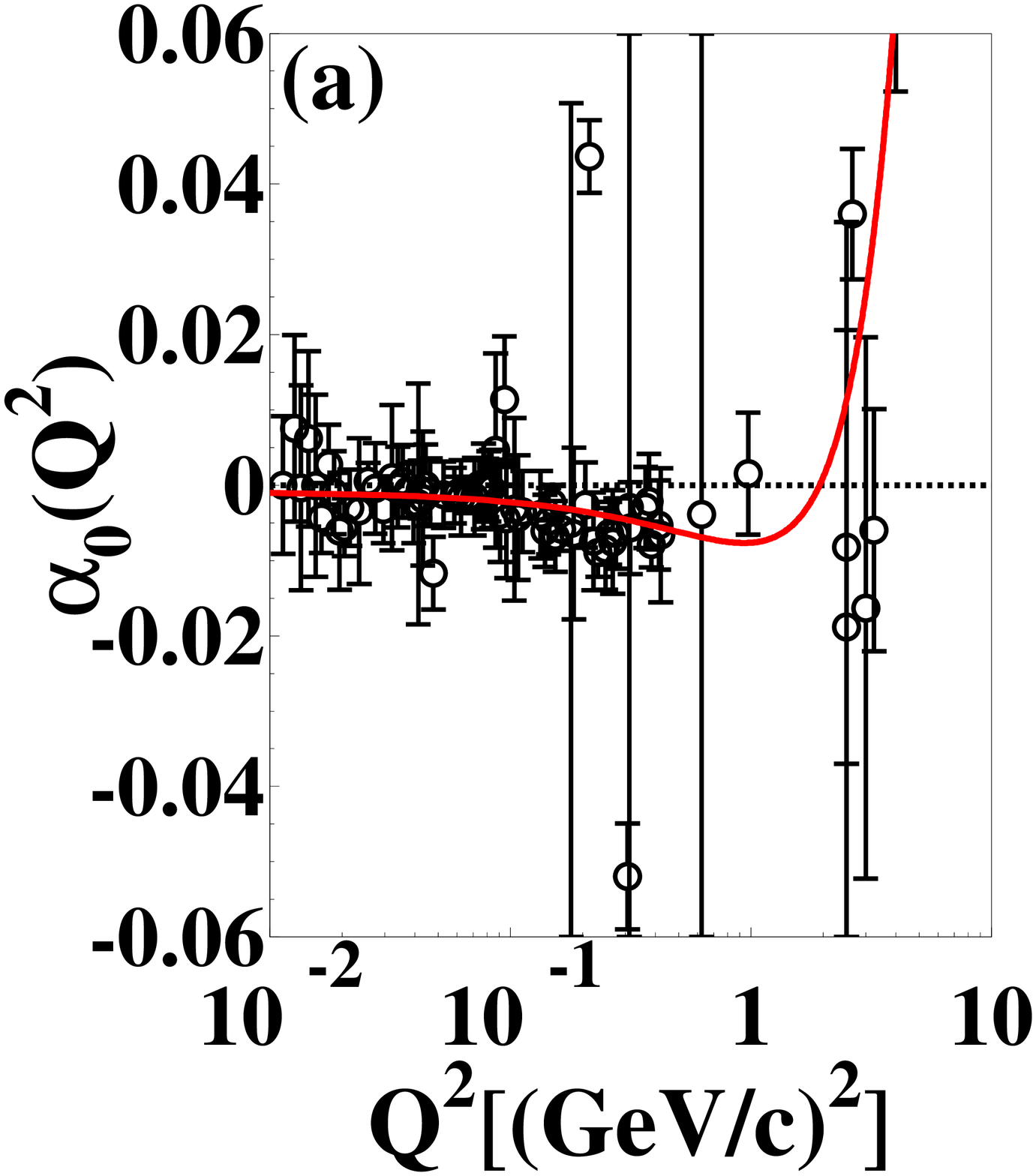} &
\includegraphics*[width=4.25cm]{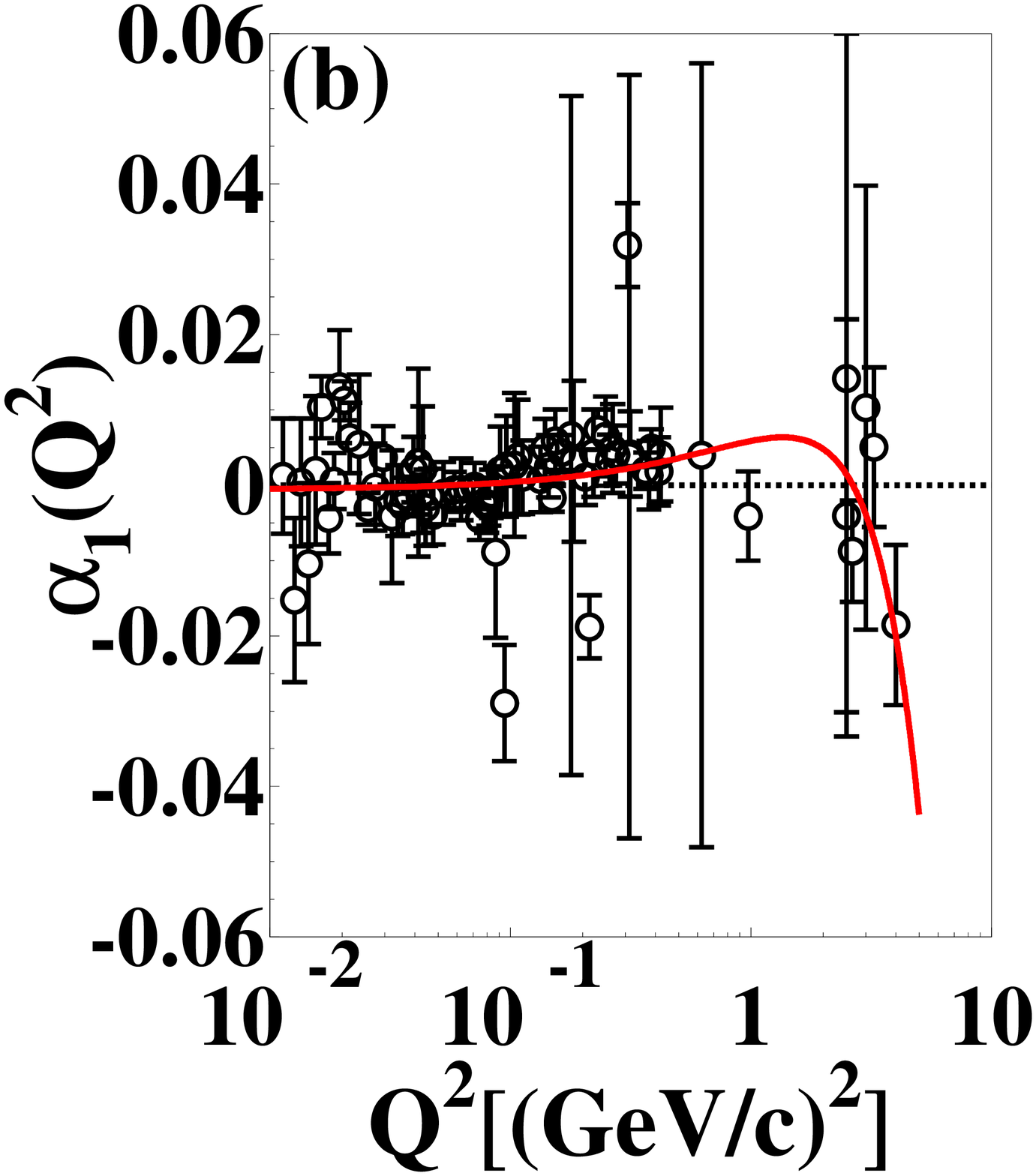} \\
\includegraphics*[width=4.25cm]{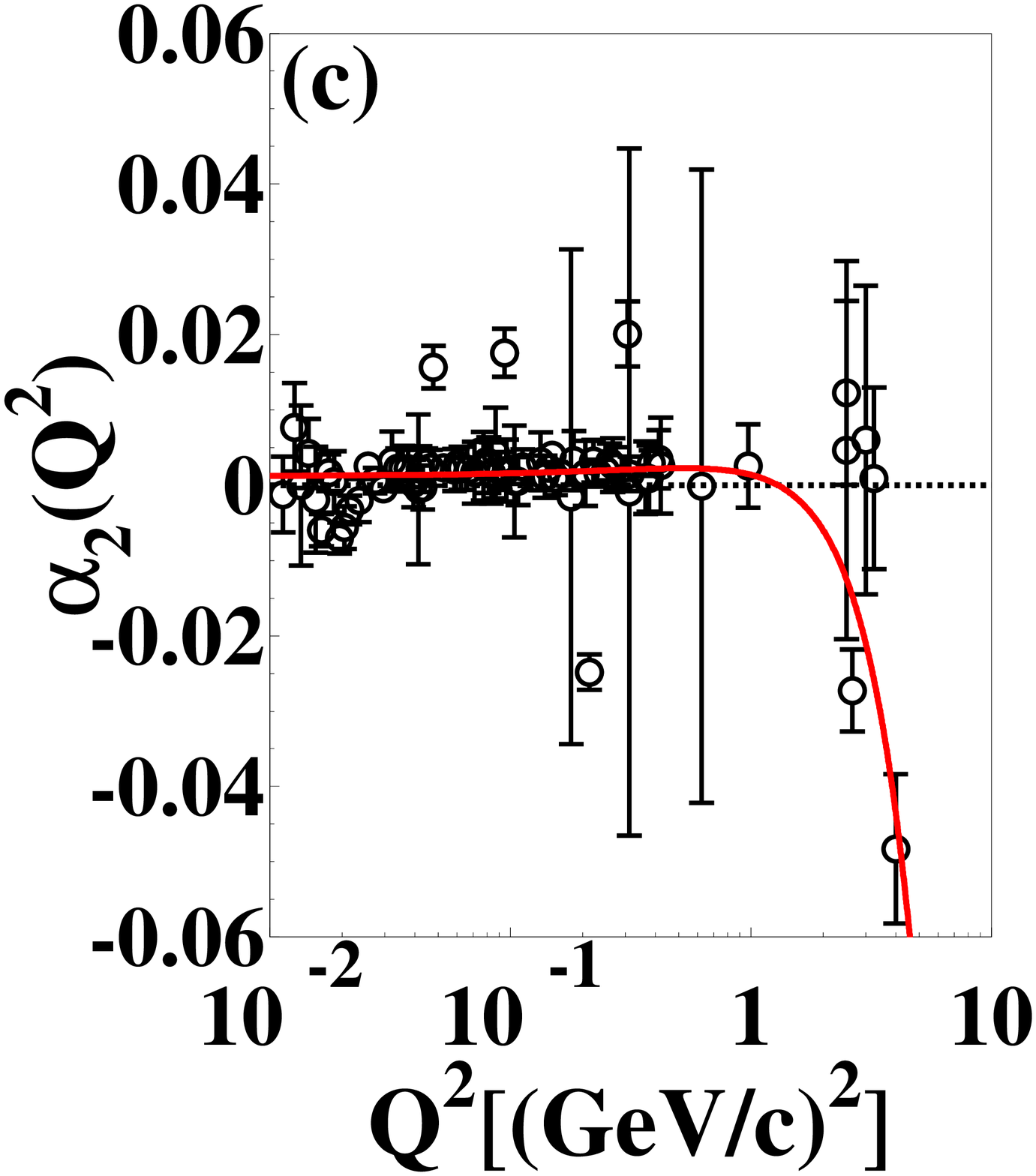} &
\includegraphics*[width=4.25cm]{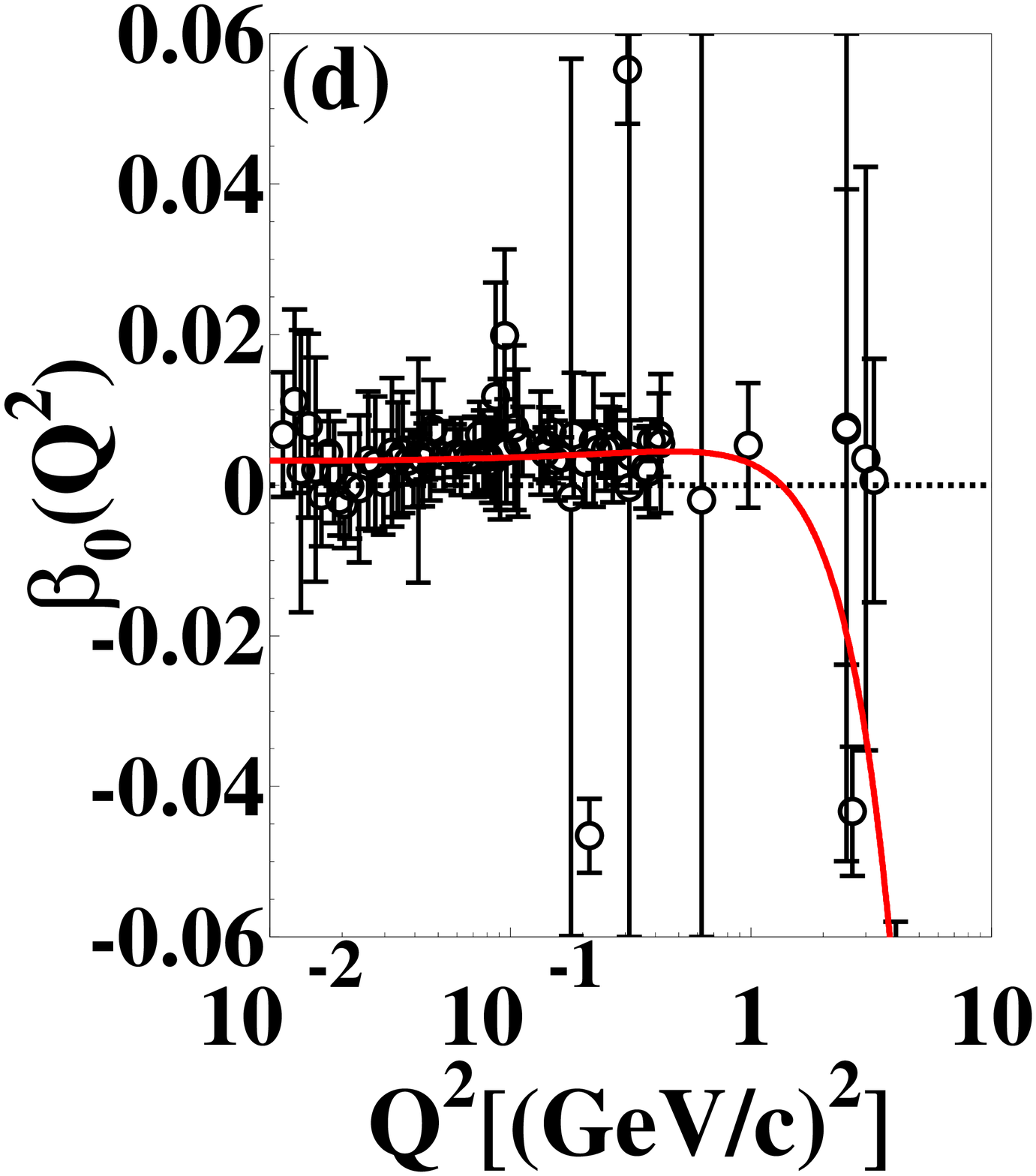} \\
\includegraphics*[width=4.25cm]{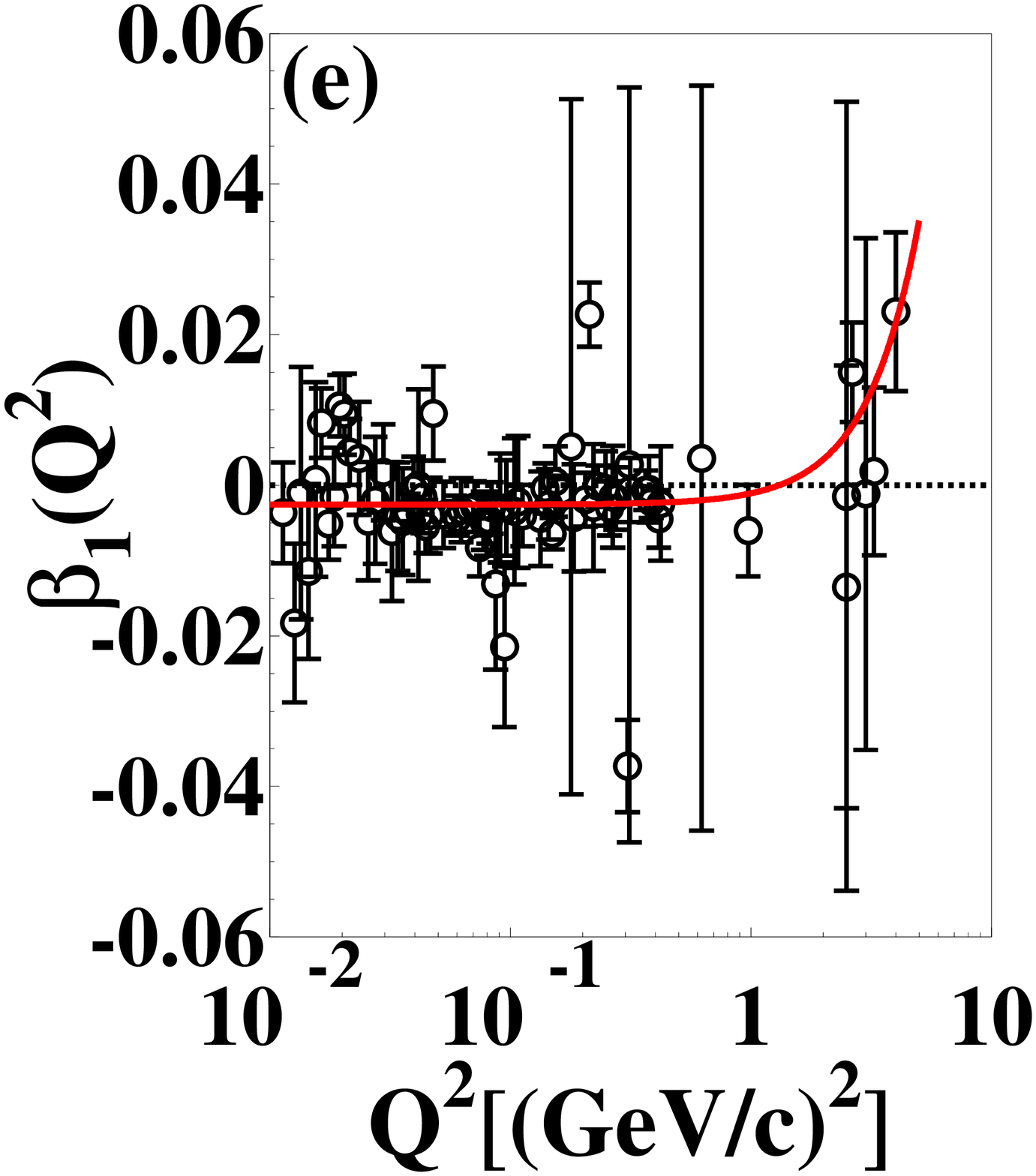} &
\includegraphics*[width=4.25cm]{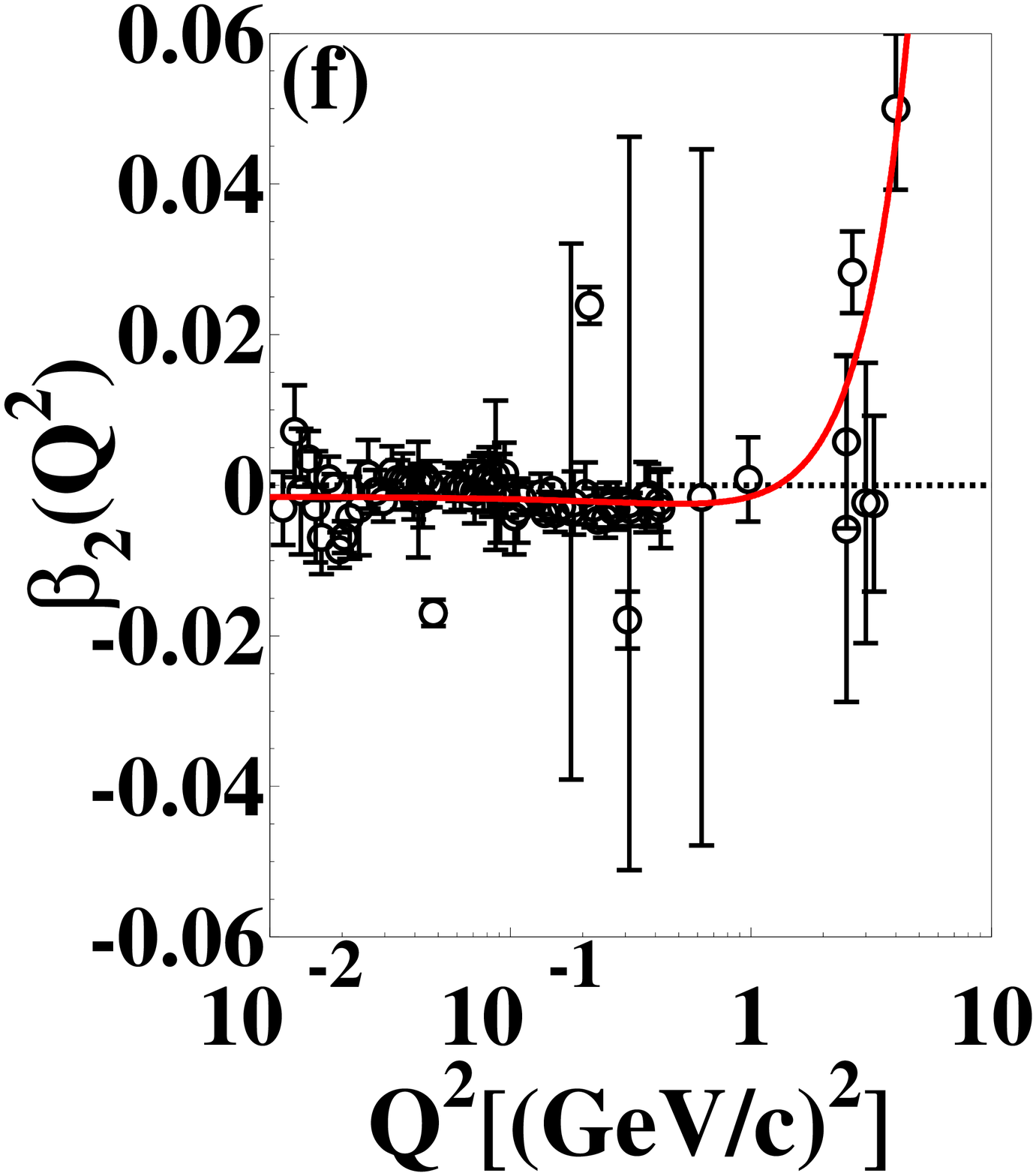} \\
\vspace{-0.5cm}
\end{tabular}
\end{center}
\caption{(Color online) The TPE amplitudes coefficients $\alpha_{(0,1,2)}(Q^2)$ and 
$\beta_{(0,1,2)}(Q^2)$ as a function of $Q^2$ as extracted from this 
work (open black circles) along with the fits (solid red lines) valid up to $Q^2 =$ 4 (GeV/c)$^2$. 
Note that $Q^2$ points which yielded large TPE amplitudes were excluded for clarity. See text for details.}
\label{fig:TPEcoffs}
\end{figure}

The TPE amplitudes coefficients $\alpha_{k}$ and $\beta_{k}$ $(k=0,1,2)$ extracted from this work 
as a function of $Q^2$ are shown in Fig.~\ref{fig:TPEcoffs}. The coefficients are at the few-percentage-points 
level, and all show hints of oscillatory behaviour below $ Q^2 =$ 1.0 (GeV/c)$^2$ with clear sign of structure 
(dips/bumps) at $Q^2 \approx$ 0.02 (GeV/c)$^2$. For $Q^2 \gtorder$ 1.0 (GeV/c)$^2$, the extractions were done using a limited 
number of $Q^2$ points, the coefficients change sign and increase in magnitude 
with increasing $Q^2$ but with larger uncertainties. 

Recently~\cite{qattan15}, the proton form factors and the TPE correction $a(Q^2)$ 
values were extracted based on the parametrization from Borisyuk and Kobushkin (BK
parametrization)~\cite{borisyuk07} where $\sigma_R$ is expressed in the following form
\begin{equation} \label{eq:Kobushkin3}
\sigma_{R} = \big(G_M^p\big)^2\big[ 1 + \frac{\varepsilon}{\tau} R^2\big] + 2 a(Q^2)(1-\varepsilon)\big(
G_M^p\big)^2.
\end{equation} 

While the values of the TPE parameter $a(Q^2)$ extracted from Mainz data~\cite{bernauer14} show hints of 
oscillatory behaviour below $Q^2 = 0.3$~GeV$^2$, the uncertainties in the extracted TPE contribution are an 
underestimate of the true uncertainties. The quoted uncertainties on the individual cross sections do not 
include correlated systematic effects, which are a significant contribution to the total uncertainty in 
their final form factor parameterization, and I do not account for any residual uncertainty in the 
normalization of the data subsets which are fit as part of the global analysis~\cite{bernauer14}. 
So while the oscillatory behaviour in the TPE amplitudes coefficients extracted in this work seems 
significant compared to the uncertainties that are shown, these uncertainties are incomplete and this 
cannot be taken as meaningful evidence for such structure.

The TPE amplitudes coefficients are then used to construct
the TPE amplitudes using Eqs.~(\ref{eq:constrain1}) and (\ref{eq:constrain4}). Out of the 93 $Q^2$ points analyzed, 
17 points yielded unexpectedly large TPE amplitudes exceeding the 10\% level. These points are: 
13 points in the range $0.2 \leq Q^2 \leq 0.7$ (GeV/c)$^2$ from Ref.~\cite{bernauer14}, $Q^2 = $ 0.779
(GeV/c)$^2$ from Ref.~\cite{janssens66}, $Q^2 = $ 2.0 (GeV/c)$^2$ from Ref.~\cite{walker94}, and 
$Q^2 = $ 2.5 and 5.0 (GeV/c)$^2$ from Ref.~\cite{andivahis94}. Note that these points are not
shown in Fig.~\ref{fig:TPEcoffs} for clarity. However, no correlation was found between the large TPE 
amplitudes values obtained using these $Q^2$ points and the large $\chi_{\nu}^2$ obtained from the $\sigma_R$ fits.

In an attempt to parametrize the $Q^2$ dependence of the TPE amplitudes coefficients 
$\alpha(\beta)_{(0,1,2)}(Q^2)$, several different functional forms were tried.
All fits yielded unexpectedly large $\chi_\nu^2$ when all the 93 $Q^2$ points were included in the fit. 
However, the best fits were achieved when ``all'' the TPE amplitudes coefficients were parametrized as a 
second-order polynomial of the form: $\alpha(\beta)_{(0,1,2)}(Q^2) = (a_0 + a_1Q^2+ a_2Q^4)$, 
and when the 17 $Q^2$ data points with large TPE amplitudes were excluded. The fits, 
valid up to $Q^2 =$ 4 (GeV/c)$^2$, are shown in Fig.~\ref{fig:TPEcoffs} as solid red lines, and the 
parameters of the fits are listed in Table~\ref{fitsparam}.
The coefficients $\alpha_2$ and $\beta_2$ have the largest $\chi_\nu^2$ values. However, calculating these two 
coefficients using the constraints: $\alpha_{2} = - (\alpha_{0}+ \alpha_{1})$ and
$\beta_{2} = - (\beta_{0} + \beta_{1})$ yielded effectively the same results as those of the fits.
  
%

\begin{table*}[!htbp]
\begin{center}
\caption[The values of the fit parameters for the TPE amplitudes coefficients $\alpha_{(0,1,2)}(Q^2)$ and 
$\beta_{(0,1,2)}(Q^2)$.] 
{The values of the fit parameters for the TPE amplitudes coefficients $\alpha_{(0,1,2)}(Q^2)$ and $\beta_{(0,1,2)}(Q^2)$.}
\begin{tabular}{l c c c c} 
\hline \hline
Coefficient & $a_0$ & $a_1$ & $a_2$ & $\chi^2_{\nu}$\\ 
\hline 
$\alpha_0(Q^2)$ & ($-$0.89$\pm$1.27)$\times 10^{-3}$ & ($-$1.45$\pm$0.80)$\times 10^{-2}$ &($+$7.75$\pm$2.59)$\times 10^{-3}$ & 1.56  \\

$\alpha_1(Q^2)$ & ($-$0.58$\pm$0.86)$\times 10^{-3}$ & ($+$1.02$\pm$0.53)$\times 10^{-2}$ &($-$3.77$\pm$0.17)$\times 10^{-3}$ & 1.36  \\

$\alpha_2(Q^2)$ & ($+$1.22$\pm$0.80)$\times 10^{-3}$ & ($+$3.95$\pm$5.47)$\times 10^{-3}$ &($-$3.78$\pm$1.87)$\times 10^{-3}$ & 1.92  \\

$\beta_0(Q^2)$ & ($+$3.19$\pm$1.43)$\times 10^{-3}$ & ($+$5.53$\pm$8.30)$\times 10^{-3}$ &($-$5.88$\pm$2.69)$\times 10^{-3}$ & 1.57  \\

$\beta_1(Q^2)$ & ($-$2.56$\pm$0.85)$\times 10^{-3}$ & ($-$0.02$\pm$4.98)$\times 10^{-3}$ &($+$1.51$\pm$1.58)$\times 10^{-3}$ & 1.28  \\

$\beta_2(Q^2)$ & ($-$1.49$\pm$0.89)$\times 10^{-3}$ & ($-$3.93$\pm$5.85)$\times 10^{-3}$ &($+$3.94$\pm$2.00)$\times 10^{-3}$ & 1.91  \\
\hline
\hline
\end{tabular}
\label{fitsparam}
\end{center}
\end{table*}
\begin{figure}[!htbp]
\begin{center}
\begin{tabular}{c c}
\includegraphics*[width=4.25cm]{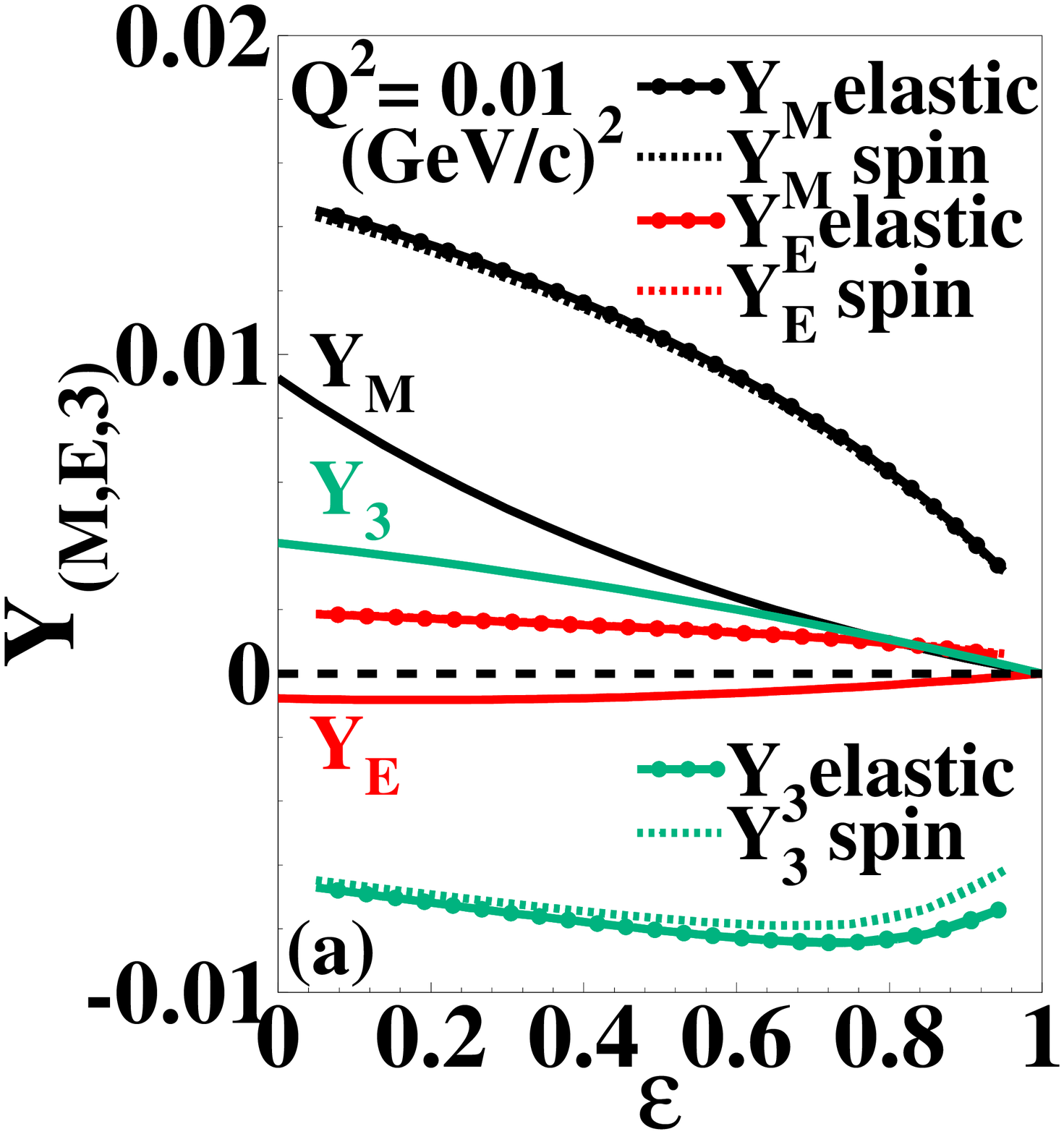} &
\includegraphics*[width=4.25cm]{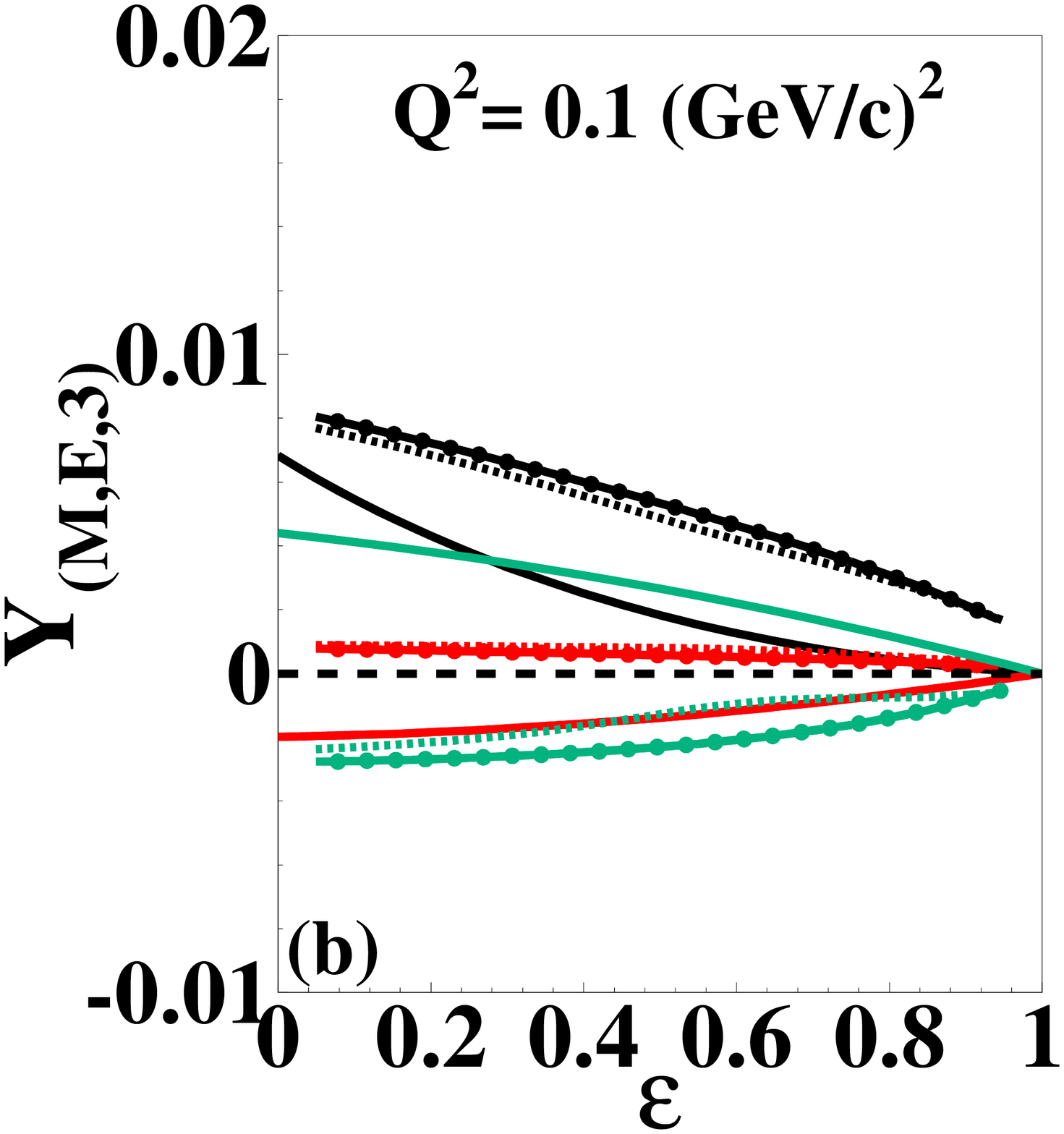} \\
\includegraphics*[width=4.25cm]{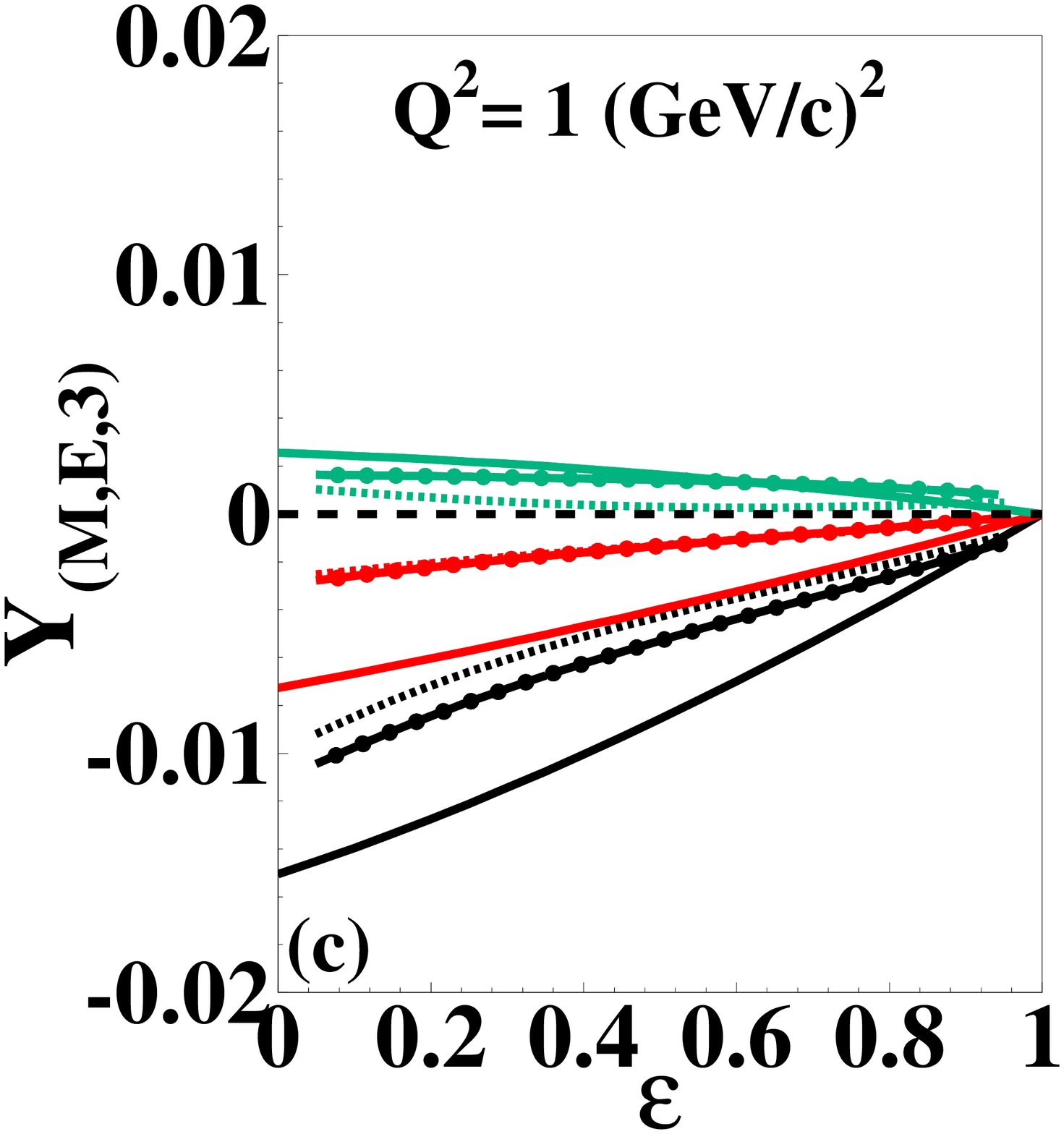} &
\includegraphics*[width=4.25cm]{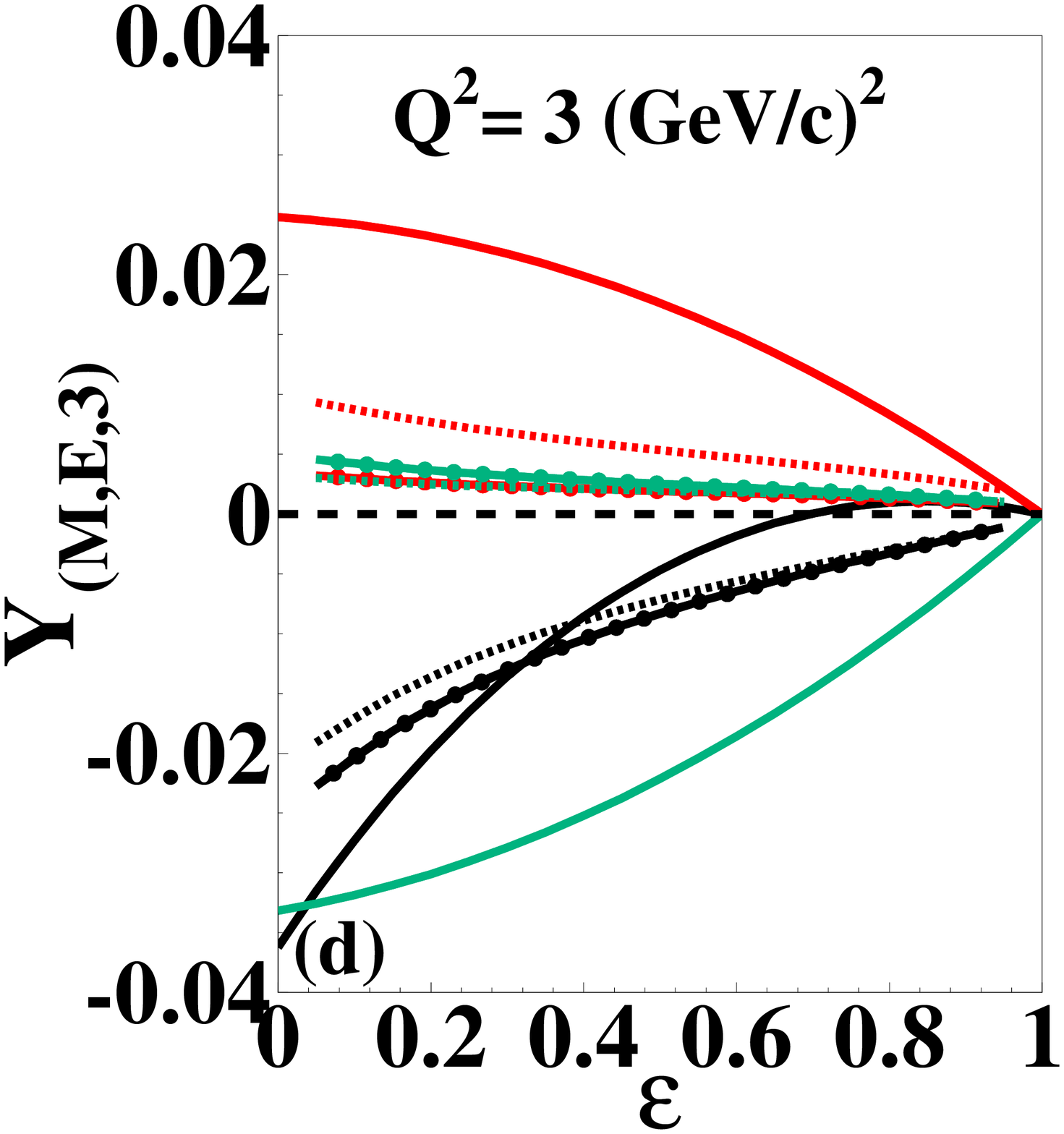} \\
\vspace{-0.5cm}
\end{tabular}
\end{center}
\caption{(Color online) The extracted TPE amplitudes from this work: $Y_M$ (solid black line), $Y_E$ (solid red line) and $Y_3$ (solid dark-green line) as a function of $\varepsilon$ at the $Q^2$ value listed in the figure. In addition, 
I compare the results to previous hadronic calculations assuming different intermediate states: 
elastic labelled as: ``$Y_M$ elastic'' (dashed-dotted black line), ``$Y_E$ elastic'' (dashed-dotted red line), ``$Y_3$ elastic'' (dashed-dotted dark-green line) from Ref.~\cite{borisyuk08}, and elastic $+$ $\pi N$ with spin 1/2 and 3/2 channels labelled as: ``$Y_M$ spin'' (long-dashed black line), ``$Y_E$ spin'' (long-dashed red line), ``$Y_3$ spin''
(long-dashed dark-green line) from Ref.~\cite{borisyuk15}.}
\label{fig:TpeAmpl}
\end{figure}
%
\begin{figure}[!htbp]
\begin{center}
\includegraphics*[width=8.2cm,height=7cm]{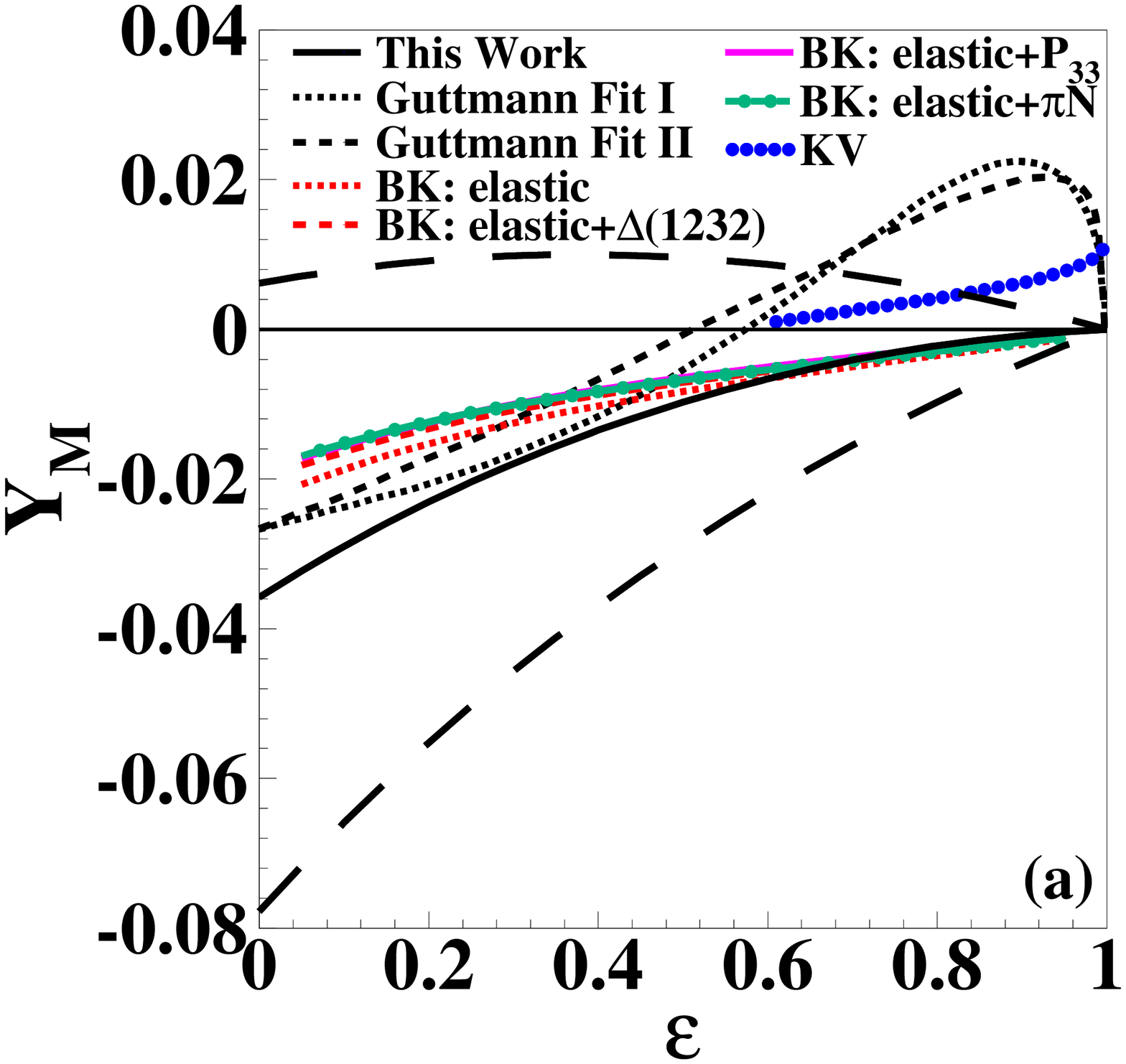} 
\includegraphics*[width=8.2cm,height=7cm]{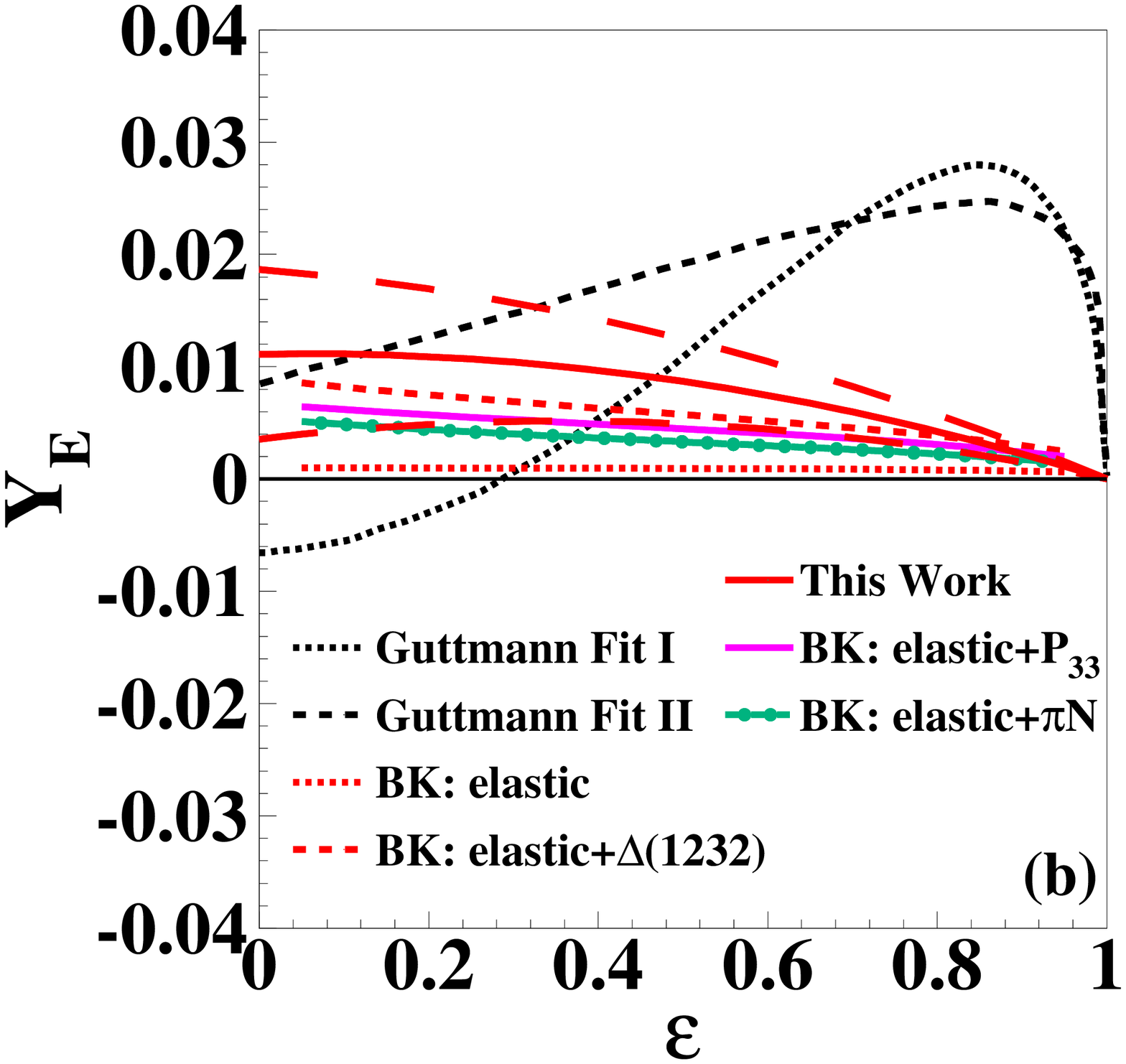}
\includegraphics*[width=8.2cm,height=7cm]{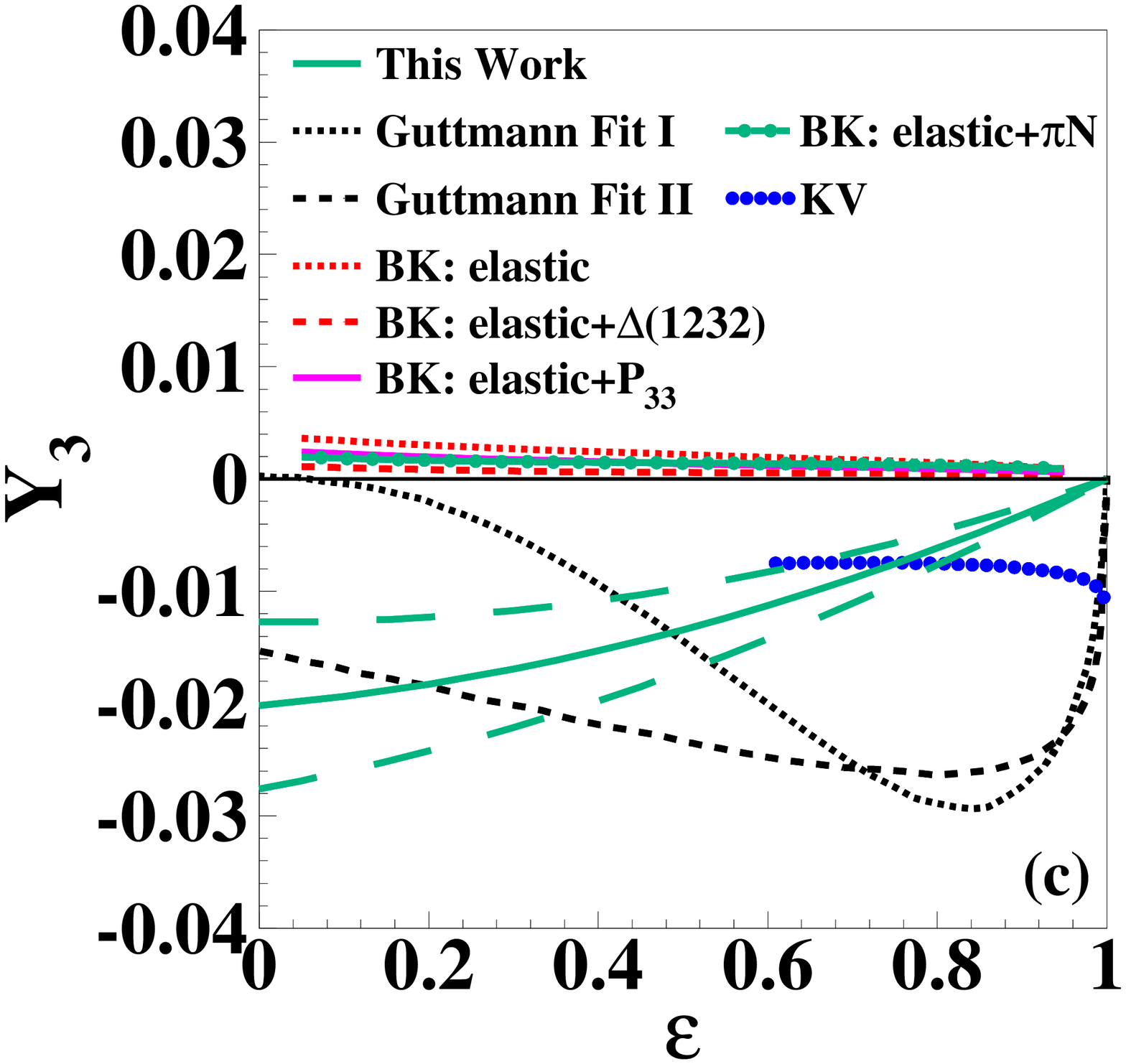} 
\vspace{-0.5cm}
\end{center}
\caption{(Color online) The extracted TPE amplitudes from this work: $Y_M$ (top) (solid black line), $Y_E$ (middle) 
(solid red line), and $Y_3$ (bottom) (solid dark-green line) as a function of $\varepsilon$ at $Q^2 = $ 2.50 
(GeV/c)$^2$. The error bands on the amplitudes are shown as very long-dashed (black,red,dark-green) lines, 
respectively. In addition, I compare the results to phenomenological extractions from Ref.~\cite{guttmann11}
``Guttmann Fit I'' (dashed black line) and ``Guttmann Fit II'' (long-dashed black line), 
and to several previous hadronic TPE predictions: ``BK: elastic''~\cite{borisyuk08} (dashed red line), 
``BK: elastic $+$ $\Delta(1232)$''~\cite{borisyuk12} (long-dashed red line), 
``BK: elastic $+$ $P_{33}$''~\cite{borisyuk14} (solid magenta line), and ``BK: elastic $+$ $\pi N$''~\cite{borisyuk15}
(dashed-dotted dark-green line). Also shown, calculations based on QCD factorization ``KV''~\cite{kivel13} 
(dotted blue line). In a leading twist QCD-type calculation the $Y_E$ amplitude cannot be calculated. See text for
details.}
\label{fig:TpeAmpl250VsEpsl}
\end{figure}

The $\varepsilon$ dependence of the TPE amplitudes as extracted from this work is shown in 
Fig.~\ref{fig:TpeAmpl} for a range of $Q^2$ values.
The amplitudes are on the few-percentage-points level, and behave roughly linearly with increasing $Q^2$. 
The amplitude $Y_M$ is the largest in magnitude. It is mainly positive at low $Q^2$ and changes 
sign and increases in magnitude with increasing $Q^2$ where it becomes non-linear. 
The amplitude $Y_3$ is also sizable and positive at low $Q^2$ and starts to decrease with increasing 
$Q^2$. At high $Q^2$ values, $Q^2 \sim$ 3.0 (GeV/c)$^2$, $Y_3$ changes sign and becomes negative and 
starts to grow in magnitude with increasing $Q^2$ where it becomes non-linear. On the other hand, 
the amplitude $Y_E$ is negligible and mainly negative at low $Q^2$. It starts to increase in magnitude with increasing $Q^2$, and then
changes sign and becomes positive at $Q^2 \sim$ 3.0 (GeV/c)$^2$ where it continues to grow in magnitude
and becomes non-linear. 

The $Y_E$ and $Y_3$ amplitudes extracted in this work differ in magnitude, and certainly have opposite sign to each other as $Q^2$ increases where they tend to partially cancel each other. This suggests that the TPE correction to $\sigma_R$ is driven mainly by $Y_M$ and to a lesser extent by 
$Y_3$. This is in agreement with the finding of Ref.~\cite{guttmann11} for the extraction of the TPE amplitudes 
at $Q^2 =$ 2.50 (GeV/c)$^2$. 

I also compare my results to several previous hadronic TPE calculations assuming different intermediate states: 
elastic labelled as ``$Y_M$ elastic'', ``$Y_E$ elastic'', ``$Y_3$ elastic''~\cite{borisyuk08}, and elastic $+$ $\pi N$ with spin 1/2 and 3/2 channels labelled as ``$Y_M$ spin'', ``$Y_E$ spin'', ``$Y_3$ spin''~\cite{borisyuk15}. 
While my results for $Y_M$ and $Y_E$ are in reasonable qualitative agreement with previous hadronic 
calculations, showing the fall and rise of both amplitudes with increasing $Q^2$, my results
for $Y_3$ have opposite sign and deviate substantially from calculations except at 
$Q^2 \sim $ 1.0 (GeV/c)$^2$.

In Fig.~\ref{fig:TpeAmpl250VsEpsl} I show the results of the $\varepsilon$ dependence of the 
TPE amplitudes at $Q^2 =$ 2.50 (GeV/c)$^2$ along with the error bands, shown as very long-dashed lines, as 
computed using the covariance matrix of the fits. I compare the results to previous phenomenological extractions of 
Ref.~\cite{guttmann11}, labelled as ``Guttmann Fit I'' and ``Guttmann Fit II''. In addition, I 
compare the results to several previous hadronic TPE calculations assuming different intermediate 
states: elastic ``BK: elastic''~\cite{borisyuk08}, elastic $+$ $\Delta(1232)$ resonance 
``BK: elastic $+$ $\Delta(1232)$''~\cite{borisyuk12}, elastic $+$ $\pi N$ ($P_{33}$ channel) 
``BK: elastic $+$ $P_{33}$''~\cite{borisyuk14}, elastic $+$ $\pi N$ (Spin 1/2 and 3/2 channels) 
``BK: elastic $+$ $\pi N$''~\cite{borisyuk15}, and to calculations based on QCD factorization within 
the SCET approach from Ref.~\cite{kivel13} ``KV''. Note that in a leading twist QCD-type calculation, 
the two amplitudes $Y_M$ and $Y_3$ can only be calculated but not $Y_E$ as the virtual photon (gluon) 
cannot flip the quark spin and calculation of the amplitude $Y_E$ requires knowledge of the quark 
transverse momenta distribution. 

For $Y_M$, and despite the large error band, my results are generally in good qualitative agreement with 
previous phenomenological extractions from Ref.~\cite{guttmann11} for $\varepsilon <$ 0.60, and deviate substantially 
above that. On the other hand, the results are in good qualitative agreement with all hadronic TPE calculations of 
Refs.~\cite{borisyuk08,borisyuk12,borisyuk14,borisyuk15} with the elastic contribution, ``BK: elastic'', 
being the closets, although all hadronic calculations are very close in value with amplitudes vanishing in the 
limit $\varepsilon \rightarrow 1$. For $Y_E$ and $Y_3$, the error bands are smaller and the amplitudes are more 
constrained than $Y_M$. For $Y_E$, my results are in very good agreement with all hadronic calculations 
as well with the inelastic contribution, ``BK: elastic $+$ $\Delta(1232)$'' being the closest, but deviate from 
phenomenological extractions for all $\varepsilon$ range. For $Y_3$, my results disagree noticeably with both 
phenomenological extractions and all hadronic calculations. On the other hand, the QCD-type calculation
within the SCET approach from Ref.~\cite{kivel13} disagree strongly with my results, previous phenomenological 
extractions, as well as hadronic TPE predictions as the amplitudes $Y_{(M,3)}$ do not vanish in the limit 
$\varepsilon \rightarrow 1$, with the absolute values of these amplitudes being much smaller than those obtained 
in Ref.~\cite{guttmann11}.
  
\begin{figure}[!htbp]
\begin{center}
\includegraphics*[width=8.2cm,height=7cm]{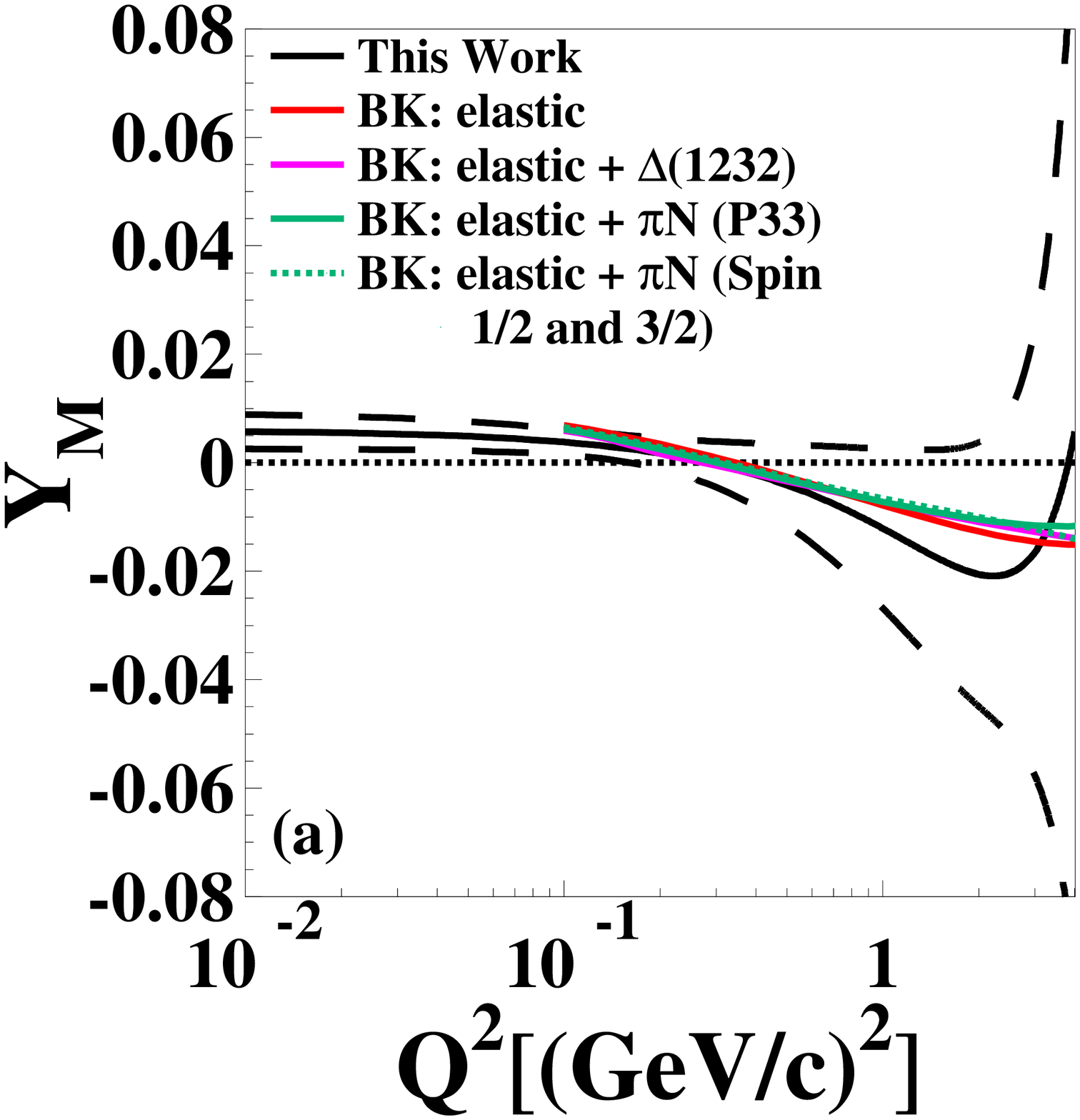} 
\includegraphics*[width=8.2cm,height=7cm]{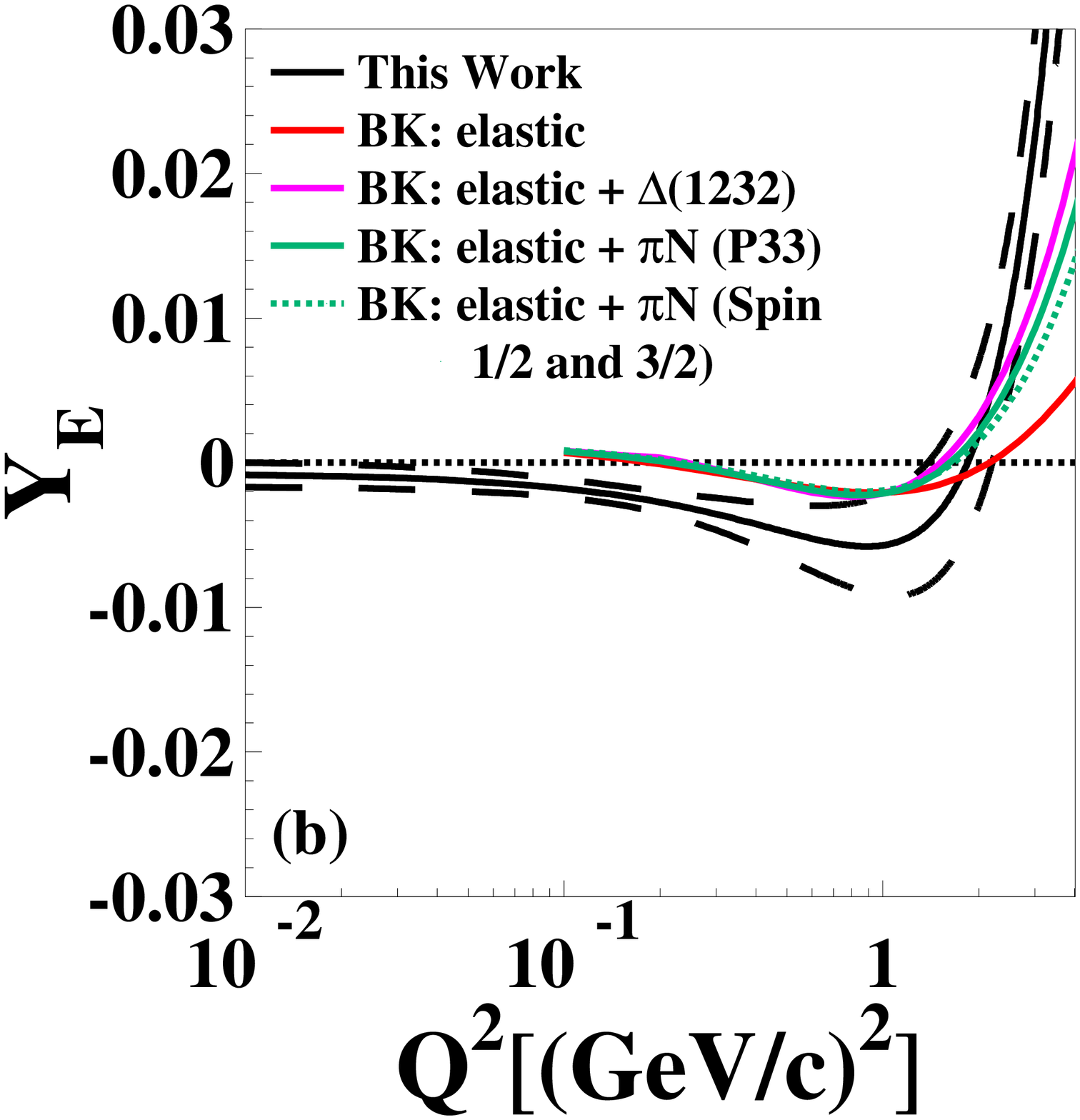}
\includegraphics*[width=8.2cm,height=7cm]{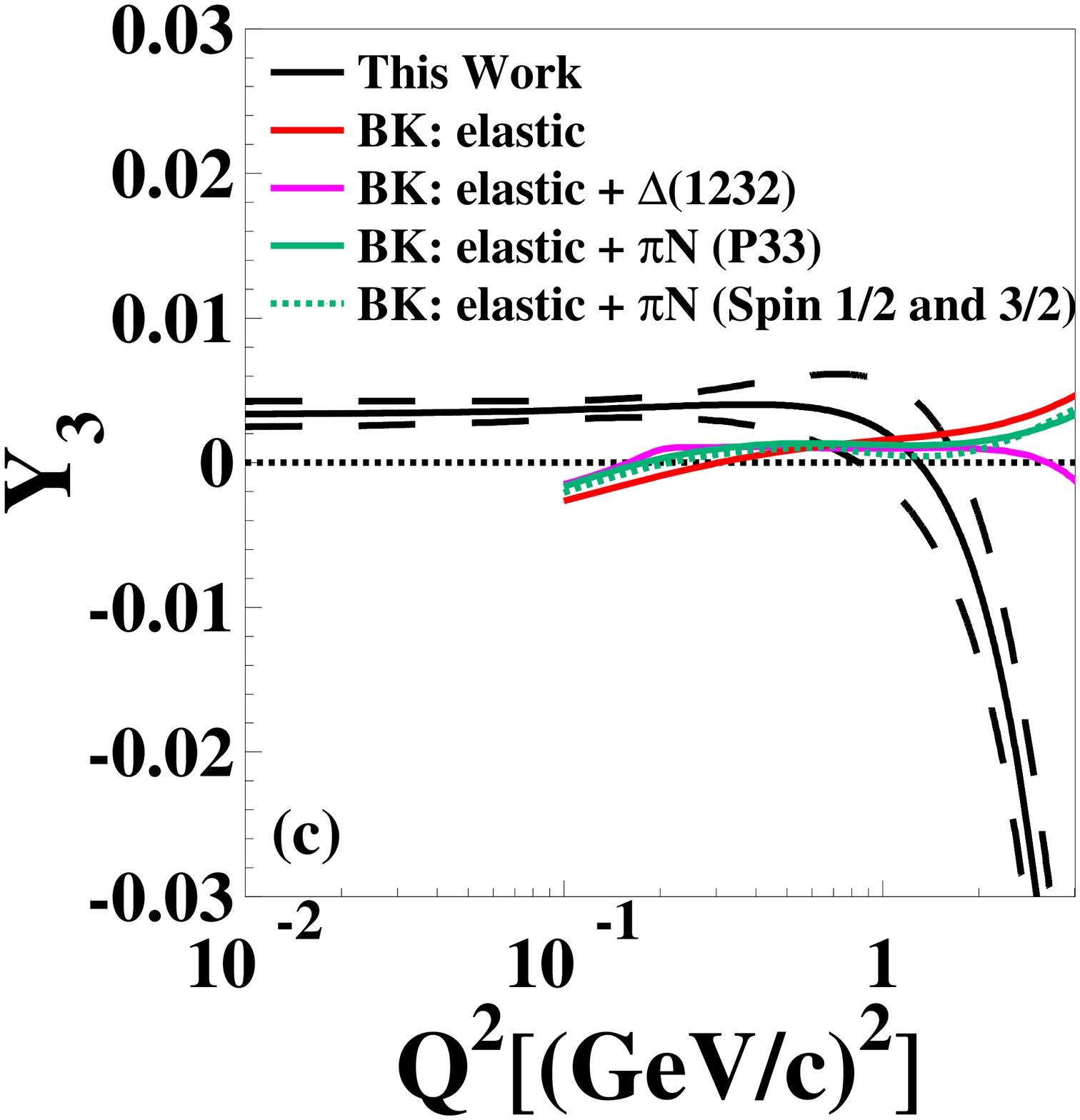} 
\vspace{-0.5cm}
\end{center}
\caption{(Color online) The extracted TPE amplitudes from this work: $Y_M$ (top), $Y_E$ (middle), and $Y_3$ 
(bottom) as a function of $Q^2$ at fixed $\varepsilon =$ 0.25 (solid black lines). The error bands on the amplitudes
are shown as very long-dashed black lines. In addition, I compare the results 
to several previous hadronic TPE calculations assuming different intermediate states: elastic 
``BK: elastic''~\cite{borisyuk08} (solid red line), elastic $+$ $\Delta$(1232) ``BK: $\Delta$ (1232)''~\cite{borisyuk12} 
(solid magenta line), elastic $+$ $\pi N$ ($P_{33}$) channel ``BK: elastic $+$ $\pi N$ ($P_{33}$)''~\cite{borisyuk14} 
(solid dark-green line), and elastic $+$ $\pi N$ (Spin 1/2 and 3/2 Channels) ``BK: elastic $+$ $\pi N$ (Spin 1/2 and 3/2)''~\cite{borisyuk15} (dashed dark-green line).}
\label{fig:TpeAmplVsQ2}
\end{figure}

Figure~\ref{fig:TpeAmplVsQ2} shows the $Q^2$ dependence of the TPE amplitudes as extracted from this work 
at a fixed $\varepsilon =$ 0.25. The error bands on the amplitudes are shown as very long-dashed black lines. 
I also compare the results to hadronic TPE calculations arising
from: elastic ``BK: elastic''~\cite{borisyuk08}, elastic $+$ $\Delta$(1232) resonance 
``BK: elastic $+$ $\Delta$(1232)''~\cite{borisyuk12}, elastic $+$ $\pi N$ ($P_{33}$) channel 
``BK: elastic $+$ $\pi N$ ($P_{33}$)''~\cite{borisyuk14}, and elastic $+$ $\pi N$ (Spin 1/2 and 3/2) channels 
``BK: elastic $+$ $\pi N$ (Spin 1/2 and 3/2)''~\cite{borisyuk15}. For $Y_M$ and $Y_E$, although my results 
fall below theoretical predictions, they generally are in reasonable qualitative agreement with 
calculations, all showing a fall and then rise of both amplitudes with increasing $Q^2$. For $Y_M$, and
at large $Q^2$ values, the error band is large and my results are closer to calculations assuming a pure proton 
in the intermediate state (elastic) suggesting that $Y_M$ is influenced mainly by the elastic contribution where 
the inelastic contributions are smaller. The amplitudes $Y_E$ and $Y_3$ are more constrained than $Y_M$ as suggested
by the computed error bands. For $Y_E$, and at large $Q^2$ values, my results are closer to calculations 
assuming inelastic contributions, elastic $+$ $\Delta(1232)$ resonance and to a lesser extent elastic $+$ $\pi N$, 
suggesting that $Y_E$ is influenced mainly by inelastic contributions at large $Q^2$ values. For $Y_3$, the amplitude 
is flat and above calculations up to $Q^2 \sim$ 1.0 (GeV/c)$^2$ where it starts to fall-off rapidly with increasing 
$Q^2$ disagreeing noticeably with theoretical predictions except for calculations assuming elastic $+$ $\Delta$(1232) 
resonance which predict a slower fall-off of the amplitude at large $Q^2$ values. The tension between the $Y_E$ and 
$Y_3$ amplitudes is obvious where they tend partially to cancel each other indicating that the TPE correction to 
$\sigma_R$ is driven mainly by $Y_M$ and to a lesser extent by $Y_3$.
 
\begin{figure}[!htbp]
\begin{center}
\begin{tabular}{c c}
\includegraphics*[width=4.25cm]{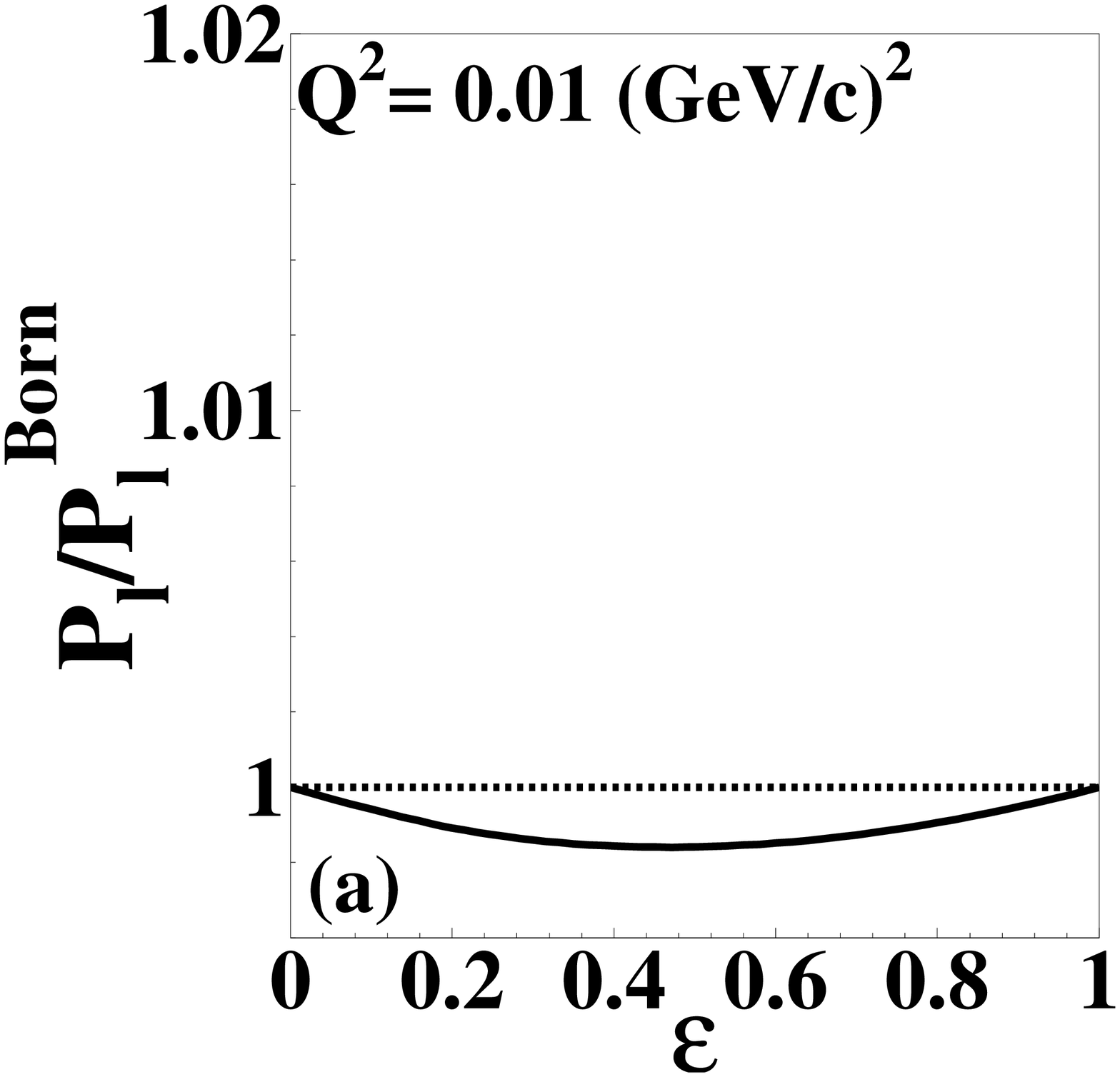} &
\includegraphics*[width=4.25cm]{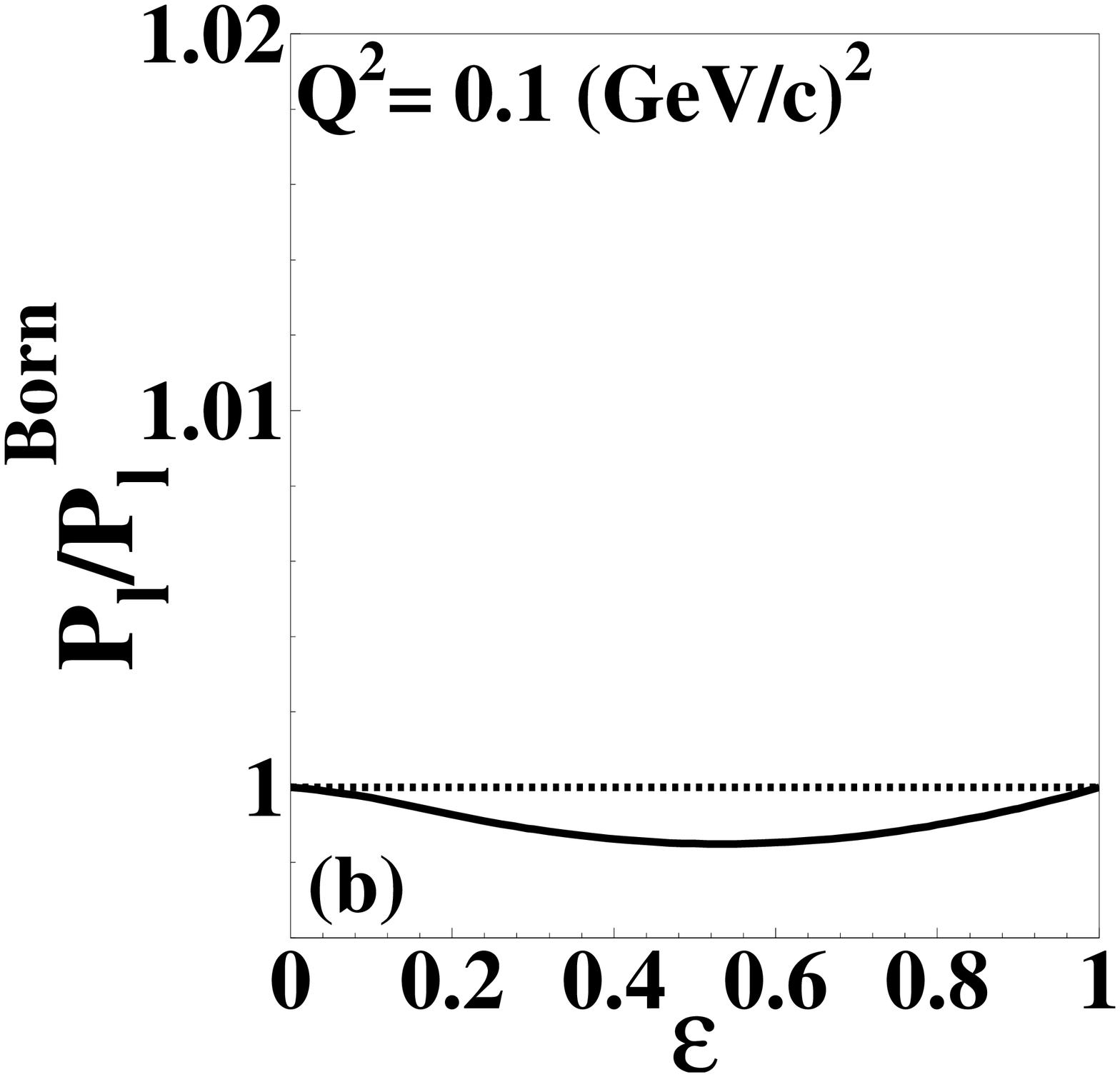} \\
\includegraphics*[width=4.25cm]{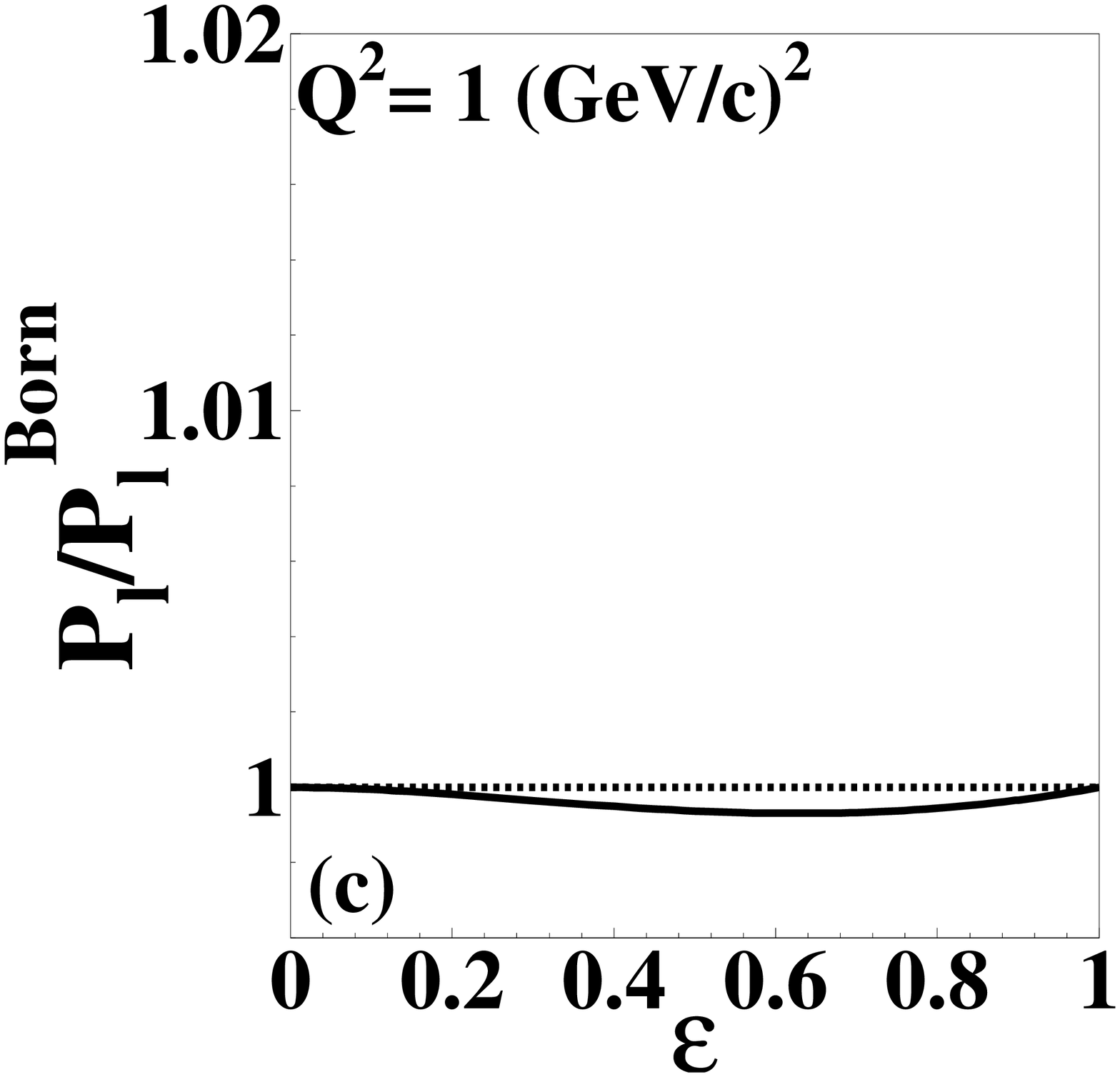} &
\includegraphics*[width=4.25cm]{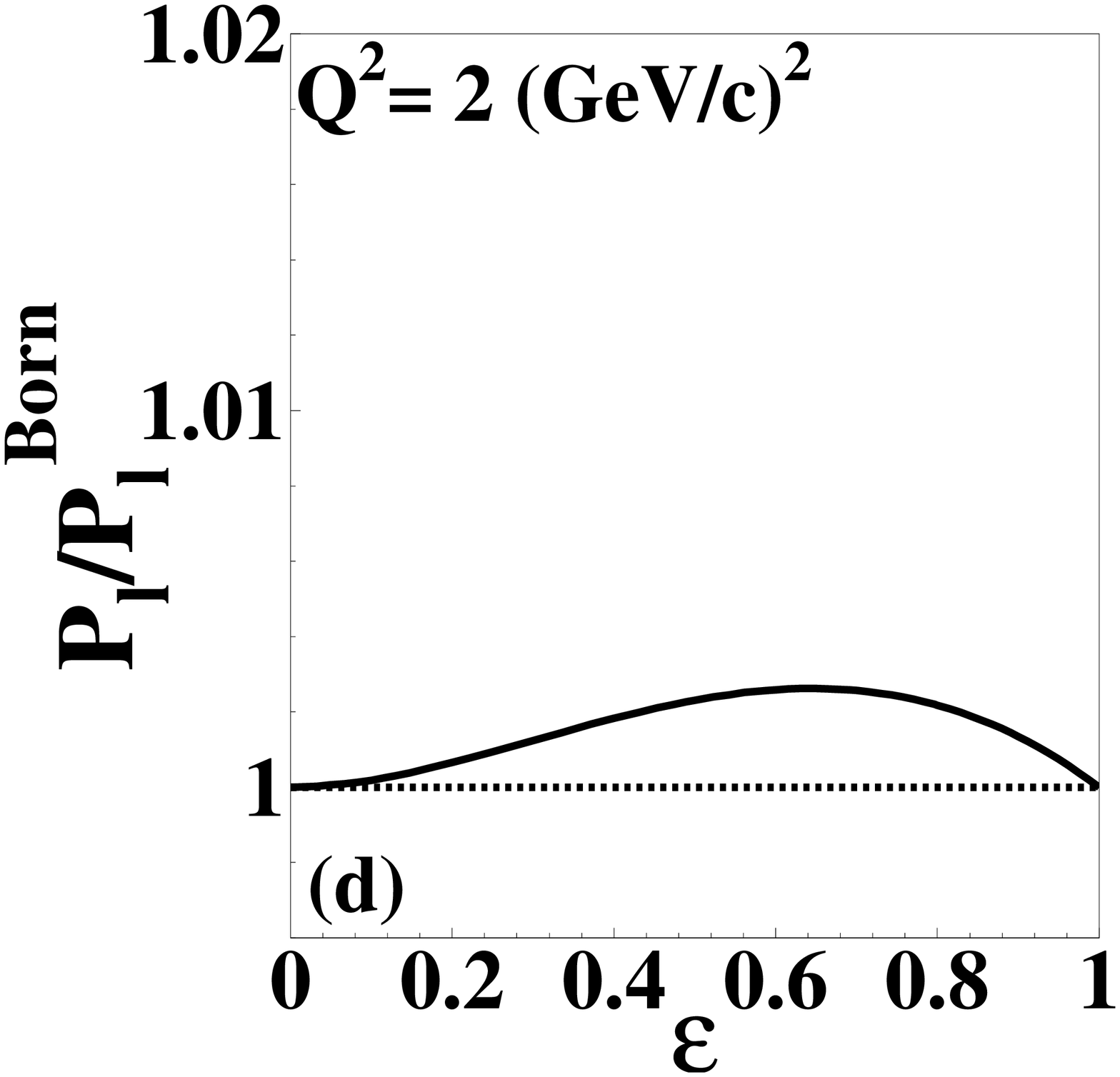} \\
\includegraphics*[width=4.25cm]{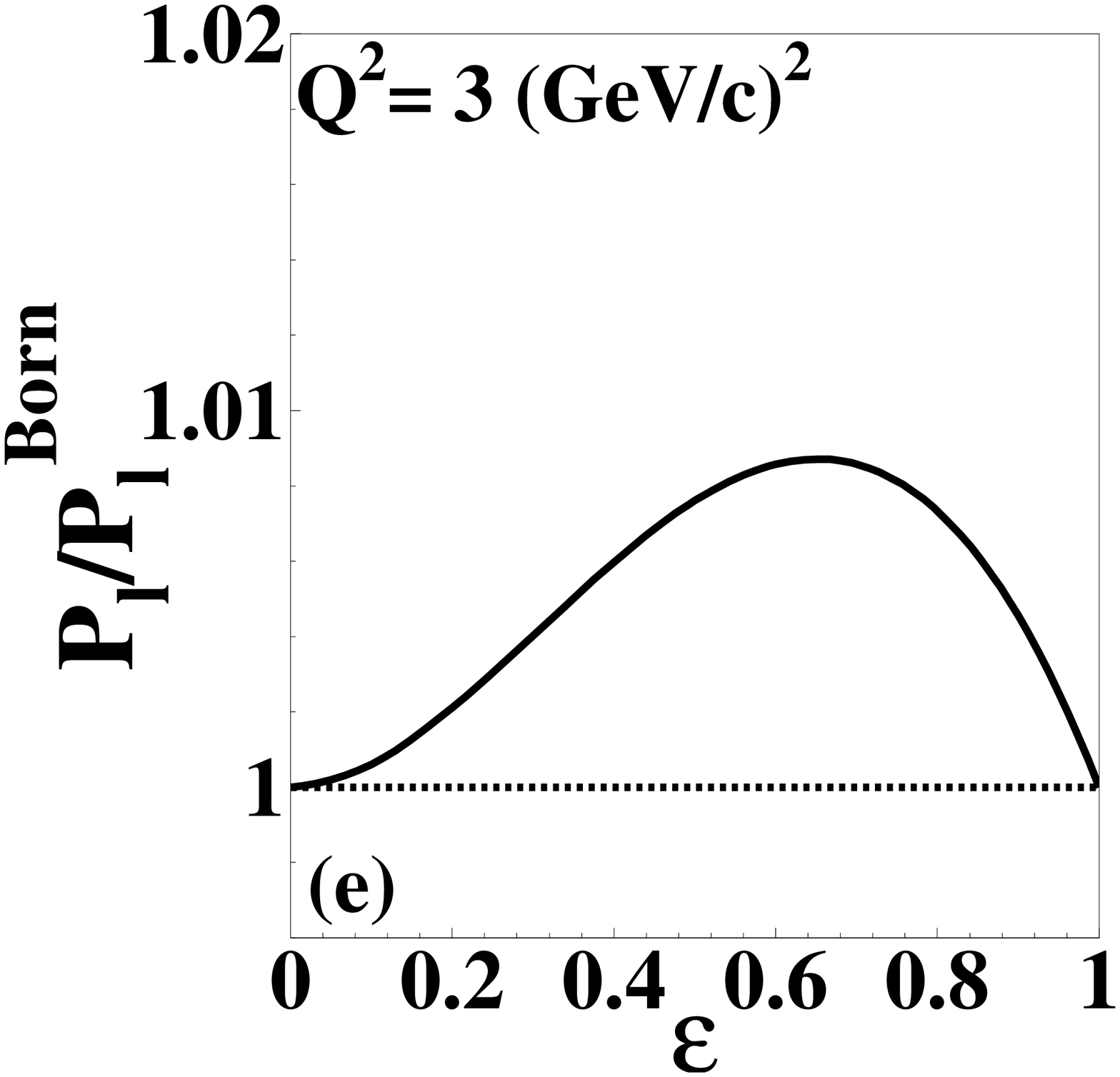} &
\includegraphics*[width=4.25cm]{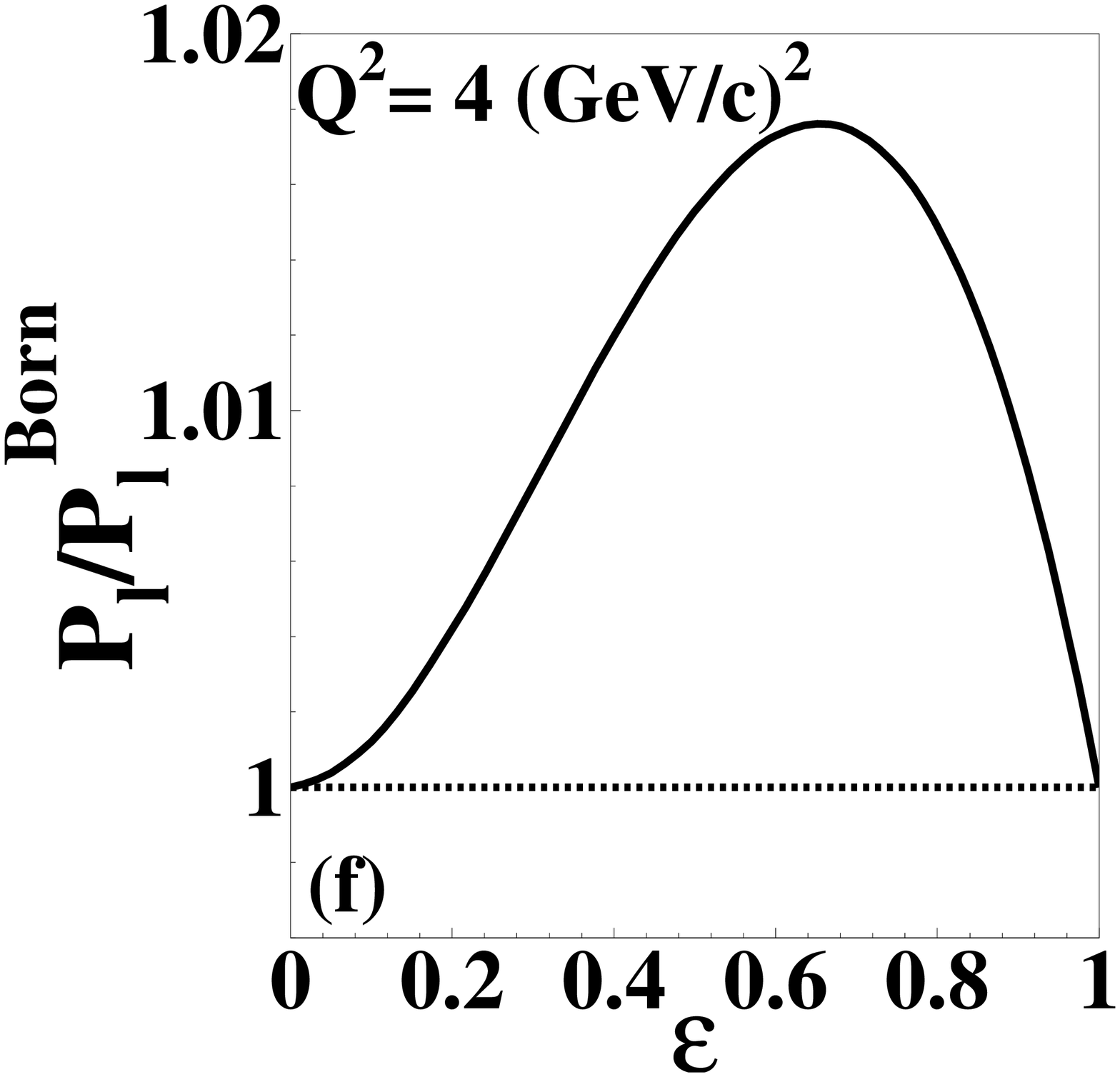} \\
\vspace{-0.5cm}
\end{tabular}
\end{center}
\caption{(Color online) The ratio $P_l/P_{l}^{\mbox{Born}}$ as a function of $\varepsilon$ as 
determined using my parametrizations of the TPE amplitudes and Eq.~(9c) at the $Q^2$ value listed in the figure (solid black line).}
\label{fig:PlPlBorn}
\end{figure}

\begin{figure}[!htbp]
\begin{center}
\includegraphics*[width=8.2cm]{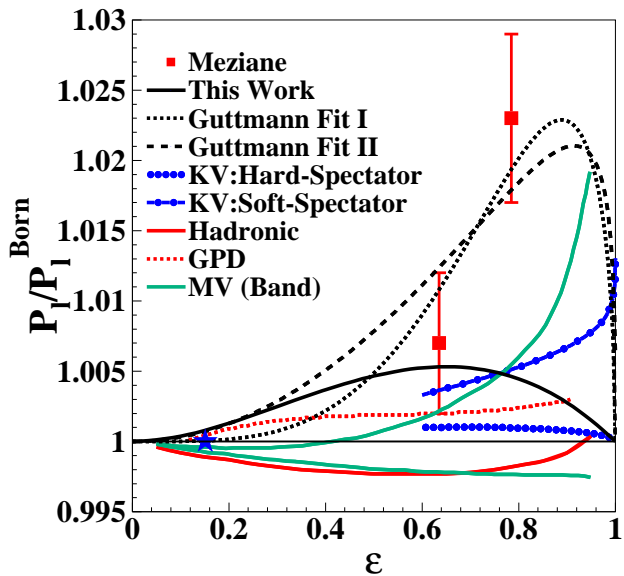}  
\vspace{-0.5cm}
\end{center}
\caption{(Color online) The extracted ratio $P_l/P_{l}^{\mbox{Born}}$ as a function of $\varepsilon$ at
$Q^2 = $ 2.50 (GeV/c)$^2$ from this work (solid black line).
Also shown the data points from the GEp-$2\gamma$ experiment Ref.~\cite{meziane11} (solid red squares). The blue 
star indicates the value of $\varepsilon$ at which the data have been normalized to unity. 
In addition, I compare the results to previous fits from Ref.~\cite{guttmann11}
``Guttmann Fit I'' (dashed black line) and ``Guttmann Fit II'' (long-dashed black line), and to 
several previous theoretical predictions: ``Hadronic''~\cite{blunden05a} (solid red line), 
``GPD''~\cite{afanasev05} (dashed red line), ``KV-Soft Spectator'' (solid blue line) and  
``KV-hard Spectator'' (dashed blue line)~\cite{kivel13}, and ``MV''~\cite{tomalak15a} prediction band 
(solid dark-green lines).}
\label{fig:PlPlBorn250}
\end{figure}

The $\varepsilon$ dependence of the ratio $P_l/P_{l}^{\mbox{Born}}$ as extracted from this work, 
using Eq.~(9c), is shown in Fig.~\ref{fig:PlPlBorn} for a range of $Q^2$ values.
At low $Q^2$, the ratio is below unity and shows little sensitivity to $\varepsilon$. 
The ratio increases with increasing $Q^2$ and changes sign, 
where it shows a sign of enhancement with $\varepsilon$ at large $Q^2$ values.
Figure~\ref{fig:PlPlBorn250} shows the $\varepsilon$ dependence of 
the ratio $P_l/P_{l}^{\mbox{Born}}$ as extracted from this work at $Q^2 =$ 2.50 (GeV/c)$^2$. 
I also compare the results to the experimental data points from the GEp-$2\gamma$ collaboration~\cite{meziane11}, 
previous fits from Ref.~\cite{guttmann11}, labelled as ``Guttmann Fit I'' and 
``Guttmann Fit II'', as well as to several theoretical predictions: hadronic TPE calculation from Ref.~\cite{blunden05a} 
``Hadronic'' where all the proton intermediate states are accounted for, the partonic model from Ref.~\cite{afanasev05} 
which accounts for hard scattering of electrons by quarks embedded in the nucleon through generalized parton 
distributions ``GPD'', calculation based on QCD factorization within the SCET approach from Ref.~\cite{kivel13} 
arising from both the soft-spectator ``KV-Soft Spectator'' and hard-spectator 
``KV-Hard Spectator'' scattering contributions, and calculation based on subtracted dispersion 
relation formalism applied to the case of a nucleon in the intermediate state  
from Ref.~\cite{tomalak15a} ``MV''. While my results show enhancement with $\varepsilon$, and 
are in good qualitative agreement with the experimental data and previous 
fits at low $\varepsilon$, they 
disagree strongly at large $\varepsilon$. On the other hand, my results disagree noticeably with theoretical 
predictions for the entire $\varepsilon$ range. The non-vanishing of the ratio 
$P_l/P_{l}^{\mbox{Born}}$ and the TPE amplitudes $Y_{(M,3)}$ at $\varepsilon =$ 1 is clearly 
a behaviour that is mainly associated with the soft-spectator contribution. 

%

\section{conclusions} \label{conclusions}
In conclusion, I improved on and extended to low- and high-$Q^2$ values the previous phenomenological 
extractions~\cite{guttmann11} of the three TPE amplitudes and the ratio $P_l/P_{l}^{\mbox{Born}}$ based
on the formalism of Guichon and Vanderhaeghen~\cite{guichon03}. Assuming that the TPE correction is responsible
mainly for the discrepancy between the cross section and polarization data measurements, and because the
recoil polarization data were confirmed {\emph{``experimentally''}} to be essentially independent of 
$\varepsilon$~\cite{meziane11}, I constrained the ratio $-\sqrt{\tau(1+\varepsilon)/(2\varepsilon)} P_{t}/P_{l}$ in Eq.~(9b)
to its $\varepsilon$-independent term (Born value) by setting the TPE contributions to zero. That allowed 
for $\sigma_{R}/(G_M^p)^2$, Eq.~(\ref{eq:Guttmann1}), to be expressed in terms of the remaining $Y_E(\varepsilon,Q^2)$ and $Y_3(\varepsilon,Q^2)$ amplitudes. As these amplitudes are functions of $\varepsilon$ and $Q^2$,
I expand each of the amplitudes $Y_E$ and $Y_3$ as a second-order polynomial to reserve as possible the linearity  
of $\sigma_R$ as well as to account for possible nonlinearities in the TPE amplitudes. Further, I 
imposed the Regge limit where the TPE correction to $\sigma_{R}$ and the TPE amplitudes 
vanishes in the limit $\varepsilon \rightarrow 1$ allowing for $\sigma_{R}/(G_M^p)^2$ to be expressed in its final
form as given by Eq.~(\ref{eq:constrain5}). By fixing $(G_M^p)^2$ and the recoil 
polarization ratio $R$ in Eq.~(\ref{eq:constrain5}), I fit the world data on $\sigma_R$ used in 
the analysis of Ref.~\cite{qattan15} and extract the TPE amplitudes coefficients $\alpha_{(0,1,2)}(Q^2)$ and 
$\beta_{(0,1,2)}(Q^2)$ which were then used to construct the three TPE amplitudes and the ratio
$P_l/P_{l}^{\mbox{Born}}$. My $\sigma_{R}$ fits described the data remarkably well with some 
deviation from linearity observed at low $\varepsilon$ for the high $Q^2$ points in agreement with
several hadronic- and pQCD-based calculations in this range.

For $(G_E^p/G_D(Q^2))^2$, my results in general are in reasonably good
agreement with extractions using calculated TPE corrections and phenomenological-based fits.
For $(G_M^p/\mu_pG_D(Q^2))^2$ and at low $Q^2$, my results are significantly above most previous fits. 
This reflects the discrepancy between the recent Mainz data which yields values of $G_M^p$ which are 
systematically 2--5\% larger than previous world data~\cite{bernauer14}.

The extracted TPE amplitudes coefficients, $\alpha_{k}$ and $\beta_{k}$ ($k=0,1,2$), are at the 
few-percentage-points level, and all show hints of oscillatory behaviour below $ Q^2 =$ 1.0 (GeV/c)$^2$ 
with clear sign of structure (dips/bumps) at $Q^2 \approx$ 0.02 (GeV/c)$^2$. For $Q^2 \gtorder$ 1.0 (GeV/c)$^2$,
the coefficients changed sign and increased in magnitude 
with increasing $Q^2$. They were all best parametrized using second-order polynomials. See Table \ref{fitsparam}.

The extracted TPE amplitudes from this work are on the few-percentage-points level, and behave roughly linearly 
with $\varepsilon$ as $Q^2$ increases where they become nonlinear at high $Q^2$ values. The amplitudes $Y_M$ and $Y_3$ 
are mainly positive at low $Q^2$ and change sign and grow in magnitude with increasing $Q^2$. On the other hand, 
the amplitude $Y_E$ is negligible and mainly negative at low $Q^2$ and changes sign and grows in magnitude 
with increasing $Q^2$. While the $Y_E$ and $Y_3$ amplitudes differ in magnitude, they certainly have 
opposite sign to each other as $Q^2$ increases where they tend to partially cancel each other. This suggests 
that the TPE correction to $\sigma_R$ is driven 
mainly by $Y_M$ and to a lesser extent by $Y_3$ in agreement with previous phenomenological
extractions~\cite{guttmann11} at $Q^2 =$ 2.50 (GeV/c)$^2$. My extractions 
for $Y_M$ and $Y_E$ are also in reasonable qualitative agreement with previous hadronic 
calculations~\cite{borisyuk08,borisyuk12,borisyuk14,borisyuk15}, all showing the fall and rise of both 
amplitudes with increasing $Q^2$. However, my $Y_3$ has opposite sign and deviates strongly from both previous 
extractions and theoretical predictions except at $Q^2 \sim $ 1.0 (GeV/c)$^2$.

I also investigated the $Q^2$ dependence of the TPE amplitudes at a fixed $\varepsilon =$ 0.25. 
Both $Y_M$ and $Y_E$ fall below theoretical predictions, but they generally are in  
reasonable qualitative agreement with hadronic theoretical predictions, all showing a fall and then 
rise of both amplitudes with increasing $Q^2$. At large $Q^2$, and while my results for $Y_M$ 
are closer to calculations assuming a pure proton in the intermediate state 
(elastic) suggesting that $Y_M$ is influenced mainly by elastic contributions, my results for
$Y_E$ are closer to calculations assuming inelastic contributions, mainly elastic $+$ $\Delta(1232)$ resonance 
and to a lesser extent elastic $+$ $\pi N$, suggesting that $Y_E$ is driven mainly by these contributions.
For $Y_3$ and at low $Q^2$, the amplitude is flat and above calculations up to 
$Q^2 \sim$ 1.0 (GeV/c)$^2$. At high $Q^2$, $Y_3$ starts to fall-off rapidly with increasing $Q^2$ 
agreeing only with calculations assuming elastic $+$ $\Delta$(1232) resonance which predict 
a slower fall-off at large $Q^2$. I also see clear tension between the $Y_E$ and $Y_3$ amplitudes  
where they tend partially to cancel each other suggesting that the TPE correction to $\sigma_R$ is 
driven mainly by $Y_M$ and to a lesser extent by $Y_3$.
 
The $\varepsilon$ dependence of the ratio $P_l/P_{l}^{\mbox{Born}}$ as extracted from this work
suggests that the ratio is below unity with little sensitivity to $\varepsilon$ at low $Q^2$ values. 
With increasing $Q^2$, the ratio increases and changes sign with a sign of enhancement with $\varepsilon$.
The results at $Q^2 =$ 2.50 (GeV/c)$^2$ show clear enhancement with $\varepsilon$, and 
are in good qualitative agreement with the experimental data from the GEp-$2\gamma$ 
collaboration~\cite{meziane11} and previous fits of Ref.~\cite{guttmann11} at low $\varepsilon$,
but they disagree noticeably at large $\varepsilon$. On the other hand, my results disagree strongly
with theoretical predictions for the entire $\varepsilon$ range.

\begin{acknowledgments}
This work was supported by Khalifa University of Science and Technology. 
I am grateful to Dr. Dmitry Borisyuk for providing me with his 
hadronic TPE calculations. I also thank Dr. D. Homouz for his technical assistance, 
and Dr. S. Phoenix for reading the manuscript and making valuable comments and suggestions.    
\end{acknowledgments}

\bibliography{longpaper_TPEparam}

\begin{thebibliography}{120}
\expandafter\ifx\csname natexlab\endcsname\relax\def\natexlab#1{#1}\fi
\expandafter\ifx\csname bibnamefont\endcsname\relax
  \def\bibnamefont#1{#1}\fi
\expandafter\ifx\csname bibfnamefont\endcsname\relax
  \def\bibfnamefont#1{#1}\fi
\expandafter\ifx\csname citenamefont\endcsname\relax
  \def\citenamefont#1{#1}\fi
\expandafter\ifx\csname url\endcsname\relax
  \def\url#1{\texttt{#1}}\fi
\expandafter\ifx\csname urlprefix\endcsname\relax\def\urlprefix{URL }\fi
\providecommand{\bibinfo}[2]{#2}
\providecommand{\eprint}[2][]{\url{#2}}

\bibitem[{\citenamefont{Sachs}(1962)}]{sachs62}
\bibinfo{author}{\bibfnamefont{R.~G.} \bibnamefont{Sachs}},
  \bibinfo{journal}{Phys. Rev.} \textbf{\bibinfo{volume}{126}},
  \bibinfo{pages}{2256} (\bibinfo{year}{1962}).
\bibitem[{\citenamefont{Rosenbluth}(1950)}]{rosenbluth50}
\bibinfo{author}{\bibfnamefont{M.~N.} \bibnamefont{Rosenbluth}},
  \bibinfo{journal}{Phys. Rev.} \textbf{\bibinfo{volume}{79}},
  \bibinfo{pages}{615} (\bibinfo{year}{1950}).
\bibitem[{\citenamefont{Dombey}(1969)}]{dombey69}
\bibinfo{author}{\bibfnamefont{N.}~\bibnamefont{Dombey}},
  \bibinfo{journal}{Rev. Mod. Phys.} \textbf{\bibinfo{volume}{41}},
  \bibinfo{pages}{236} (\bibinfo{year}{1969}).
\bibitem[{\citenamefont{Akhiezer and Rekalo}(1974)}]{akhiezer74}
\bibinfo{author}{\bibfnamefont{A.~I.} \bibnamefont{Akhiezer}} \bibnamefont{and}
  \bibinfo{author}{\bibfnamefont{M.~P.} \bibnamefont{Rekalo}},
  \bibinfo{journal}{Sov. J. Part. Nucl.} \textbf{\bibinfo{volume}{4}},
  \bibinfo{pages}{277} (\bibinfo{year}{1974}).
\bibitem[{\citenamefont{Arnold et~al.}(1981)\citenamefont{Arnold, Carlson, and Gross}}]{arnold81}
\bibinfo{author}{\bibfnamefont{R.~G.} \bibnamefont{Arnold}},
  \bibinfo{author}{\bibfnamefont{C.~E.} \bibnamefont{Carlson}},
  \bibnamefont{and} \bibinfo{author}{\bibfnamefont{F.}~\bibnamefont{Gross}},
  \bibinfo{journal}{Phys. Rev.} \textbf{\bibinfo{volume}{C23}},
  \bibinfo{pages}{363} (\bibinfo{year}{1981}).
\bibitem[{\citenamefont{Qattan et~al.}(2005)}]{qattan05}
\bibinfo{author}{\bibfnamefont{I.~A.} \bibnamefont{Qattan}}
  \bibnamefont{\emph{et al.}}, \bibinfo{journal}{Phys. Rev. Lett.}
  \textbf{\bibinfo{volume}{94}}, \bibinfo{pages}{142301}
  (\bibinfo{year}{2005}).
\bibitem[{\citenamefont{Arrington et~al.}(2007{\natexlab{a}})\citenamefont{Arrington, Roberts, and Zanotti}}]{arrington07a}
\bibinfo{author}{\bibfnamefont{J.}~\bibnamefont{Arrington}},
  \bibinfo{author}{\bibfnamefont{C.}~\bibnamefont{Roberts}}, \bibnamefont{and}
  \bibinfo{author}{\bibfnamefont{J.}~\bibnamefont{Zanotti}},
  \bibinfo{journal}{J. Phys.} \textbf{\bibinfo{volume}{G34}},
  \bibinfo{pages}{S23} (\bibinfo{year}{2007}{\natexlab{a}}).
\bibitem[{\citenamefont{Perdrisat et~al.}(2007)\citenamefont{Perdrisat, Punjabi, and Vanderhaeghen}}]{perdrisat07}
\bibinfo{author}{\bibfnamefont{C.}~\bibnamefont{Perdrisat}},
  \bibinfo{author}{\bibfnamefont{V.}~\bibnamefont{Punjabi}}, \bibnamefont{and}
  \bibinfo{author}{\bibfnamefont{M.}~\bibnamefont{Vanderhaeghen}},
  \bibinfo{journal}{Prog. Part. Nucl. Phys.} \textbf{\bibinfo{volume}{59}},
  \bibinfo{pages}{694} (\bibinfo{year}{2007}).
\bibitem[{\citenamefont{Guichon and Vanderhaeghen}(2003)}]{guichon03}
\bibinfo{author}{\bibfnamefont{P.~A.~M.} \bibnamefont{Guichon}}
  \bibnamefont{and}
  \bibinfo{author}{\bibfnamefont{M.}~\bibnamefont{Vanderhaeghen}},
  \bibinfo{journal}{Phys. Rev. Lett.} \textbf{\bibinfo{volume}{91}},
  \bibinfo{pages}{142303} (\bibinfo{year}{2003}).
\bibitem[{\citenamefont{Arrington}(2003)}]{arrington03a}
\bibinfo{author}{\bibfnamefont{J.}~\bibnamefont{Arrington}},
  \bibinfo{journal}{Phys. Rev.} \textbf{\bibinfo{volume}{C68}},
  \bibinfo{pages}{034325} (\bibinfo{year}{2003}).
\bibitem[{\citenamefont{Arrington}(2004)}]{arrington04a}
\bibinfo{author}{\bibfnamefont{J.}~\bibnamefont{Arrington}},
  \bibinfo{journal}{Phys. Rev.} \textbf{\bibinfo{volume}{C69}},
  \bibinfo{pages}{022201 (R)} (\bibinfo{year}{2004}).
\bibitem[{\citenamefont{Blunden et~al.}(2003)\citenamefont{Blunden,Melnitchouk, and Tjon}}]{blunden03}
\bibinfo{author}{\bibfnamefont{P.~G.} \bibnamefont{Blunden}},
  \bibinfo{author}{\bibfnamefont{W.}~\bibnamefont{Melnitchouk}},
  \bibnamefont{and} \bibinfo{author}{\bibfnamefont{J.~A.} \bibnamefont{Tjon}},
  \bibinfo{journal}{Phys. Rev. Lett.} \textbf{\bibinfo{volume}{91}},
  \bibinfo{pages}{142304} (\bibinfo{year}{2003}).
\bibitem[{\citenamefont{Blunden et~al.}(2005)\citenamefont{Blunden, Melnitchouk, and Tjon}}]{blunden05a}
\bibinfo{author}{\bibfnamefont{P.~G.} \bibnamefont{Blunden}},
  \bibinfo{author}{\bibfnamefont{W.}~\bibnamefont{Melnitchouk}},
  \bibnamefont{and} \bibinfo{author}{\bibfnamefont{J.~A.} \bibnamefont{Tjon}},
  \bibinfo{journal}{Phys. Rev.} \textbf{\bibinfo{volume}{C72}},
  \bibinfo{pages}{034612} (\bibinfo{year}{2005}).
\bibitem[{\citenamefont{Kondratyuk et~al.}(2005)\citenamefont{Kondratyuk, Blunden, Melnitchouk, and Tjon}}]{kondratyuk05}
\bibinfo{author}{\bibfnamefont{S.}~\bibnamefont{Kondratyuk}},
  \bibinfo{author}{\bibfnamefont{P.~G.} \bibnamefont{Blunden}},
  \bibinfo{author}{\bibfnamefont{W.}~\bibnamefont{Melnitchouk}},
  \bibnamefont{and} \bibinfo{author}{\bibfnamefont{J.~A.} \bibnamefont{Tjon}},
  \bibinfo{journal}{Phys. Rev. Lett.} \textbf{\bibinfo{volume}{95}},
  \bibinfo{pages}{172503} (\bibinfo{year}{2005}).
\bibitem[{\citenamefont{Kondratyuk et~al.}(2007)\citenamefont{Kondratyuk and Blunden}}]{kondratyuk07}
\bibinfo{author}{\bibfnamefont{S.}~\bibnamefont{Kondratyuk}}
\bibnamefont{and} \bibinfo{author}  \bibinfo{author}{\bibfnamefont{P.~G.} \bibnamefont{Blunden}},
  \bibinfo{journal}{Phys. Rev.} \textbf{\bibinfo{volume}{C75}},
  \bibinfo{pages}{038201} (\bibinfo{year}{2007}).
\bibitem[{\citenamefont{Kivel and Vanderhaeghen}(2009)}]{kivel09}
\bibinfo{author}{\bibfnamefont{N.}~\bibnamefont{Kivel}} \bibnamefont{and}
  \bibinfo{author}{\bibfnamefont{M.}~\bibnamefont{Vanderhaeghen}},
  \bibinfo{journal}{Phys. Rev. Lett.} \textbf{\bibinfo{volume}{103}},
  \bibinfo{pages}{092004} (\bibinfo{year}{2009}).
\bibitem[{\citenamefont{Kivel and Vanderhaeghen}(2011)}]{kivel11}
\bibinfo{author}{\bibfnamefont{N.}~\bibnamefont{Kivel}} \bibnamefont{and}
  \bibinfo{author}{\bibfnamefont{M.}~\bibnamefont{Vanderhaeghen}},
  \bibinfo{journal}{Phys. Rev.} \textbf{\bibinfo{volume}{D83}},
  \bibinfo{pages}{093005} (\bibinfo{year}{2011}).
\bibitem[{\citenamefont{Kivel and Vanderhaeghen}(2013)}]{kivel13}
\bibinfo{author}{\bibfnamefont{N.}~\bibnamefont{Kivel}} \bibnamefont{and}
  \bibinfo{author}{\bibfnamefont{M.}~\bibnamefont{Vanderhaeghen}},
  \bibinfo{journal}{J. High Energy Physics} \textbf{\bibinfo{volume}{1304}},
  \bibinfo{pages}{29} (\bibinfo{year}{2013}).
\bibitem[{\citenamefont{Tomalak and Vanderhaeghen}(2015)}]{tomalak15a}
\bibinfo{author}{\bibfnamefont{O.}~\bibnamefont{Tomalak}} \bibnamefont{and}
  \bibinfo{author}{\bibfnamefont{M.}~\bibnamefont{Vanderhaeghen}},
\bibinfo{journal}{Eur. Phys. J.} \textbf{\bibinfo{volume}{A51}},
 \bibinfo{pages}{24} (\bibinfo{year}{2015}).
\bibitem[{\citenamefont{Tomalak and Vanderhaeghen}(2015)}]{tomalak15b}
\bibinfo{author}{\bibfnamefont{O.}~\bibnamefont{Tomalak}} \bibnamefont{and}
  \bibinfo{author}{\bibfnamefont{M.}~\bibnamefont{Vanderhaeghen}},
  (\bibinfo{year}{2015}), \eprint{arXiv:1508.03759 [hep-ph]}.
\bibitem[{\citenamefont{Lorenz et~al.}(2015)\citenamefont{Lorenz, MeiBner, Hammer, and Dong}}]{lorenz15}
\bibinfo{author}{\bibfnamefont{I.~T.} \bibnamefont{Lorenz}},
  \bibinfo{author}{\bibfnamefont{Ulf-G.} \bibnamefont{MeiBner}},
  \bibinfo{author}{\bibfnamefont{H.-W.} \bibnamefont{Hammer}},
  \bibnamefont{and} \bibinfo{author}{\bibfnamefont{Y.-B.} \bibnamefont{Dong}},
  \bibinfo{journal}{Phys. Rev.} \textbf{\bibinfo{volume}{D91}},
  \bibinfo{pages}{014023} (\bibinfo{year}{2015}).
\bibitem[{\citenamefont{Chen et~al.}(2004)\citenamefont{Chen, Afanasev, Brodsky, Carlson, and Vanderhaeghen}}]{chen04}
\bibinfo{author}{\bibfnamefont{Y.~C.} \bibnamefont{Chen}},
  \bibinfo{author}{\bibfnamefont{A.}~\bibnamefont{Afanasev}},
  \bibinfo{author}{\bibfnamefont{S.~J.} \bibnamefont{Brodsky}},
  \bibinfo{author}{\bibfnamefont{C.~E.} \bibnamefont{Carlson}},
  \bibnamefont{and}
  \bibinfo{author}{\bibfnamefont{M.}~\bibnamefont{Vanderhaeghen}},
  \bibinfo{journal}{Phys. Rev. Lett.} \textbf{\bibinfo{volume}{93}},
  \bibinfo{pages}{122301} (\bibinfo{year}{2004}).
\bibitem[{\citenamefont{Afanasev et~al.}(2005)\citenamefont{Afanasev, Brodsky, Carlson, Chen, and Vanderhaeghen}}]{afanasev05}
  \bibinfo{author}{\bibfnamefont{A.}~\bibnamefont{Afanasev}},
  \bibinfo{author}{\bibfnamefont{S.~J.} \bibnamefont{Brodsky}},
  \bibinfo{author}{\bibfnamefont{C.~E.} \bibnamefont{Carlson}},
  \bibinfo{author}{\bibfnamefont{Y.~C.} \bibnamefont{Chen}},
  \bibnamefont{and}
  \bibinfo{author}{\bibfnamefont{M.}~\bibnamefont{Vanderhaeghen}},
  \bibinfo{journal}{Phys. Rev.} \textbf{\bibinfo{volume}{D93}},
  \bibinfo{pages}{013008} (\bibinfo{year}{2005}).
\bibitem[{\citenamefont{Bystritskiy et~al.}(2007)\citenamefont{Bystritskiy, Kuraev, and Tomasi-Gustafsson}}]{bystritskiy07}
  \bibinfo{author}{\bibfnamefont{Y.~M.} \bibnamefont{Bystritskiy}},
  \bibinfo{author}{\bibfnamefont{E.~A.} \bibnamefont{Kuraev}},
  \bibnamefont{and}
  \bibinfo{author}{\bibfnamefont{E.}~\bibnamefont{Tomasi-Gustafsson}},
  \bibinfo{journal}{Phys. Rev.} \textbf{\bibinfo{volume}{C75}},
  \bibinfo{pages}{015207} (\bibinfo{year}{2007}).
\bibitem[{\citenamefont{Tomasi-Gustafsson and Gakh}(2005)}]{gustafsson05}
\bibinfo{author}{\bibfnamefont{E.}~\bibnamefont{Tomasi-Gustafsson}}
  \bibnamefont{and} \bibinfo{author}{\bibfnamefont{G.~I.} \bibnamefont{Gakh}},
  \bibinfo{journal}{Phys. Rev. C} \textbf{\bibinfo{volume}{72}},
  \bibinfo{pages}{015209} (\bibinfo{year}{2005}).
\bibitem[{\citenamefont{Borisyuk and Kobushkin}(2006)}]{borisyuk06}
\bibinfo{author}{\bibfnamefont{D.}~\bibnamefont{Borisyuk}} \bibnamefont{and}
  \bibinfo{author}{\bibfnamefont{A.}~\bibnamefont{Kobushkin}},
  \bibinfo{journal}{Phys. Rev.} \textbf{\bibinfo{volume}{C74}},
  \bibinfo{pages}{065203} (\bibinfo{year}{2006}).
\bibitem[{\citenamefont{Borisyuk and Kobushkin}(2007)}]{borisyuk07}
\bibinfo{author}{\bibfnamefont{D.}~\bibnamefont{Borisyuk}} \bibnamefont{and}
  \bibinfo{author}{\bibfnamefont{A.}~\bibnamefont{Kobushkin}},
  \bibinfo{journal}{Phys. Rev.} \textbf{\bibinfo{volume}{C75}},
  \bibinfo{pages}{038202} (\bibinfo{year}{2007}).
\bibitem[{\citenamefont{Borisyuk and Kobushkin}(2008)}]{borisyuk08}
\bibinfo{author}{\bibfnamefont{D.}~\bibnamefont{Borisyuk}} \bibnamefont{and}
  \bibinfo{author}{\bibfnamefont{A.}~\bibnamefont{Kobushkin}},
  \bibinfo{journal}{Phys. Rev.} \textbf{\bibinfo{volume}{C78}},
  \bibinfo{pages}{025208} (\bibinfo{year}{2008}).
\bibitem[{\citenamefont{Borisyuk and Kobushkin}(2009)}]{borisyuk09}
\bibinfo{author}{\bibfnamefont{D.}~\bibnamefont{Borisyuk}} \bibnamefont{and}
  \bibinfo{author}{\bibfnamefont{A.}~\bibnamefont{Kobushkin}},
  \bibinfo{journal}{Phys. Rev.} \textbf{\bibinfo{volume}{D79}},
  \bibinfo{pages}{034001} (\bibinfo{year}{2009}).
\bibitem[{\citenamefont{Borisyuk and Kobushkin}(2012)}]{borisyuk12}
\bibinfo{author}{\bibfnamefont{D.}~\bibnamefont{Borisyuk}} \bibnamefont{and}
  \bibinfo{author}{\bibfnamefont{A.}~\bibnamefont{Kobushkin}},
  \bibinfo{journal}{Phys. Rev.} \textbf{\bibinfo{volume}{C86}},
  \bibinfo{pages}{055204} (\bibinfo{year}{2012}).
\bibitem[{\citenamefont{Borisyuk and Kobushkin}(2014)}]{borisyuk14}
\bibinfo{author}{\bibfnamefont{D.}~\bibnamefont{Borisyuk}} \bibnamefont{and}
  \bibinfo{author}{\bibfnamefont{A.}~\bibnamefont{Kobushkin}},
  \bibinfo{journal}{Phys. Rev.} \textbf{\bibinfo{volume}{C89}},
  \bibinfo{pages}{025204} (\bibinfo{year}{2014}).
\bibitem[{\citenamefont{Borisyuk and Kobushkin}(2015)}]{borisyuk15}
\bibinfo{author}{\bibfnamefont{D.}~\bibnamefont{Borisyuk}} \bibnamefont{and}
  \bibinfo{author}{\bibfnamefont{A.}~\bibnamefont{Kobushkin}},
  \bibinfo{journal}{Phys. Rev.} \textbf{\bibinfo{volume}{C92}},
  \bibinfo{pages}{035204} (\bibinfo{year}{2015}).
\bibitem[{\citenamefont{Zhou et~al.}(2007)}]{zhou07}
\bibinfo{author}{\bibfnamefont{Hai Qing}~\bibnamefont{Zhou}},
\bibinfo{author}{\bibfnamefont{Chung Wen}~\bibnamefont{Kao}},
\bibnamefont{and} \bibinfo{author}{\bibfnamefont{Shin Nan}~\bibnamefont{Yang}},
  \bibinfo{journal}{Phys. Rev. Lett.} \textbf{\bibinfo{volume}{99}},
  \bibinfo{pages}{262001} (\bibinfo{year}{2007}), \bibnamefont{and}
[Erratum-ibid. \bibinfo{journal}{Phys. Rev. Lett.} \textbf{\bibinfo{volume}{100}},
  \bibinfo{pages}{059903} (\bibinfo{year}{2008})].
\bibitem[{\citenamefont{Zhou}(2009)}]{zhou09}
\bibinfo{author}{\bibfnamefont{Hai-Qing}~\bibnamefont{Zhou}},
  \bibinfo{journal}{Chin. Phys. Lett.} \textbf{\bibinfo{volume}{26}},
  \bibinfo{pages}{061201} (\bibinfo{year}{2009}).
\bibitem[{\citenamefont{Zhou and Yang}(2015)}]{zhou15}
\bibinfo{author}{\bibfnamefont{Hai-Qing}~\bibnamefont{Zhou}},
\bibnamefont{and} \bibinfo{author}{\bibfnamefont{Shin Nan}~\bibnamefont{Yang}},
  \bibinfo{journal}{Eur. Phys. J.} \textbf{\bibinfo{volume}{A51}},
  \bibinfo{pages}{105} (\bibinfo{year}{2015}).
\bibitem[{\citenamefont{Graczyk}(2015)}]{graczyk15}
  \bibinfo{author}{\bibfnamefont{Krzysztof~M.} \bibnamefont{Graczyk}},
  \bibnamefont{and} \bibinfo{author}{\bibfnamefont{Cezary}~ \bibnamefont{Juszczak}},
  \bibinfo{journal}{J. Phys.} \textbf{\bibinfo{volume}{G42}},
  \bibinfo{pages}{034019} (\bibinfo{year}{2015}).
\bibitem[{\citenamefont{Graczyk}(2013)}]{graczyk13}
  \bibinfo{author}{\bibfnamefont{Krzysztof~M.} \bibnamefont{Graczyk}},
    \bibinfo{journal}{Phys. Rev.} \textbf{\bibinfo{volume}{C88}},
  \bibinfo{pages}{065205} (\bibinfo{year}{2013}).
\bibitem[{\citenamefont{Graczyk}(2011)}]{graczyk11}
  \bibinfo{author}{\bibfnamefont{Krzysztof~M.} \bibnamefont{Graczyk}},
    \bibinfo{journal}{Phys. Rev.} \textbf{\bibinfo{volume}{C84}},
  \bibinfo{pages}{034314} (\bibinfo{year}{2011}).
\bibitem[{\citenamefont{Braun et~al.}(2006)\citenamefont{Braun, Lenz, and Wittmann}}]{braun06}
  \bibinfo{author}{\bibfnamefont{V.~M.} \bibnamefont{Braun}},
  \bibinfo{author}{\bibfnamefont{A.}~\bibnamefont{Lenz}},
  \bibnamefont{and}
  \bibinfo{author}{\bibfnamefont{M.}~\bibnamefont{Wittmann}},
  \bibinfo{journal}{Phys. Rev.} \textbf{\bibinfo{volume}{D73}},
  \bibinfo{pages}{094019} (\bibinfo{year}{2006}).
\bibitem[{\citenamefont{Qattan}(2005)}]{qattanphd}
\bibinfo{author}{\bibfnamefont{I.~A.} \bibnamefont{Qattan}}, Ph.D. thesis,
  \bibinfo{school}{Northwestern University} (\bibinfo{year}{2005}),
  \bibinfo{note}{arXiv:nucl-ex/0610006}.
\bibitem[{\citenamefont{Tvaskis et~al.}(2006)}]{tvaskis06}
\bibinfo{author}{\bibfnamefont{V.}~\bibnamefont{Tvaskis}} \bibnamefont{\emph{et al.}},
  \bibinfo{journal}{Phys. Rev.} \textbf{\bibinfo{volume}{C73}},
  \bibinfo{pages}{025206} (\bibinfo{year}{2006}).
\bibitem[{\citenamefont{Borisyuk and Kobushkin}(2007)}]{borisyuk07b}
\bibinfo{author}{\bibfnamefont{D.}~\bibnamefont{Borisyuk}} \bibnamefont{and}
  \bibinfo{author}{\bibfnamefont{A.}~\bibnamefont{Kobushkin}},
  \bibinfo{journal}{Phys. Rev.} \textbf{\bibinfo{volume}{C76}},
  \bibinfo{pages}{022201(R)} (\bibinfo{year}{2007}).
\bibitem[{\citenamefont{Borisyuk and Kobushkin}(2011)}]{borisyuk11}
\bibinfo{author}{\bibfnamefont{D.}~\bibnamefont{Borisyuk}} \bibnamefont{and}
  \bibinfo{author}{\bibfnamefont{A.}~\bibnamefont{Kobushkin}},
  \bibinfo{journal}{Phys. Rev.} \textbf{\bibinfo{volume}{D83}},
  \bibinfo{pages}{057501} (\bibinfo{year}{2011}).
\bibitem[{\citenamefont{Chen et~al.}(2007)\citenamefont{Chen, Kao, and Yang}}]{chen07}
\bibinfo{author}{\bibfnamefont{Y.-C.} \bibnamefont{Chen}},
  \bibinfo{author}{\bibfnamefont{C.-W.} \bibnamefont{Kao}}, \bibnamefont{and}
  \bibinfo{author}{\bibfnamefont{S.-N.} \bibnamefont{Yang}},
  \bibinfo{journal}{Phys. Lett.} \textbf{\bibinfo{volume}{B652}},
  \bibinfo{pages}{269} (\bibinfo{year}{2007}).
\bibitem[{\citenamefont{Arrington}(2005)}]{arrington05}
\bibinfo{author}{\bibfnamefont{J.}~\bibnamefont{Arrington}},
  \bibinfo{journal}{Phys. Rev.} \textbf{\bibinfo{volume}{C71}},
  \bibinfo{pages}{015202} (\bibinfo{year}{2005}).
\bibitem[{\citenamefont{Guttmann et~al.}(2011)}]{guttmann11}
\bibinfo{author}{\bibfnamefont{J.}~\bibnamefont{Guttmann}},
  \bibinfo{author}{\bibfnamefont{N.} \bibnamefont{Kivel}}, 
  \bibinfo{author}{\bibfnamefont{M.} \bibnamefont{Meziane}},
\bibnamefont{and}
  \bibinfo{author}{\bibfnamefont{M.} \bibnamefont{Vanderhaeghen}},
  \bibinfo{journal}{Eur. Phys. J.} \textbf{\bibinfo{volume}{A47}},
  \bibinfo{pages}{77} (\bibinfo{year}{2011}).
\bibitem[{\citenamefont{Rekalo and Tomasi-Gustafsson}(2004)}]{rekalo04}
\bibinfo{author}{\bibfnamefont{M.~P.} \bibnamefont{Rekalo}} 
\bibnamefont{and} \bibinfo{author}{\bibfnamefont{E.}~\bibnamefont{Tomasi-Gustafsson}},
  \bibinfo{journal}{Eur. Phys. J.} \textbf{\bibinfo{volume}{A22}},
  \bibinfo{pages}{331} (\bibinfo{year}{2004}).
\bibitem[{\citenamefont{Qattan and Alsaad}(2011)}]{qattan11a}
\bibinfo{author}{\bibfnamefont{I.~A.} \bibnamefont{Qattan}} \bibnamefont{and}
  \bibinfo{author}{\bibfnamefont{A.}~\bibnamefont{Alsaad}},
  \bibinfo{journal}{Phys. Rev.} \textbf{\bibinfo{volume}{C83}},
  \bibinfo{pages}{054307} (\bibinfo{year}{2011}), [Erratum-ibid. 
  \textbf{\bibinfo{volume}{C84}},~\bibinfo{pages}{029905} (\bibinfo{year}{2011})].
\bibitem[{\citenamefont{Qattan, Alsaad, and Arrington}(2011)}]{qattan11b}
\bibinfo{author}{\bibfnamefont{I.~A.} \bibnamefont{Qattan}},
\bibinfo{author}{\bibfnamefont{A.~} \bibnamefont{Alsaad}}, \bibnamefont{and}
  \bibinfo{author}{\bibfnamefont{J.}~\bibnamefont{Arrington}},
  \bibinfo{journal}{Phys. Rev.} \textbf{\bibinfo{volume}{C84}},
  \bibinfo{pages}{054317} (\bibinfo{year}{2011}).
\bibitem[{\citenamefont{Qattan and Arrington}(2012)}]{qattan12}
\bibinfo{author}{\bibfnamefont{I.~A.} \bibnamefont{Qattan}},
\bibnamefont{and} \bibinfo{author}{\bibfnamefont{J.}~\bibnamefont{Arrington}},
  \bibinfo{journal}{Phys. Rev.} \textbf{\bibinfo{volume}{C86}},
  \bibinfo{pages}{065210} (\bibinfo{year}{2012}).
\bibitem[{\citenamefont{Qattan and Arrington}(2014)}]{qattan14}
\bibinfo{author}{\bibfnamefont{I.~A.} \bibnamefont{Qattan}},
\bibnamefont{and} \bibinfo{author}{\bibfnamefont{J.}~\bibnamefont{Arrington}},
  \bibinfo{journal}{Eur. Phys. J. WOC} \textbf{\bibinfo{volume}{66}},
  \bibinfo{pages}{06020} (\bibinfo{year}{2014}).
\bibitem[{\citenamefont{Qattan, Arrington, and Alsaad}(2015)}]{qattan15}
\bibinfo{author}{\bibfnamefont{I.~A.} \bibnamefont{Qattan}},
\bibinfo{author}{\bibfnamefont{J.}~\bibnamefont{Arrington}},
\bibnamefont{and} \bibinfo{author}{\bibfnamefont{A.~} \bibnamefont{Alsaad}},
  \bibinfo{journal}{Phys. Rev.} \textbf{\bibinfo{volume}{C91}},
  \bibinfo{pages}{065203} (\bibinfo{year}{2015}).
\bibitem[{\citenamefont{Arrington et~al.}(2011)\citenamefont{Arrington, Blunden, and Melnitchouk}}]{arrington11b}
\bibinfo{author}{\bibfnamefont{J.}~\bibnamefont{Arrington}},
  \bibinfo{author}{\bibfnamefont{P.}~\bibnamefont{Blunden}}, \bibnamefont{and}
  \bibinfo{author}{\bibfnamefont{W.}~\bibnamefont{Melnitchouk}},
  \bibinfo{journal}{Prog.Part.Nucl.Phys.} \textbf{\bibinfo{volume}{66}},
  \bibinfo{pages}{782} (\bibinfo{year}{2011}).
\bibitem[{\citenamefont{Carlson and Vanderhaeghen}(2007)}]{carlson07}
\bibinfo{author}{\bibfnamefont{C.~E.} \bibnamefont{Carlson}} \bibnamefont{and}
  \bibinfo{author}{\bibfnamefont{M.}~\bibnamefont{Vanderhaeghen}},
  \bibinfo{journal}{Ann. Rev. Nucl. Part. Sci.} \textbf{\bibinfo{volume}{57}},
  \bibinfo{pages}{171} (\bibinfo{year}{2007}).
\bibitem[{\citenamefont{Christy et~al.}(2004)}]{christy04}
\bibinfo{author}{\bibfnamefont{M.~E.} \bibnamefont{Christy}}
  \bibnamefont{\emph{et al.}}, \bibinfo{journal}{Phys. Rev.}
  \textbf{\bibinfo{volume}{C70}}, \bibinfo{pages}{015206} (\bibinfo{year}{2004}).
\bibitem[{\citenamefont{Meziane et~al.}(2011)}]{meziane11}
\bibinfo{author}{\bibfnamefont{M.}~\bibnamefont{Meziane}} \bibnamefont{\emph{et al.}},
  \bibinfo{journal}{Phys. Rev. Lett.} \textbf{\bibinfo{volume}{106}},
  \bibinfo{pages}{132501} (\bibinfo{year}{2011}).
\bibitem[{\citenamefont{Arrington}(2004)}]{arrington04b}
\bibinfo{author}{\bibfnamefont{J.}~\bibnamefont{Arrington}},
  \bibinfo{journal}{Phys. Rev.} \textbf{\bibinfo{volume}{C69}},
  \bibinfo{pages}{032201 (R)} (\bibinfo{year}{2004}).
\bibitem[{\citenamefont{Adikaram et~al.}(2015)}]{adikaram15}
\bibinfo{author}{\bibfnamefont{D.~} \bibnamefont{Adikaram}} 
  \bibnamefont{\emph{et al.}} (CLAS Collaboration),
  \bibinfo{journal}{Phys. Rev. Lett.} \textbf{\bibinfo{volume}{114}},
  \bibinfo{pages}{062003} (\bibinfo{year}{2015}).
\bibitem[{\citenamefont{Rachek et~al.}(2015)}]{rachek15}
\bibinfo{author}{\bibfnamefont{I.~A.} \bibnamefont{Rachek}}
  \bibnamefont{\emph{et al.}}, 
  \bibinfo{journal}{Phys. Rev. Lett.} \textbf{\bibinfo{volume}{114}},
  \bibinfo{pages}{062005} (\bibinfo{year}{2015}).
\bibitem[{\citenamefont{Andivahis et~al.}(1994)}]{andivahis94}
\bibinfo{author}{\bibfnamefont{L.}~\bibnamefont{Andivahis}}
  \bibnamefont{\emph{et al.}}, \bibinfo{journal}{Phys. Rev.}
  \textbf{\bibinfo{volume}{D50}}, \bibinfo{pages}{5491} (\bibinfo{year}{1994}).
\bibitem[{\citenamefont{Walker et~al.}(1994)}]{walker94}
\bibinfo{author}{\bibfnamefont{R.~C.} \bibnamefont{Walker}}
  \bibnamefont{\emph{et al.}}, \bibinfo{journal}{Phys. Rev.}
  \textbf{\bibinfo{volume}{D49}}, \bibinfo{pages}{5671} (\bibinfo{year}{1994}).
\bibitem[{\citenamefont{Qattan et~al.}(2016)\citenamefont{Qattan,Arrington,Alsaad}}]{suppl2016}
\bibinfo{howpublished}{See Supplemental Material at [to be inserted] for the proton form factors, 
and the TPE amplitudes coefficients from this work}.
\bibitem[{\citenamefont{Arrington et~al.}(2007)\citenamefont{Arrington,Melnitchouk, and Tjon}}]{arrington07}
\bibinfo{author}{\bibfnamefont{J.}~\bibnamefont{Arrington}},
  \bibinfo{author}{\bibfnamefont{W.}~\bibnamefont{Melnitchouk}},
  \bibnamefont{and} \bibinfo{author}{\bibfnamefont{J.~A.} \bibnamefont{Tjon}},
  \bibinfo{journal}{Phys. Rev.} \textbf{\bibinfo{volume}{C76}},
  \bibinfo{pages}{035205} (\bibinfo{year}{2007}).
\bibitem[{\citenamefont{Venkat et~al.}(2011)\citenamefont{Venkat,Arrington,Miller, and Zhan}}]{venkat11}
\bibinfo{author}{\bibfnamefont{S.}~\bibnamefont{Venkat}},
  \bibinfo{author}{\bibfnamefont{J.}~\bibnamefont{Arrington}},
  \bibinfo{author}{\bibfnamefont{G.~A.} \bibnamefont{Miller}}, \bibnamefont{and}
  \bibinfo{author}{\bibfnamefont{X.}~\bibnamefont{Zhan}},
  \bibinfo{journal}{Phys. Rev.} \textbf{\bibinfo{volume}{C83}},
  \bibinfo{pages}{015203} (\bibinfo{year}{2011}).
\bibitem[{\citenamefont{Alberico et~al.}(2009)\citenamefont{Alberico,Bilenky,Giunti, and Graczyk}}]{alberico09}
\bibinfo{author}{\bibfnamefont{W.}~\bibnamefont{Alberico}},
  \bibinfo{author}{\bibfnamefont{S.~M.} \bibnamefont{Bilenky}},
  \bibinfo{author}{\bibfnamefont{C.}~\bibnamefont{Giunti}}, \bibnamefont{and}
  \bibinfo{author}{\bibfnamefont{K.~M.} \bibnamefont{Graczyk}},
  \bibinfo{journal}{Phys. Rev.} \textbf{\bibinfo{volume}{C79}},
  \bibinfo{pages}{065204} (\bibinfo{year}{2009}).
\bibitem[{\citenamefont{Bernauer et~al.}(2014)}]{bernauer14}
\bibinfo{author}{\bibfnamefont{J.~C.} \bibnamefont{Bernauer}}
  \bibnamefont{\emph{et al.}} (A1 Collaboration),  
  \bibinfo{journal}{Phys. Rev.} \textbf{\bibinfo{volume}{C90}},
  \bibinfo{pages}{015206} (\bibinfo{year}{2014}). 
\bibitem[{\citenamefont{Puckett et~al.}(2012)}]{puckett12}
\bibinfo{author}{\bibfnamefont{A.~J.~R.} \bibnamefont{Puckett}} \bibnamefont{\emph{et al.}},
  \bibinfo{journal}{Phys. Rev.} \textbf{\bibinfo{volume}{C85}},
  \bibinfo{pages}{045203} (\bibinfo{year}{2012}).
\bibitem[{\citenamefont{Janssens et~al.}(1966)\citenamefont{Janssens,Hofstadter,Hughes, and Yearian}}]{janssens66}
\bibinfo{author}{\bibfnamefont{T.}~\bibnamefont{Janssens}},
  \bibinfo{author}{\bibfnamefont{R.}~\bibnamefont{Hofstadter}},
  \bibinfo{author}{\bibfnamefont{E.~B.} \bibnamefont{Hughes}},
  \bibnamefont{and} \bibinfo{author}{\bibfnamefont{M.~R.} \bibnamefont{Yearian}},
  \bibinfo{journal}{Phys. Rev.} \textbf{\bibinfo{volume}{142}},
  \bibinfo{pages}{922} (\bibinfo{year}{1966}).
\end{thebibliography}

\end{document}